\documentclass[aps,prb,twocolumn,10pt,superscriptaddress,showpacs]{revtex4-2}
\usepackage{amsmath,amssymb,graphics,epsfig,epstopdf,color,verbatim,ulem,braket,tabularx,multirow,array,diagbox,colortbl,hhline}
\usepackage[colorlinks,linkcolor=blue,citecolor=blue,urlcolor=blue]{hyperref}
\usepackage{physics}
\usepackage{bm}
\usepackage{blindtext}
\graphicspath{{Figures/}}

\begin{document}

\title{Accelerating ground-state auxiliary-field quantum Monte Carlo simulations
\\ by {\it delayed update} and {\it block force-bias update}}

\author{Hao Du}
\affiliation{Institute of Modern Physics, Northwest University, Xi'an 710127, China}

\author{Yuan-Yao He}
\email{heyuanyao@nwu.edu.cn}
\affiliation{Institute of Modern Physics, Northwest University, Xi'an 710127, China}
\affiliation{Shaanxi Key Laboratory for Theoretical Physics Frontiers, Xi'an 710127, China}
\affiliation{Peng Huanwu Center for Fundamental Theory, Xi'an 710127, China}
\affiliation{Hefei National Laboratory, Hefei 230088, China}

\begin{abstract}
Ground-state auxiliary-field quantum Monte Carlo (AFQMC) methods have become key numerical tools for studying quantum phases and phase transitions in interacting many-fermion systems. Despite the broad applicability, the efficiency of these algorithms is often limited by the bottleneck associated with the {\it local update} of the field configuration. In this work, we propose two novel update schemes, the {\it delayed update} and {\it block force-bias update}, both of which can generally and efficiently accelerate ground-state AFQMC simulations. The {\it delayed update}, with a predetermined delay rank, is an elegantly improved version of the {\it local update}, accelerating the process by replacing multiple vector-vector outer products in the latter with a single matrix-matrix multiplication. The {\it block force-bias update} is a block variant of the conventional force-bias update, which is a highly efficient scheme for dilute systems but suffers from the low acceptance ratio in lattice models. Our modified scheme maintains the high efficiency while offering flexible tuning of the acceptance ratio, controlled by the block size, for any desired fermion filling. We apply these two update schemes to both the standard and spin-orbit coupled two-dimensional Hubbard models, demonstrating their speedup over the {\it local update} with respect to the delay rank and block size. We also explore their efficiencies across varying system sizes and model parameters. Our results identify a speedup of $\sim$$8$ for systems with $1600$ lattice sites. Furthermore, we have investigated the broader applications as well as an application diagram of these update schemes to general correlated fermion systems. 
\end{abstract}

\date{\today}

\maketitle

\section{Introduction}
\label{sec:intro}

Strongly correlated fermion systems, ranging from interacting lattice models~\cite{Arovas2022,Qin2022} to correlated electron materials~\cite{Kotliar2006,Dagotto2008,Basov2011}, are at the forefront of modern research in condensed matter physics~\cite{Shaginyan2016,Armitage2020}. Owing to the nonperturbative nature, these systems typically demand highly accurate or even unbiased theoretical methods to precisely capture and characterize their complex and intricate phenomena~\cite{Rohringer2018}. Among these methods, the path-integral auxiliary-field quantum Monte Carlo (AFQMC)~\cite{Blankenbecler1981,Scalapino1981,Sugiyama1986,Sorella1989,White1989,Assaad2008,Shiwei1995,Shiwei1997,Shiwei1999,Shiwei2000,Yuanyao2019,Shiwei2003,Shihao2021,Shiwei2019Review} decouples the two-body interactions into free fermions coupled to auxiliary fields via the Hubbard-Stratonovich transformations~\cite{Hirsch1983,Assaad1998,Shihao2013,WangDa2014}, and then computes fermionic observables by Monte Carlo sampling of the field configurations~\cite{Assaad2008,Shiwei2019Review}. Specifically, AFQMC methods encompass the numerical exact Determinant Quantum Monte Carlo (DQMC)~\cite{Blankenbecler1981,Scalapino1981,Sugiyama1986,Sorella1989,White1989,Assaad2008}, and the subsequent Constrained-Path Quantum Monte Carlo (CPQMC)~\cite{Shiwei1995,Shiwei1997,Shiwei1999,Shiwei2000,Yuanyao2019,Shiwei2003,Shihao2021,Shiwei2019Review} which employs the constrained-path or phaseless approximations to control the sign or phase problem. Both DQMC and CPQMC have distinct algorithms designed for finite-temperature and ground-state simulations, respectively. The combined use of these methods offers a unified and powerful AFQMC framework~\cite{Assaad2008,Shiwei2019Review} for studying arbitrary correlated fermion systems with two-body interactions, including lattice models~\cite{LeBlanc2015,Qin2016a,*Qin2016b,Zheng2017,Qin2020,Haoxu2022,Haoxu2024,Vitali2019,*Vitali2022,Hongxia2019,*Hongxia2020}, dilute Fermi gas or electron gas~\cite{Carlson2011,Shihao2015,Ettore2017,Yuanyao2022,Joonho2021}, {\it ab initio} quantum chemistry~\cite{Shihao2021,Joonho2022,Williams2020,Motta2017}, and realistic materials~\cite{Fengjie2015,Siyuan2021,*Siyuan2023,Aouina2023,Yubo2024}. The remaining challenges primarily lie in two aspects, namely the accuracy of CPQMC methods and the computational efficiency of both DQMC and CPQMC approaches. 

While significant progress has been achieved in improving the accuracy of CPQMC methods over the past decade~\cite{Yuanyao2019,Xiao2023,Fengjie2015,Siyuan2021,*Siyuan2023,Aouina2023,Yubo2024,Qin2016a,*Qin2016b,Zheng2017,Qin2020,Haoxu2022,Haoxu2024,Shihao2017,Motta2017,Xiao2021,*Zhiyu2023,Vitali2019,*Vitali2022,Hongxia2019,*Hongxia2020}, advancements in the overall efficiency of the AFQMC framework at a fundamental algorithmic level have remained relatively limited since the 1990s~\cite{Xiaoyan2017,Yuanyao2019L,Sun2024,Fanjie2025}. The major difficulty in boosting the efficiency originates from the fact that all the aforementioned AFQMC methods involve matrices, with the corresponding field configuration weight being a matrix determinant~\cite{Assaad2008,Shiwei2019Review}. These mathematical objects are quite hard to be manipulated in numerics. This issue is especially prominent in the configuration update, which is undoubtedly the dominating portion of any Monte Carlo simulation. Correspondingly, AFQMC simulations are mostly restricted to the {\it local update}, and the universally applicable global update algorithm, as widely used in many other Monte Carlo algorithms~\cite{Swendsen1987,Wolff1989,Heringa1998,Rieger1999,Henk2002,Sandvik1999,Sandvik2002}, is still competely absent in this framework. The {\it local update} involves the successive flipping of auxiliary fields during the Markov chain process used in DQMC~\cite{White1989,Assaad2008}, whereas in CPQMC adopting the branching random walk formalism, it corresponds to the generation of fields with a constrained probability~\cite{Shiwei1997,Yuanyao2019}. Moreover, the {\it local update} part typically accounts for over $80\%$ of the total computational time in an AFQMC simulation~\cite{Sun2024}, and most of that is consumed to update the single-particle Green's function matrix. This provides a clear and valuable opportunity for generally improving the efficiency of AFQMC methods by accelerating the configuration update process.

Building on the insight, the Hirsch-Fye QMC (a specialized finite-temperature DQMC method for the Anderson impurity model) first demonstrated~\cite{Alvarez2008,Nukala2009} that the low-rank update of the Green's function, following each accepted local update, can be delayed and instead substituted with a single matrix operation once the acceptance number reaches a predetermined delay rank. In the context of Hubbard model, it corresponds to replacing multiple vector-vector outer products (DGER or ZGERU)~\cite{lapack99} with a single matrix-matrix multiplication (DGEMM or ZGEMM), leading to a significant reduction in the prefactor of the computational scaling of the local update. More recently, Sun {\it et al.}~\cite{Sun2024} extended the above delayed update scheme to DQMC simulations of general lattice models, and achieved substantial speedup over the local update, particularly within the finite-temperature algorithm. For the ground-state DQMC (also known as Projector Quantum Monte Carlo, PQMC)~\cite{Sugiyama1986,Sorella1989,White1989}, a corresponding delayed update scheme was also discussed in Ref.~\cite{Sun2024}. However, it follows the finite-temperature framework and retains the computational cost of $\mathcal{O}(N_s^3)$, which practically exceeds the $\mathcal{O}(N_sN_e^2)$ complexity of the local update in the PQMC algorithm (with $N_s$ representing the system size and $N_e$ as the number of fermions). Consequently, its efficiency is evidently limited to the special case of near half-filling~\cite{Sun2024}, where $N_e$ is comparable to $N_s$. Therefore, developing a generally applicable delayed update scheme, that can accelerate the local update while preserving the $\mathcal{O}(N_sN_e^2)$ complexity, for the ground-state AFQMC remains an open challenge.

Another important update scheme used in the ground-state AFQMC simulations is the force-bias update~\cite{Shihao2015,Shihao2021}. This method updates all the auxiliary fields on a given imaginary-time slice simultaneously, with a probability density involving a dynamic force bias, which must be computed each time before the sampling. It is then followed by the calculation of the determinant ratio (only in PQMC) and the core propagating matrix (in both PQMC and ground-state CPQMC) in the simulations. The overall complexity of this update scheme remains $\mathcal{O}(N_sN_e^2)$, but it completely avoids the low-rank matrix operations, thereby superior to the local update. Practically, the force-bias update has been extensively used as a highly efficient algorithm for dilute systems, including interacting Fermi gas~\cite{Shihao2015,Ettore2017,Yuanyao2022}, as well as molecules and realistic materials~\cite{Shiwei2003,Shihao2021,Shiwei2019Review}. However, in PQMC, its acceptance ratio is highly sensitive to the fermion filling of the system and becomes vanishingly small with growing system size for lattice models at dense fillings. Therefore, this conventional force-bias update scheme must be modified to maintain its efficiency and be applicable to simulations of general correlated fermion lattice models.

In this paper, we aim to tackle the challenges outlined above and accelerate the current ground-state AFQMC simulations. For the delayed update, we propose an alternative scheme to that in Ref.~\cite{Sun2024}, one that integrates into the ground-state AFQMC framework and preserves the computational complexity of $\mathcal{O}(N_sN_e^2)$. For the force-bias update, we develop a block variant that decomposes a single update in the conventional algorithm into multiple successive attempts, with the block size serving as the control parameter for the overall acceptance ratio. Then we systematically test these two new update algorithms in PQMC simulations of the two-dimensional (2D) Hubbard model, both in its standard form and with Rashba spin-orbit coupling (SOC). Their speedups over the local update from our results demonstrate the general efficiency of both update schemes in the lattice models across different system sizes and model parameters. Although the new update algorithms are implemented in PQMC method due to its simplicity and unbiased nature, their applications to the ground-state CPQMC method are straightforward and are also discussed in the context.

The remainder of this paper is structured as follows. In Sec.~\ref{sec:modelmethod}, we describe the lattice models used for our tests, and the formalism of PQMC method with the local update and conventional force-bias update. In Sec.~\ref{sec:NewUpdates}, we present the algorithmic details of our introduced developments: the delayed update and the block force-bias update. Section~\ref{sec:Results} is dedicated to our numerical results of the efficiency tests for both update schemes as well as physical results for the lattice models. In Sec.~\ref{sec:discussion}, we present discussions on broader applications of the update schemes in ground-state AFQMC simulations for general correlated fermion systems. Finally, we draw conclusions of our work in Sec.~\ref{sec:Summary}. The Appendixes provide PQMC formalism for spin-coupled systems, formula derivations, and the detailed data of several figures in this work.

\section{Hubbard Models and PQMC Method}
\label{sec:modelmethod}

In this section, we first describe the 2D standard Hubbard and spin-orbit coupled Hubbard (SOC-Hubbard) models in Sec.~\ref{sec:TheModel}. Then, we present the framework of PQMC method (Sec.~\ref{sec:PQMC}), the general update scheme (Sec.~\ref{sec:GeneralUpdate}), and the local update as well as conventional force-bias update (Sec.~\ref{sec:LocalFrcbias}).

\subsection{The 2D Hubbard models}
\label{sec:TheModel}

The standard Hubbard model on square lattice read as
\begin{equation}\begin{aligned}
\label{eq:2DHubbard}
\hat{H}
=\sum_{\mathbf{k}\sigma}\varepsilon_{\mathbf{k}}c_{\mathbf{k}\sigma}^+c_{\mathbf{k}\sigma}^{}
+ U\sum_{i}\Big(\hat{n}_{i\uparrow}\hat{n}_{i\downarrow} - \frac{\hat{n}_{i\uparrow} + \hat{n}_{i\downarrow}}{2}\Big),
\end{aligned}\end{equation}
and the SOC-Hubbard model~\cite{Peter2017,Yufeng2024} is described by
\begin{equation}\begin{aligned}
\label{eq:2DSocHubbard}
\hat{H}=&\sum_{\mathbf{k} \sigma}\varepsilon_{\mathbf{k}} c_{\mathbf{k}\sigma}^+c_{\mathbf{k}\sigma}^{} + \sum_{\mathbf{k}} (\mathcal{L}_{\mathbf{k}}c_{\mathbf{k}\downarrow}^+c_{\mathbf{k}\uparrow}^{} +\mathrm{H.c.}) \\
&\hspace{0.3cm}+U\sum_{i}\Big(\hat{n}_{i\uparrow}\hat{n}_{i\downarrow} - \frac{\hat{n}_{i\uparrow}+\hat{n}_{i\downarrow}}{2}\Big),
\end{aligned}\end{equation}
where $\sigma$ ($=\uparrow$ or $\downarrow$) denotes the spin and $\hat{n}_{i\sigma}=c_{i\sigma}^+c_{i\sigma}^{}$ is the density operator. The model parameters include the nearest-neighbor hopping strength $t$, the Rashba SOC strength $\lambda$, and the on-site Coulomb interaction $U$. Under periodic boundary conditions, we have the kinetic energy dispersion $\varepsilon_{\mathbf{k}}=-2t(\cos k_x + \cos k_y)$ and the SOC term $\mathcal{L}_{\mathbf{k}}=2\lambda(\sin k_y-i\sin k_x)$, where the momentum $k_x,k_y$ are defined in units of $2\pi/L$ with $L$ as the linear system size. We denote the fermion filling as $n=N_e/N_s$, with $N_s=L^2$ representing the system size and $N_e=N_{\uparrow}+N_{\downarrow}$ as the total number of fermions. In ground-state AFQMC simulations, both $N_s$ and $N_e$ (or, equivalently, $N_{\uparrow}$ and $N_{\downarrow}$) are the direct input parameters. 

We primarily use the above lattice models as test platforms to illustrate and compare the efficiency of different update schemes in PQMC simulations. Since the configuration update has little dependence on whether the interaction is repulsive or attractive, we focus on the attractive case ($U<0$) in our calculations. In this regime, both models in Eqs.~(\ref{eq:2DHubbard}) and (\ref{eq:2DSocHubbard}) are free from the minus sign problem for arbitrary filling within the PQMC method. This allows us to test the update schemes away from the half-filling, which extends beyond the scope of the previous study~\cite{Sun2024}. For $U<0$, the models (\ref{eq:2DHubbard}) and (\ref{eq:2DSocHubbard}) both exhibit superconducting ground states, with the former featuring spin-singlet $s$-wave pairing~\cite{Paiva2004,Fontenele2022} and the latter displaying a mixed-parity pairing order~\cite{Peter2017,Yufeng2024}. 

\subsection{The formalism of PQMC method}
\label{sec:PQMC}

To ensure clarity, we review the formalism of PQMC method~\cite{Sugiyama1986,Sorella1989,White1989} along with its local and force-bias updates (in following subsections), in the framework of spin-decoupled system with the standard Hubbard model in Eq.~(\ref{eq:2DHubbard}) as an representative. We then address the differences of the corresponding PQMC formalism for spin-coupled system, such as the SOC Hubbard model (\ref{eq:2DSocHubbard}), in Appendix~\ref{sec:AppendixA} to maintain the integrity of the approach.

In PQMC, the ground state of the interacting lattice model is obtained via performing the imaginary-time projection on an initial wave function $|\psi_{T}\rangle$ as
\begin{equation}\begin{aligned}
\label{eq:Projection}
|\Psi_{G}\rangle \propto \lim_{\Theta\to\infty} e^{-\Theta\hat{H}}|\psi_{T}\rangle,
\end{aligned}\end{equation}
where $\Theta$ is the projection parameter. Then the ground-state observables can be computed using the formula
\begin{equation}\begin{aligned}
\label{eq:GSObs}
\langle\hat{O}\rangle
=\frac{\langle\Psi_{G}|\hat{O}|\Psi_{G}\rangle}{\langle\Psi_{G}|\Psi_{G}\rangle}
=\lim_{\Theta\to\infty}\frac{\langle\psi_{T}|e^{-\Theta\hat{H}}\hat{O}e^{-\Theta\hat{H}}|\psi_{T}\rangle}{\langle\psi_{T}|e^{-2\Theta\hat{H}}|\psi_{T}\rangle}.
\end{aligned}\end{equation}
Practically, $|\psi_{T}\rangle$ should not be orthogonal to the many-body ground state, i.e., $\langle\Psi_{G}|\psi_{T}\rangle\ne0$, so that the projection with a large but finite $\Theta$ can asymptotically yield the ground state. It is operationally convenient to choose $|\psi_{T}\rangle$ to be a Slater determinant expressed as
\begin{equation}\begin{aligned}
\label{eq:TrialWvfc}
|\psi_{T}\rangle = \prod_{m=1}^{N_{\uparrow}}(\boldsymbol{c}_{\uparrow}^+\mathbf{P}_{\uparrow})_m|0\rangle \ \otimes\ \prod_{m=1}^{N_{\downarrow}}(\boldsymbol{c}_{\downarrow}^+\mathbf{P}_{\downarrow})_m|0\rangle,
\end{aligned}\end{equation}
where $\boldsymbol{c}_{\sigma}^+=(c_{1\sigma}^+,c_{2\sigma}^+,\cdots,c_{N_s\sigma}^+)$, and $\mathbf{P}_{\sigma}$ is an $N_s\times N_{\sigma}$ matrix. It is well suited for spin-decoupled systems, such as the standard Hubbard model~(\ref{eq:2DHubbard}). In numerical calculations, $|\psi_{T}\rangle$ is typically chosen as the ground-state Slater determinant wave function of the noninteracting part of $\hat{H}$ or its mean-field Hamiltonian~\cite{Vitali2019,Otsuka2016}. Alternative choices for $|\psi_{T}\rangle$ include the multideterminant wave function~\cite{Shihao2014} and the Gutzwiller projected variational wave function~\cite{Chang2024}.

To proceed with Eq.~(\ref{eq:GSObs}), we start from the discretization of $2\Theta=M\Delta\tau$ for the ``effective'' partition function as $Z=\langle\psi_{T}|e^{-2\Theta\hat{H}}|\psi_{T}\rangle=\langle\psi_{T}|(e^{-\Delta\tau\hat{H}})^M|\psi_{T}\rangle$, and then apply the Trotter-Suzuki decomposition for $e^{-\Delta\tau\hat{H}}$, such as the second-order formula
\begin{equation}\begin{aligned}
\label{eq:AsymTrot}
e^{-\Delta\tau\hat{H}}=e^{-\Delta\tau\hat{H}_I}e^{-\Delta\tau\hat{H}_0}+\mathcal{O}[(\Delta\tau)^2],
\end{aligned}\end{equation}
where $\hat{H}=\hat{H}_0+\hat{H}_I$, with $\hat{H}_0$ and $\hat{H}_I$ representing the kinetic and interaction parts, respectively. The Trotter error $\mathcal{O}[(\Delta\tau)^2]$ originates from $[\hat{H}_0,\hat{H}_I]\ne0$, and can be eliminated by extrapolating several calculations towards $\Delta\tau\to0$. Then the Hubbard-Stratonovich (HS) transformation, generally expressed as $e^{-\Delta\tau\hat{H}_I}=\sum_{\mathbf{x}}p(\mathbf{x})\hat{B}_I(\mathbf{x})$, is applied to transform the many-body operator $e^{-\Delta\tau\hat{H}_I}$ into a single-particle propagator $\hat{B}_I(\mathbf{x})$, which is formulated as free fermions coupled to auxiliary fields $\mathbf{x}=(x_1,x_2,\cdots,x_{N_f})$ (with $N_f$ comparable to $N_s$). Combining these two steps, the $e^{-\Delta\tau\hat{H}}$ operator at the $\ell$-th time slice can be written as
\begin{equation}\begin{aligned}
\label{eq:ExpDtH}
e^{-\Delta\tau\hat{H}(\ell)} 
\simeq \sum_{\mathbf{x}_{\ell}}p(\mathbf{x}_{\ell})\hat{B}(\mathbf{x}_{\ell}),
\end{aligned}\end{equation}
where $\hat{B}(\mathbf{x}_{\ell})=\hat{B}_I(\mathbf{x}_{\ell})e^{-\Delta\tau\hat{H}_0}$, and the Trotter error is indicated by the ``$\simeq$'' symbol from now on. Then, it is evident that the expectation in $Z=\langle\psi_{T}|(e^{-\Delta\tau\hat{H}})^M|\psi_{T}\rangle$ involves only single-particle propagators, and thus it can be evaluated explicitly to yield $Z\simeq\sum_{\mathbf{X}}W(\mathbf{X})$ with $W(\mathbf{X})$ as the configuration weight, given by~\cite{Assaad2008}
\begin{equation}\begin{aligned}
\label{eq:ConfgWeight}
W(\mathbf{X}) 
&= P(\mathbf{X})\langle\psi_T|\hat{B}(\mathbf{x}_M)\cdots\hat{B}(\mathbf{x}_{\ell})\cdots\hat{B}(\mathbf{x}_1)|\psi_T\rangle \\
&= P(\mathbf{X})\prod_{\sigma=\uparrow,\downarrow}\det(\mathbf{P}_{\sigma}^+\mathbf{B}_{M}^{\sigma}\cdots\mathbf{B}_{\ell}^{\sigma}\cdots\mathbf{B}_{1}^{\sigma}\mathbf{P}_{\sigma}),
\end{aligned}\end{equation}
where $\mathbf{X}=(\mathbf{x}_1,\cdots,\mathbf{x}_{\ell},\cdots,\mathbf{x}_M)$ is a complete auxiliary-field configuration, and $P(\mathbf{X})=\prod_{\ell=1}^M p(\mathbf{x}_{\ell})$ is a probability density. The propagator matrix $\mathbf{B}_{\ell}^{\sigma}$ takes the form 
\begin{equation}\begin{aligned}
\label{eq:BlMat}
\mathbf{B}_{\ell}^{\sigma}=e^{\mathbf{H}_I^{\sigma}(\mathbf{x}_{\ell})}e^{-\Delta\tau\mathbf{H}_0^{\sigma}}
\end{aligned}\end{equation}
where $\mathbf{H}_0^{\sigma}$ is the hopping matrix of the kinetic part as $\hat{H}_0=\sum_{ij,\sigma}(\mathbf{H}_0^{\sigma})_{ij}c_{i\sigma}^+c_{j\sigma}^{}$, and the interaction-dependent matrix $\mathbf{H}_I^{\sigma}(\mathbf{x}_{\ell})$ is the  embedded in the $\hat{B}_I(\mathbf{x}_{\ell})$ operator as $\hat{B}_I(\mathbf{x}_{\ell})=\exp\{\sum_{ij,\sigma}[\mathbf{H}_I^{\sigma}(\mathbf{x}_{\ell})]_{ij}c_{i\sigma}^+c_{j\sigma}^{}\}$. All of $\mathbf{B}_{\ell}^{\sigma}$, $\mathbf{H}_0^{\sigma}$, and $\mathbf{H}_I^{\sigma}(\mathbf{x})$ are $N_s\times N_s$ matrices. 

After the above procedures, Eq.~(\ref{eq:GSObs}) can be reformulated as
\begin{equation}\begin{aligned}
\label{eq:GSObsQMC}
\langle\hat{O}\rangle
= \frac{\sum_{\mathbf{X}}W(\mathbf{X})O(\mathbf{X})}{\sum_{\mathbf{X}^{\prime}}W(\mathbf{X}^{\prime})}
= \sum_{\mathbf{X}}\omega(\mathbf{X})O(\mathbf{X}),
\end{aligned}\end{equation}
where $\omega(\mathbf{X})$ is the normalized probability density used for importance sampling, as $\omega(\mathbf{X})=W(\mathbf{X})/\sum_{\mathbf{X}^{\prime}}W(\mathbf{X}^{\prime})$. The measurement of $\hat{O}$ for a given configuration $\mathbf{X}$, denoted as $O(\mathbf{X})$, has the expression
\begin{equation}\begin{aligned}
\label{eq:Oxformula}
O(\mathbf{X}) 
= \frac{\langle\psi_T|\prod_{\ell_2=\ell+1}^M \hat{B}(\mathbf{x}_{\ell_2}) \hat{O} \prod_{\ell_1=1}^{\ell}\hat{B}(\mathbf{x}_{\ell_1})|\psi_T\rangle}{\langle\psi_T|\prod_{\ell^{\prime}=1}^M \hat{B}(\mathbf{x}_{\ell^{\prime}})|\psi_T\rangle}.
\end{aligned}\end{equation}
In practical simulations, the measurement does not need to be taken exactly at the midpoint of the path [i.e., $\ell\Delta\tau=\Theta$ in Eq.~(\ref{eq:Oxformula}) following Eq.~(\ref{eq:GSObs})]. As long as both $\ell\Delta\tau$ and $(M-\ell)\Delta\tau$ are sufficiently large to ensure convergence to the ground state in the projection [Eq.~(\ref{eq:Projection})], it is reasonable to perform the measurement using Eq.~(\ref{eq:Oxformula}). A key quantity is the static (equal-time) single-particle Green's function matrix $\mathbf{G}^{\sigma}(\ell\Delta\tau,\ell\Delta\tau)$, whose elements correspond to the observable $\hat{O}=c_{i\sigma}^{}c_{j\sigma}^{+}$, and it can be evaluated~\cite{Assaad2008} with Eq.~(\ref{eq:Oxformula}) as
\begin{equation}\begin{aligned}
\label{eq:GrFMat}
\mathbf{G}^{\sigma}(\ell\Delta\tau,\ell\Delta\tau)=\mathbf{1}_{N_s}-\mathbf{R}^{\sigma}(\mathbf{L}^{\sigma}\mathbf{R}^{\sigma})^{-1}\mathbf{L}^{\sigma},
\end{aligned}\end{equation}
where $\mathbf{L}^{\sigma}=\mathbf{P}_{\sigma}^+\mathbf{B}_{M}^{\sigma}\mathbf{B}_{M-1}^{\sigma}\cdots\mathbf{B}_{\ell+1}^{\sigma}$ is an $N_{\sigma}\times N_s$ matrix, $\mathbf{R}^{\sigma}=\mathbf{B}_{\ell}^{\sigma}\cdots\mathbf{B}_2^{\sigma}\mathbf{B}_1^{\sigma}\mathbf{P}_{\sigma}$ is an $N_s\times N_{\sigma}$ matrix, and $\mathbf{1}_{N_s}$ as the $N_s\times N_s$ identity matrix. Then the measurement of all the static single-particle and two-particle observables can be calculated directly or via Wick decomposition from $\mathbf{G}^{\sigma}(\ell\Delta\tau,\ell\Delta\tau)$. Equation~(\ref{eq:GSObsQMC}) is then used to compute the Monte Carlo average, followed by standard statistical analysis to estimate the associated uncertainty.

We next discuss the realization of above PQMC formalism in the standard Hubbard model~(\ref{eq:2DHubbard}). For the Hubbard interaction $\hat{H}_I=U\sum_{i}[\hat{n}_{i\uparrow}\hat{n}_{i\downarrow}-(\hat{n}_{i\uparrow}+\hat{n}_{i\downarrow})/2]$, there are many applicable HS transformations~\cite{Hirsch1983,Assaad1998,Shihao2013,WangDa2014}, and we adopt the one decoupled into spin-$\hat{s}^z$ channel with discrete auxiliary fields~\cite{Hirsch1983} (denoted as HS-$\hat{s}^z$) reading as
\begin{equation}\begin{aligned}
\label{eq:HSspinDecomp}
e^{-\Delta\tau U \big(\hat{n}_{i\uparrow} \hat{n}_{i\downarrow} - \frac{\hat{n}_{i\uparrow} + \hat{n}_{i\downarrow}}{2}\big) }
= \frac{1}{2}\sum_{x_{i}=\pm1}e^{{\rm i}\gamma x_{i}(\hat{n}_{i\uparrow}-\hat{n}_{i\downarrow})},
\end{aligned}\end{equation}
with the coupling coefficient $\gamma=\cos^{-1}(e^{\Delta\tau U/2})$ for $U<0$. This formula decouples the interaction into the spin-$\hat{s}^z$ channel. Our previous studies~\cite{Yufeng2024,Xie2025} have shown that such HS transformation into spin channels can substantially reduce the statistical fluctuations of pairing-related observables, comparing to that decoupled into charge-density channel~\cite{Hirsch1983}. With Eq.~(\ref{eq:HSspinDecomp}), the probability density $p(\mathbf{x}_{\ell})$ in Eq.~(\ref{eq:ExpDtH}) is a constant as $p(\mathbf{x}_{\ell})=1/2^{N_s}$, and the propagator $\hat{B}_I(\mathbf{x}_{\ell})$ can be expressed as
\begin{equation}\begin{aligned}
\label{eq:BIOpForHubbard}
\hat{B}_I(\mathbf{x}_{\ell})
= {\rm exp}\Big[{\rm i}\gamma\sum_{i=1}^{N_s} x_{\ell,i}(\hat{n}_{i\uparrow}-\hat{n}_{i\downarrow})\Big]
= \prod_{i=1}^{N_s}\hat{b}_i(x_{\ell,i}),
\end{aligned}\end{equation}
with $\hat{b}_i(x_{\ell,i})=e^{{\rm i}\gamma x_{\ell,i}(\hat{n}_{i\uparrow}-\hat{n}_{i\downarrow})}$. As a result, the corresponding matrix $\mathbf{H}_I^{\sigma}(\mathbf{x}_{\ell})$ becomes diagonal as $\mathbf{H}_I^{\sigma}(\mathbf{x}_{\ell})={\rm i}\gamma f_{\sigma}\times{\rm Diag}(x_{\ell,1},x_{\ell,2},\cdots,x_{\ell,N_s})$ with $f_{\uparrow}=+1$ and $f_{\downarrow}=-1$. Then the equality $\mathbf{B}_{\ell}^{\downarrow}=(\mathbf{B}_{\ell}^{\uparrow})^*$ follows from Eq.~(\ref{eq:BlMat}), given the relations $\mathbf{H}_I^{\downarrow}(\mathbf{x}_{\ell})=[\mathbf{H}_I^{\uparrow}(\mathbf{x}_{\ell})]^*$ and $\mathbf{H}_0^{\downarrow}=\mathbf{H}_0^{\uparrow}$ (both are real). Consequently, under additional conditions of spin balance ($N_{\uparrow}=N_{\downarrow}$) and $\mathbf{P}_{\downarrow}=(\mathbf{P}_{\uparrow})^*$ in $|\psi_{T}\rangle$ [see Eq.~(\ref{eq:TrialWvfc})], the spin-up and spin-down determinants in $W(\mathbf{X})$ from Eq.~(\ref{eq:ConfgWeight}) form a complex conjugate pair, which ensures the absence of the sign problem for the model~(\ref{eq:2DHubbard}) with $U<0$. In this situation, we only need to involve the spin-up channel in the simulation, as all matrices in spin-down sector can be readily obtained by taking the complex conjugate of their spin-up counterparts. 

We further remark on several implementation issues in PQMC simulations for the model~(\ref{eq:2DHubbard}). {\it First}, we use the asymmetric Trotter-Suzuki decomposition [see Eq.~(\ref{eq:AsymTrot})] for its simplicity to demonstrate the algorithmic details. But it is straightforward to apply the symmetric one, i.e., $e^{-\Delta\tau\hat{H}}=e^{-\frac{\Delta\tau}{2}\hat{H}_0}e^{-\Delta\tau\hat{H}_I}e^{-\frac{\Delta\tau}{2}\hat{H}_0}+\mathcal{O}[(\Delta\tau)^3]$ (see details in Appendix~\ref{sec:AppendixB}). {\it Second}, a numerical stablization procedure is applied during propagation (sweeping the time slices from $\ell=1$ to $\ell=M$ and back in reverse) periodically, i.e., every $k_{\tau}$ time slices. It eliminates the numetical instability and essentially performs the reorthogonalization of single-particle orbitals~\cite{Shiwei2019Review}. For example, within QR algorithm, $\mathbf{R}^{\sigma}$ and $\mathbf{L}^{\sigma}$ in Eq.~(\ref{eq:GrFMat}) are factorized as $\mathbf{R}^{\sigma}=\mathbf{U}_{R}^{\sigma}\mathbf{S}_{R}$ and $\mathbf{L}^{\sigma}=\mathbf{S}_{L}\mathbf{U}_{L}^{\sigma}$, where $\mathbf{U}_{R}^{\sigma}$ ($\mathbf{U}_{L}^{\sigma}$) is an $N_s\times N_{\sigma}$ ($N_{\sigma}\times N_s$) matrix with orthonormal columns (rows), and $\mathbf{S}_{R}$, $\mathbf{S}_{L}$ are $N_{\sigma}\times N_{\sigma}$ square matrices. Then the $\mathbf{G}^{\sigma}(\ell\Delta\tau,\ell\Delta\tau)$ matrix in Eq.~(\ref{eq:GrFMat}) can be written as
\begin{equation}\begin{aligned}
\label{eq:GrFMatNew}
\mathbf{G}^{\sigma}(\ell\Delta\tau,\ell\Delta\tau)=\mathbf{1}_{N_s}-\mathbf{U}_{R}^{\sigma}\mathbf{T}^{\sigma}\mathbf{U}_{L}^{\sigma},
\end{aligned}\end{equation}
where $\mathbf{T}^{\sigma}=(\mathbf{U}_{L}^{\sigma}\mathbf{U}_{R}^{\sigma})^{-1}$ is an $N_{\sigma}\times N_{\sigma}$ square matrix. {\it Third}, we keep track of the matrices $\mathbf{U}_{L}^{\sigma}$, $\mathbf{U}_{R}^{\sigma}$ and $\mathbf{T}^{\sigma}$ in PQMC simulations, and the propagation of $\mathbf{L}^{\sigma}$ and $\mathbf{R}^{\sigma}$ are replaced respectively by $\mathbf{U}_{L}^{\sigma}$ and $\mathbf{U}_{R}^{\sigma}$. Furthermore, $\mathbf{T}^{\sigma}$ is upgraded upon accepting a configuration update, and $\mathbf{G}^{\sigma}(\ell\Delta\tau,\ell\Delta\tau)$ only needs to be computed [using Eq.~(\ref{eq:GrFMatNew})] at each measurement step. {\it Fourth}, the fast Fourier transform (FFT) is applied to $\mathbf{U}_{L}^{\sigma}$ and $\mathbf{U}_{R}^{\sigma}$ during propagation to switch between real and momentum spaces, ensuring that the one-body propagator matrices are diagonal, i.e., $e^{\pm\Delta\tau\mathbf{H}_0}$ in momentum space and $e^{\pm\mathbf{H}_I^{\sigma}(\mathbf{x}_{\ell})}$ in real space. As a result, the computational complexity of evaluating the products $e^{\pm\Delta\tau\mathbf{H}_0}\mathbf{U}_{R}^{\sigma}$ and $\mathbf{U}_{L}^{\sigma}e^{\pm\Delta\tau\mathbf{H}_0}$ is significantly reduced from $\mathcal{O}(N_s^2N_{\sigma})$ using standard matrix-matrix multiplication to $\mathcal{O}(N_sN_{\sigma}\ln N_s)$ via FFT~\cite{FFTnote}. This approach also facilitates easy access to the Green's function in momentum space~\cite{Yuanyao2022,Song2025}. With all above techniques, the core components of PQMC algorithm (matrix propagation, numerical stablization, and configuration update) has the overall computational complexity of $\mathcal{O}(\Theta N_sN_e^2)$ (considering $N_{\sigma}=N_e/2$), i.e., scaling linearly with $\Theta$ and $N_s$ (since $\ln N_s\ll N_s$)~\cite{Shihao2015}. Besides, each measurement costs at least $\mathcal{O}(N_s^2N_e)$ due to the calculation of the full $\mathbf{G}^{\sigma}(\ell\Delta\tau,\ell\Delta\tau)$ matrix. Furthermore, we note that PQMC algorithm can suffer from the infinite variance problem~\cite{Shihao2016Inf,Wan2025}, even in sign-problem-free simulations. Nevertheless, this issue mainly concerns the measurement of physical observables, while its solution involving a bridge link method does not affect the auxiliary-field update process discussed in this work.

\subsection{General view of the configuration update}
\label{sec:GeneralUpdate}

Considersing the absence of a global update (or space-time cluster update) scheme in PQMC method, here we present the general paradigm for configuration updates on a specific time slice. This includes the calculation of the weight ratio and the update of $\mathbf{T}^{\sigma}=(\mathbf{U}_{L}^{\sigma}\mathbf{U}_{R}^{\sigma})^{-1}$ and $\mathbf{U}_{R}$ matrices during the update. 

We consider updating a configuration $\mathbf{X}$ to a new one as $\mathbf{X}^{\prime}$, and they differ only at the $\ell$-th time slice, i.e., $\mathbf{x}_{\ell}$ in $\mathbf{X}$ and $\mathbf{x}_{\ell}^{\prime}$ in $\mathbf{X}^{\prime}$. We apply a generalized Metropolis acceptance probability from the detailed balance as
\begin{equation}\begin{aligned}
\label{eq:DetailBalance}
\mathcal{A}(\mathbf{X}\to\mathbf{X}^{\prime})
=\min\Big\{1,\frac{W(\mathbf{X}^{\prime})}{W(\mathbf{X})}\frac{\mathcal{P}(\mathbf{X}^{\prime}\to\mathbf{X})}{\mathcal{P}(\mathbf{X}\to\mathbf{X}^{\prime})}\Big\},
\end{aligned}\end{equation}
where $\mathcal{P}(\mathbf{X}\to\mathbf{X}^{\prime})$ is the prior probability for proposing the update of the fields, and it can be carefully designed to achieve high acceptance ratio. So our focus is to compute the weight ratio $W(\mathbf{X}^{\prime})/W(\mathbf{X})$. From Eq.~(\ref{eq:ConfgWeight}), the fraction $P(\mathbf{X}^{\prime})/P(\mathbf{X})$ can be readily obtained, and the ratio of the determinants as $R=r_{\uparrow}r_{\downarrow}$ is the key with $r_{\sigma}$ given by
\begin{equation}\begin{aligned}
\label{eq:DetRatio00}
r_{\sigma} 
&= \frac{\det[\mathbf{P}_{\sigma}^+\mathbf{B}_{M}^{\sigma}\cdots\mathbf{B}_{\ell+1}^{\sigma}(\mathbf{B}_{\ell}^{\sigma})^{\prime}\mathbf{B}_{\ell-1}^{\sigma}\cdots\mathbf{B}_{1}^{\sigma}\mathbf{P}_{\sigma}]}{{\rm det}(\mathbf{P}_{\sigma}^+\mathbf{B}_{M}^{\sigma}\cdots\mathbf{B}_{\ell+1}^{\sigma}\mathbf{B}_{\ell}^{\sigma}\mathbf{B}_{\ell-1}^{\sigma}\cdots\mathbf{B}_{1}^{\sigma}\mathbf{P}_{\sigma})} \\
&= \frac{\det[\mathbf{L}^{\sigma}(\mathbf{1}_{N_s}+\boldsymbol{\Delta})\mathbf{R}^{\sigma}]}{{\rm det}(\mathbf{L}^{\sigma}\mathbf{R}^{\sigma})},
\end{aligned}\end{equation}
where $(\mathbf{B}_{\ell}^{\sigma})^{\prime}=e^{\mathbf{H}_I^{\sigma}(\mathbf{x}_{\ell}^{\prime})}e^{-\Delta\tau\mathbf{H}_0^{\sigma}}$, $\mathbf{B}_{\ell}^{\sigma}=e^{\mathbf{H}_I^{\sigma}(\mathbf{x}_{\ell})}e^{-\Delta\tau\mathbf{H}_0^{\sigma}}$, and the matrices $\mathbf{L}^{\sigma}$, $\mathbf{R}^{\sigma}$ defined in Eq.~(\ref{eq:GrFMat}) are used. We have introduced $\boldsymbol{\Delta}^{\sigma}=e^{\mathbf{H}_I^{\sigma}(\mathbf{x}_{\ell}^{\prime})}e^{-\mathbf{H}_I^{\sigma}(\mathbf{x}_{\ell})}-\mathbf{1}_{N_s}$, which accounts for $(\mathbf{B}_{\ell}^{\sigma})^{\prime}=(\mathbf{1}_{N_s}+\boldsymbol{\Delta}^{\sigma})\mathbf{B}_{\ell}^{\sigma}$. With further simplification (see Appendix~\ref{sec:AppendixC}), the determinant ratio in Eq.~(\ref{eq:DetRatio00}) can be finally expressed as
\begin{equation}\begin{aligned}
\label{eq:DetRatio10}
r_{\sigma} 
&= \frac{{\rm det}\big[\mathbf{U}_L^{\sigma}(\mathbf{1}_{N_s}+\boldsymbol{\Delta}^{\sigma})\mathbf{U}_R^{\sigma}\big]}{{\rm det}\big(\mathbf{U}_L^{\sigma}\mathbf{U}_R^{\sigma}\big)} \\
&= {\rm det}\big(\mathbf{1}_{N_{\sigma}}+\mathbf{U}_L^{\sigma}\boldsymbol{\Delta}^{\sigma}\mathbf{U}_R^{\sigma}\mathbf{T}^{\sigma}\big),
\end{aligned}\end{equation}
where $\mathbf{T}^{\sigma}=(\mathbf{U}_{L}^{\sigma}\mathbf{U}_{R}^{\sigma})^{-1}$, and $\mathbf{1}_{N_{\sigma}}$ is the $N_{\sigma}\times N_{\sigma}$ identity matrix. Once the above proposed update is accepted, we need to renew both $\mathbf{U}_{R}^{\sigma}$ and $\mathbf{T}^{\sigma}$ matrices using
\begin{equation}\begin{aligned}
\label{eq:UpdateLTRMat}
(\mathbf{U}_{R}^{\sigma})^{\prime} 
&= \big(\mathbf{1}_{N_s}+\boldsymbol{\Delta}^{\sigma}\big)\mathbf{U}_{R}^{\sigma}, \\
(\mathbf{T}^{\sigma})^{\prime} 
&= \mathbf{T}^{\sigma}\big(\mathbf{1}_{N_{\sigma}}+\mathbf{U}_L^{\sigma}\boldsymbol{\Delta}^{\sigma}\mathbf{U}_R^{\sigma}\mathbf{T}^{\sigma}\big)^{-1},
\end{aligned}\end{equation}
with its proof summarized in Appendix~\ref{sec:AppendixC}. Equations~(\ref{eq:DetRatio10}) and~(\ref{eq:UpdateLTRMat}) are the key results of the configuration update process. In the following discussions of specific update schemes, we concentrate on specializations and simplifications of these equations.

\subsection{The local update and full force-bias update}
\label{sec:LocalFrcbias}

In the {\it local update}, we typically update the auxiliary fields $\mathbf{x}_{\ell}=(x_{\ell,1},x_{\ell,2},\cdots,x_{\ell,N_s})$ sequentially, element by element. Thus, we do not need to involve the prior probability $\mathcal{P}(\mathbf{X}\to\mathbf{X}^{\prime})$ in Eq.~(\ref{eq:DetailBalance}). For the $i$-th lattice site, the single-step local update flips $x_{\ell,i}$ (from $+1$ to $-1$, or vice versa), and the $\boldsymbol{\Delta}^{\sigma}$ matrix in Eqs.~(\ref{eq:DetRatio10}) and~(\ref{eq:UpdateLTRMat}) only has one nonzero element, i.e., $\Delta_{i,i}^{\sigma}=e^{+{\rm i}\gamma f_{\sigma}(x_{\ell,i}^{\prime}-x_{\ell,i})}-1$ ($f_{\uparrow}=+1$ and $f_{\downarrow}=-1$). We introduce the row vector $\mathbf{u}_i^{\rm T}=(\mathbf{U}_R^{\sigma})_{i{\rm-row}}$ (``$i$-row'' as the $i$-th row, and the superscript ``T'' meaning transpose) and the column vector $\mathbf{v}_i=\Delta_{i,i}^{\sigma}(\mathbf{U}_L^{\sigma})_{i{\rm-col}}$ (``$i$-col'' as the $i$-th column). Then the ratio in Eq.~(\ref{eq:DetRatio10}) can be expressed as
\begin{equation}\begin{aligned}
\label{eq:DetRatiolocal}
r_{i,\sigma} = {\rm det}\big(\mathbf{1}_{N_{\sigma}} + \mathbf{v}_i\mathbf{y}^{\rm T}\big) = 1 + \mathbf{y}^{\rm T}\mathbf{v}_i,
\end{aligned}\end{equation}
with $\mathbf{y}^{\rm T}=\mathbf{u}_i^{\rm T}\mathbf{T}^{\sigma}$ as a row vector. If the update is accepted, $\mathbf{T}^{\sigma}$ is upgraded to $(\mathbf{T}^{\sigma})^{\prime}$ following Eq.~(\ref{eq:UpdateLTRMat}) as
\begin{equation}\begin{aligned}
\label{eq:UpdateLTRlocal}
(\mathbf{T}^{\sigma})^{\prime} 
= \mathbf{T}^{\sigma}\big(\mathbf{1}_{N_{\sigma}} + \mathbf{v}_i\mathbf{y}^{\rm T}\big)^{-1}
= \mathbf{T}^{\sigma} - \mathbf{x}\mathbf{y}^{\rm T},
\end{aligned}\end{equation}
with $\mathbf{x}=(\mathbf{T}^{\sigma}\mathbf{v}_i)/r_{i,\sigma}$ as a column vector. The Sherman-Morrison formula is applied in the second equality of Eq.~(\ref{eq:UpdateLTRlocal}) (see Appendix~\ref{sec:AppendixC}). The calculations of $\mathbf{x}$ and $\mathbf{y}^{\rm T}$ involve the matrix-vector multiplications (ZGEMV)~\cite{lapack99}, while $\mathbf{x}\mathbf{y}^{\rm T}$ corresponds to a vector-vector outer product (ZGERU). They consist of the leading computational complexity of the update process, as $\mathcal{O}(3N_{\sigma}^2)$. These operations related to vectors are responsible for the low efficiency of the local update. Besides, updating $\mathbf{U}_{R}^{\sigma}$ matrix, as $(\mathbf{U}_{R}^{\sigma})^{\prime}=\big(\mathbf{1}_{N_s}+\boldsymbol{\Delta}^{\sigma}\big)\mathbf{U}_{R}^{\sigma}$, only changes its $i$-th row. For the whole $\ell$-th time slice, we simply iterate the site index from $i=1$ to $i=N_s$ and repeat the above calculations, which overall contributes to a $\mathcal{O}(3N_sN_{\sigma}^2)$ complexity. This local update scheme is also referred to as the {\it fast update}~\cite{Sun2024}.

The {\it force-bias update} originates from the insight of an alternative way to compute $W(\mathbf{X}^{\prime})/W(\mathbf{X})$. We introduce the Slater determinant wave functions
\begin{equation}\begin{aligned}
\label{eq:FullFrcbsPhiLPhiR}
\langle\psi_l| &= \langle\psi_T|\hat{B}(\mathbf{x}_M)\hat{B}(\mathbf{x}_{M-1})\cdots\hat{B}(\mathbf{x}_{\ell+1}),  \\
|\psi_r\rangle &= [\hat{B}_I(\mathbf{x}_{\ell})]^{-1}\hat{B}(\mathbf{x}_{\ell})\cdots\hat{B}(\mathbf{x}_2)\hat{B}(\mathbf{x}_1)|\psi_T\rangle,
\end{aligned}\end{equation}
and from Eq.~(\ref{eq:ConfgWeight}), $W(\mathbf{X}^{\prime})/W(\mathbf{X})$ can be reformulated as
\begin{equation}\begin{aligned}
\frac{W(\mathbf{X}^{\prime})}{W(\mathbf{X})}
= \frac{\langle\psi_l|\hat{B}_I(\mathbf{x}_{\ell}^{\prime})|\psi_r\rangle}{\langle\psi_l|\psi_r\rangle} \times \Big(\frac{\langle\psi_l|\hat{B}_I(\mathbf{x}_{\ell})|\psi_r\rangle}{\langle\psi_l|\psi_r\rangle}\Big)^{-1},
\end{aligned}\end{equation}
in which the ratio $\langle\psi_l|\hat{B}_I(\mathbf{x})|\psi_r\rangle/\langle\psi_l|\psi_r\rangle$ (with $\mathbf{x}=\mathbf{x}_{\ell}$ or $\mathbf{x}_{\ell}^{\prime}$) can be taken as the measurement of the observable $\hat{O}=\hat{B}_I(\mathbf{x})$, similar to that in Eq.~(\ref{eq:Oxformula}). For the model~(\ref{eq:2DHubbard}) with the $\hat{B}_I(\mathbf{x})$ in Eq.~(\ref{eq:BIOpForHubbard}), the coupling coefficient follows $\gamma=\cos^{-1}(e^{\Delta\tau U/2})=\sqrt{-\Delta\tau U}+\mathcal{O}[(-\Delta\tau U)^{3/2}]$, and thus the ratio can be computed as
\begin{equation}\begin{aligned}
\label{eq:RatioApprox}
\frac{\langle\psi_l|\hat{B}_I(\mathbf{x})|\psi_r\rangle}{\langle\psi_l|\psi_r\rangle} = e^{{\rm i}\gamma\sum_{i=1}^{N_s} x_{i}(\bar{n}_{i\uparrow}-\bar{n}_{i\downarrow})} + \mathcal{O}(\Delta\tau U f_{\mathbf{x}}),
\end{aligned}\end{equation}
where the force bias $\bar{n}_{i\sigma}=\langle\hat{n}_{i\sigma}\rangle$ is defined as
\begin{equation}\begin{aligned}
\label{eq:FrcBias}
\bar{n}_{i\sigma} = \frac{\langle\psi_l|\hat{n}_{i\sigma}|\psi_r\rangle}{\langle\psi_l|\psi_r\rangle},
\end{aligned}\end{equation}
and $f_{\mathbf{x}}=(\langle\hat{v}^2\rangle-\langle\hat{v}\rangle^2)/2$ is half the variance of the operator $\hat{v}={\rm i}\sum_{i=1}^{N_s} x_i(\hat{n}_{i\uparrow}-\hat{n}_{i\downarrow})$ (see Appendix~\ref{sec:AppendixC}). The $\bar{n}_{i\sigma}$ in Eq.~(\ref{eq:FrcBias}) is actually a measurement of the $\hat{n}_{\sigma}$ operator. Then, a straightforward choice for the prior probability in Eq.~(\ref{eq:DetailBalance}) is $\mathcal{P}(\mathbf{X}\to\mathbf{X}^{\prime})=e^{{\rm i}\gamma\sum_{i=1}^{N_s} x_{\ell,i}^{\prime}(\bar{n}_{i\uparrow}-\bar{n}_{i\downarrow})}$, which only depends on the new configuration $\mathbf{X}^{\prime}$. This induces the relation
\begin{equation}\begin{aligned}
\label{eq:AcceptRatio}
\frac{W(\mathbf{X}^{\prime})}{W(\mathbf{X})}\frac{\mathcal{P}(\mathbf{X}^{\prime}\to\mathbf{X})}{\mathcal{P}(\mathbf{X}\to\mathbf{X}^{\prime})}
= 1 + \mathcal{O}\big[\Delta\tau U (f_{\mathbf{x}_{\ell}^{\prime}}-f_{\mathbf{x}_{\ell}})\big],
\end{aligned}\end{equation}
which explicitly illustrates that the acceptance ratio applying Eq.~(\ref{eq:DetailBalance}) for this update asymptotically reaches unity towards the $\Delta\tau\to0$ limit. 

In practical simulations, the force-bias update proceeds through the following steps. {\it First}, according to Eq.~(\ref{eq:GrFMatNew}), the force bias $\bar{n}_{i\sigma}$ in Eq.~(\ref{eq:FrcBias}) can be evaluated from the diagonal element of the corresponding Green's function, i.e., $\bar{n}_{i\sigma}=[\mathbb{U}_{R}^{\sigma}(\mathbf{U}_{L}^{\sigma}\mathbb{U}_{R}^{\sigma})^{-1}\mathbf{U}_{L}^{\sigma}]_{ii}$ with $\mathbb{U}_{R}^{\sigma}=e^{-\mathbf{H}_I^{\sigma}(\mathbf{x}_{\ell})}\mathbf{U}_{R}^{\sigma}$ (Note that the calculation of $\mathbb{U}_{R}^{\sigma}$ can be integrated into the propagation of $\mathbf{U}_{R}^{\sigma}$ matrix). This step accounts for a $\mathcal{O}(2N_sN_{\sigma}^2+N_sN_{\sigma}+N_{\sigma}^3)$ complexity. {\it Second}, the new auxiliary fields $\mathbf{x}_{\ell}^{\prime}$ are sampled using $\mathcal{P}(\mathbf{X}\to\mathbf{X}^{\prime})=\prod_i\bar{b}_i(x_{\ell,i}^{\prime})$ with $\bar{b}_i(x_{\ell,i}^{\prime})=e^{{\rm i}\gamma x_{\ell,i}^{\prime}(\bar{n}_{i\uparrow}-\bar{n}_{i\downarrow})}$. More specifically, we construct a normalized probability density function $\lambda(x_{\ell,i}^{\prime})=\bar{b}_i(x_{\ell,i}^{\prime})/\sum_{x_{\ell,i}^{\prime}}\bar{b}_i(x_{\ell,i}^{\prime})$, and then sample $x_{\ell,i}^{\prime}$ from $\lambda(x_{\ell,i}^{\prime})$ for all the components (from $i=1$ to $i=N_s$). {\it Third}, the acceptance probability in Eq.~(\ref{eq:DetailBalance}) is evaluated to determine whether to accept the update. The ratio of the prior probability is evident as $\mathcal{P}(\mathbf{X}^{\prime}\to\mathbf{X})/\mathcal{P}(\mathbf{X}\to\mathbf{X}^{\prime})=\prod_i e^{{\rm i}\gamma (x_{\ell,i} - x_{\ell,i}^{\prime})(\bar{n}_{i\uparrow}-\bar{n}_{i\downarrow})}$. For the weight ratio $W(\mathbf{X}^{\prime})/W(\mathbf{X})$, we can directly compute $\det[\mathbf{U}_L^{\sigma}(\mathbf{1}_{N_s}+\boldsymbol{\Delta}^{\sigma})\mathbf{U}_R^{\sigma}]=\det[\mathbf{U}_L^{\sigma}(\mathbf{U}_R^{\sigma})^{\prime}]$ to evaluate the $r_{\sigma}$ using the first equality of Eq.~(\ref{eq:DetRatio10}), since ${\rm det}(\mathbf{U}_L^{\sigma}\mathbf{U}_R^{\sigma})$ is stored and updated along with the propagation and update. Note that the matrix inverse $(\mathbf{T}^{\sigma})^{\prime}=[\mathbf{U}_L^{\sigma}(\mathbf{U}_R^{\sigma})^{\prime}]^{-1}$ and its determinant can be computed simultaneously using ZGETRF~\cite{lapack99}. This step has a $\mathcal{O}(N_sN_{\sigma}^2+N_sN_{\sigma}+N_{\sigma}^3)$ complexity. Combining all three procedures, the overall computational scaling of the force-bias update for a single time slice is $\mathcal{O}(3N_sN_{\sigma}^2+2N_sN_{\sigma}+2N_{\sigma}^3)$.

From above description, the {\it force-bias update} simultaneously updates all the auxiliary fields on a given time slice within a single Metropolis acceptance-rejection step. Thus, it can be regarded as a real-space global update algorithm for PQMC method. It can achieve high efficiency due to the fact that it completely avoids any operations involving vectors. However, as shown by Eqs.~(\ref{eq:RatioApprox}) and (\ref{eq:AcceptRatio}), the acceptance ratio of this update scheme also depends on $f_{\mathbf{x}}$ (with $\mathbf{x}=\mathbf{x}_{\ell}$ or $\mathbf{x}_{\ell}^{\prime}$), which is a function of the fermion filling $n$ and system size $N_s$ (see Appendix~\ref{sec:AppendixC}). For dilute systems (with tiny $n$), $f_{\mathbf{x}}$ remains relatively small even for large $N_s$, allowing the force-bias update to sustain a high acceptance ratio. Nevertheless, at dense fillings, $f_{\mathbf{x}}$ becomes significant and grows with increasing $N_s$, leading to a progressively lower acceptance ratio that eventually approaches zero. A similar issue arises for both dilute and dense systems when the interaction strength $U$ is very large. While reducing the imaginary-time step $\Delta\tau$ can help alleviate this problem, it substantially increases the computational cost of the simulations. Throughout the remainder of this work, we refer to this conventional update scheme as the full force-bias update. 

\section{The delayed update and block force-bias update}
\label{sec:NewUpdates}

In this section, we present our new developments of the delayed update (Sec.~\ref{sec:Delayed}) and block force-bias update (Sec.~\ref{sec:BlockForcebias}) schemes, following the notations and formulas introduced in Sec.~\ref{sec:GeneralUpdate} and Sec.~\ref{sec:LocalFrcbias}. 

\subsection{The delayed update}
\label{sec:Delayed}

In local update scheme, upgrading the $\mathbf{T}^{\sigma}$ matrix using Eq.~(\ref{eq:UpdateLTRlocal}) is the most time-consuming step. Accordingly, the delayed update is specifically designed to mitigate the cost of this operation. It computes the determinant ratio $r_{i,\sigma}$ as usual, but delays the upgrade of $\mathbf{T}^{\sigma}$ matrix until a predetermined delay rank $n_d$ is reached. 

In following illustration of our delayed update scheme, we use $(\mathbf{T}^{\sigma})_{[0]}$ and $(\mathbf{T}^{\sigma})_{[k]}$ (with $k\ge1$) to denote the initial $\mathbf{T}^{\sigma}$ matrix before the update and the new $\mathbf{T}^{\sigma}$ after the $k$-th accepted update, respectively. The target of the delayed update is to achieve the relation
\begin{equation}\begin{aligned}
\label{eq:DelayTMat00}
(\mathbf{T}^{\sigma})_{[k]} = (\mathbf{T}^{\sigma})_{[0]} - \sum_{m=1}^k \mathbf{x}_m\mathbf{y}_m^{\rm T},
\end{aligned}\end{equation}
for which we need to figure out the explicit expressions of the column vector $\mathbf{x}_m$ and the row vector $\mathbf{y}_m^{\rm T}$ using $\mathbf{U}_L^{\sigma}$, $\mathbf{U}_R^{\sigma}$ and $(\mathbf{T}^{\sigma})_{[0]}$. For $k=1$ case, the delayed update of $\mathbf{T}^{\sigma}$ in Eq.~(\ref{eq:DelayTMat00}) degenerates into the local update as in Eq.~(\ref{eq:UpdateLTRlocal}). More generally, after $(k-1)$ accepted update moves, the current $\mathbf{T}^{\sigma}$ matrix becomes $(\mathbf{T}^{\sigma})_{[k-1]}$. We proceed to examine the update of the auxiliary field at the $i_k$-th lattice site (denoted as $x_{\ell,i_k}$) with the following procedures. We first flip $x_{\ell,i_k}$, and compute the corresponding determinant ratio as $r_{i_k,\sigma}=1+\mathbf{y}_k^{\rm T}\mathbf{v}_{i_k}$, where $\mathbf{y}_k^{\rm T}=\mathbf{u}_{i_k}^{\rm T}(\mathbf{T}^{\sigma})_{[k-1]}$ and $\mathbf{x}_k=(\mathbf{T}^{\sigma})_{[k-1]}\mathbf{v}_{i_k}/r_{i_k,\sigma}$, with $\mathbf{u}_{i_k}^{\rm T}=(\mathbf{U}_R^{\sigma})_{i_k{\rm-row}}$ and $\mathbf{v}_{i_k}=\Delta_{i_k,i_k}^{\sigma}(\mathbf{U}_L^{\sigma})_{i_k{\rm-col}}$. Applying a relation analogous to Eq.~(\ref{eq:DelayTMat00}) between $(\mathbf{T}^{\sigma})_{[k-1]}$ and $(\mathbf{T}^{\sigma})_{[0]}$, we can reach a recursive formula for the ratio as
\begin{equation}\begin{aligned}
\label{eq:DelayRatio}
r_{i_k,\sigma} = 1 + \mathbf{u}_{i_k}^{\rm T}(\mathbf{T}^{\sigma})_{[0]}\mathbf{v}_{i_k} - \sum_{m=1}^{k-1}\big(\mathbf{u}_{i_k}^{\rm T}\mathbf{x}_m\big)\big(\mathbf{y}_m^{\rm T}\mathbf{v}_{i_k}\big).
\end{aligned}\end{equation}
Then, we evaluate the ratio of the configuration weights as $W(\mathbf{X}^{\prime})/W(\mathbf{X})=[P(\mathbf{X}^{\prime})/P(\mathbf{X})]\times r_{i_k,\uparrow}r_{i_k,\downarrow}$, and decide whether to accept this proposed update according to the detailed balance condition in Eq.~(\ref{eq:DetailBalance}) [Note that, in both the local and delayed updates, we choose the prior probability that satisfies $\mathcal{P}(\mathbf{X}\to\mathbf{X}^{\prime})=\mathcal{P}(\mathbf{X}^{\prime}\to\mathbf{X})$]. If this update move is accepted, it is counted as the $k$-th accepted update, and the corresponding vectors $\mathbf{x}_k$ and $\mathbf{y}_k^{\rm T}$ in Eq.~(\ref{eq:DelayTMat00}) need to be calculated and stored. Similarly, we can express $\mathbf{x}_k$ and $\mathbf{y}_k^{\rm T}$ using $\mathbf{U}_L^{\sigma}$, $\mathbf{U}_R^{\sigma}$ and $(\mathbf{T}^{\sigma})_{[0]}$ as
\begin{equation}\begin{aligned}
\label{eq:DelayRgenk}
\mathbf{x}_k &= \frac{1}{r_{i_k,\sigma}}\Big[ (\mathbf{T}^{\sigma})_{[0]}\mathbf{v}_{i_k} - \sum_{m=1}^{k-1}\mathbf{x}_m\big(\mathbf{y}_m^{\rm T}\mathbf{v}_{i_k}\big) \Big], \\
\mathbf{y}_k^{\rm T} &= \mathbf{u}_{i_k}^{\rm T}(\mathbf{T}^{\sigma})_{[0]} - \sum_{m=1}^{k-1}\big(\mathbf{u}_{i_k}^{\rm T}\mathbf{x}_m\big)\mathbf{y}_m^{\rm T},
\end{aligned}\end{equation}
where $c_m=(\mathbf{u}_{i_k}^{\rm T}\mathbf{x}_m)$ and $d_m=(\mathbf{y}_m^{\rm T}\mathbf{v}_{i_k})$ are scalars obtained from inner products between the vectors, and both $(\mathbf{T}^{\sigma})_{[0]}\mathbf{v}_{i_k}$ and $\mathbf{u}_{i_k}^{\rm T}(\mathbf{T}^{\sigma})_{[0]}$ involve matrix-vector multiplications. On the other hand, if the proposed update for $x_{\ell,i_k}$ is rejected, we simply proceed to examine the update of $x_{\ell,i_k+1}$, during which Eq.~(\ref{eq:DelayRatio}) is used to compute $r_{i_k+1,\sigma}$.
Comparing to the local update, this delayed update scheme requires additional calculations of $c_m$, $d_m$, and $c_m\mathbf{y}_m^{\rm T}$, $d_m\mathbf{x}_m$ (with $m\in[1,k-1]$), which feature a $\mathcal{O}[4(k-1)N_{\sigma}]$ complexity for $\mathbf{x}_k$ and $\mathbf{y}_k^{\rm T}$ vectors. Besides, the update of $\mathbf{U}_{R}^{\sigma}$ following an accepted flip of $x_{\ell,i_k}$, as $(\mathbf{U}_{R}^{\sigma})^{\prime}=\big(\mathbf{1}_{N_s}+\boldsymbol{\Delta}^{\sigma}\big)\mathbf{U}_{R}^{\sigma}$, only modifies the $i_k$-th row of the $\mathbf{U}_{R}^{\sigma}$ matrix. As a result, it does not interfere with subsequent updates that involve other rows of $\mathbf{U}_{R}^{\sigma}$.

In practical simulations, there are two different strategies to set the predetermined delay rank $n_d$ in above delayed update scheme for ground-state AFQMC methods. {\it First}, we can set $n_d$ as the total number of proposed update moves, while the number of accepted updates is $n_{\rm A}$, with $n_{\rm A}\le n_d$ and $R=n_{\rm A}/n_d$ representing the acceptance ratio. Once $n_d$ update attempts have been processed, the $\mathbf{T}^{\sigma}$ matrix is updated using an improved version of Eq.~(\ref{eq:DelayTMat00}) as 
\begin{equation}\begin{aligned}
\label{eq:FinalDelay}
(\mathbf{T}^{\sigma})_{[n_{\rm A}]} = (\mathbf{T}^{\sigma})_{[0]} - \sum_{m=1}^{n_{\rm A}} \mathbf{x}_m\mathbf{y}_m^{\rm T} = (\mathbf{T}^{\sigma})_{[0]} - \mathbf{X}\mathbf{Y},
\end{aligned}\end{equation}
where $\mathbf{X}=(\mathbf{x}_1|\mathbf{x}_2|\cdots|\mathbf{x}_{n_{\rm A}})$ is an $N_{\sigma}\times n_{\rm A}$ matrix, and $\mathbf{Y}=(\mathbf{y}_1|\mathbf{y}_2|\cdots|\mathbf{y}_{n_{\rm A}})^{\rm T}$ is an $n_{\rm A}\times N_{\sigma}$ matrix. This update can then be efficiently evaluated using a matrix-matrix multiplication for $\mathbf{X}\mathbf{Y}$. Moreover, in this case, the matrix-vector multiplications $(\mathbf{T}^{\sigma})_{[0]}\mathbf{v}_{i}$ (for $\mathbf{x}_m$) and $\mathbf{u}_{i}^{\rm T}(\mathbf{T}^{\sigma})_{[0]}$ (for $\mathbf{y}_m^{\rm T}$) in Eq.~(\ref{eq:DelayRgenk}) can be prepared in advance for each delayed update, and replaced by matrix-matrix multiplications as $(\mathbf{T}^{\sigma})_{[0]}\mathbf{V}$ and $\mathbf{U}(\mathbf{T}^{\sigma})_{[0]}$, where $\mathbf{V}$ and $\mathbf{U}$ are $N_{\sigma}\times n_d$ and $n_d\times N_{\sigma}$ matrices, respectively, constructed from the $n_d$ column (row) vectors of $\mathbf{v}_{i}$ ($\mathbf{u}_{i}^{\rm T}$) involved in the delayed update. Nevertheless, this improvement can introduce redundant calculations of $\mathbf{u}_{i}^{\rm T}(\mathbf{T}^{\sigma})_{[0]}$ for rejected updates [note that $(\mathbf{T}^{\sigma})_{[0]}\mathbf{v}_{i}$ is required in all update moves to evaluate $r_{i,\sigma}$]. Therefore, this strategy is better suited for AFQMC simulations in which the update has a high acceptance ratio. These matrix multiplications, including $(\mathbf{T}^{\sigma})_{[0]}\mathbf{V}$, $\mathbf{U}(\mathbf{T}^{\sigma})_{[0]}$, and $\mathbf{XY}$, have a leading computational complexity of $\mathcal{O}[(2n_d+n_{\rm A})N_{\sigma}^2]$. In addition, the extra calculations involving $\sum_{m}c_m\mathbf{y}_m^{\rm T}$ and $\sum_{m}d_m\mathbf{x}_m$ in Eq.~(\ref{eq:DelayRgenk}) contribute to a subleading cost of $\mathcal{O}(2n_{\rm A}(n_{\rm A}-1)N_{\sigma})$~\cite{SubleadNote}. {\it Second}, we can set $n_d$ as the accumulated number of accepted moves, which is also used in the delayed update for the finite-temperature AFQMC methods~\cite{Sun2024}. Once $n_d$ accepted updates have been accumulated, the $\mathbf{T}^{\sigma}$ matrix is updated using $(\mathbf{T}^{\sigma})_{[n_d]} = (\mathbf{T}^{\sigma})_{[0]} - \mathbf{X}\mathbf{Y}$, where $\mathbf{X}$ and $\mathbf{Y}$ are $N_{\sigma}\times n_d$ and $n_d \times N_{\sigma}$ matrices, respectively, constructed in the same way as in the first case. However, in this case, computing $(\mathbf{T}^{\sigma})_{[0]}\mathbf{v}_{i_k}$ and $\mathbf{u}_{i_k}^{\rm T}(\mathbf{T}^{\sigma})_{[0]}$ via matrix-vector multiplications is unavoidable, which is typically less efficient. Accordingly, the computational complexity is $\mathcal{O}(2n_{\rm try} N_{\sigma}^2+n_d N_{\sigma}^2)$ with $n_{\rm try}$ denoting the number of attempted moves in a single-step delayed update. In parallel, the evaluations of $\sum_{m}c_m\mathbf{y}_m^{\rm T}$ and $\sum_{m}d_m\mathbf{x}_m$ in Eq.~(\ref{eq:DelayRgenk}) requires a subleading cost of $\mathcal{O}(2n_d(n_d-1)N_{\sigma})$~\cite{SubleadNote}. 

For both strategies, the update of all auxiliary fields on a given time slice can be divided into multiple successive delayed updates, such that $\sum n_d=N_s$ in the first strategy and $\sum n_{\rm try}=N_s$ in the second. Consequently, the total computational cost of our delayed update scheme scales as $\mathcal{O}(N_s N_{\sigma}^2)$, matching that of the conventional local update scheme. In our PQMC simulations, we observe that, applying the discrete HS transformation with a two-component field [e.g., Eq.~(\ref{eq:HSspinDecomp})] for the Hubbard interaction, consistently yields an acceptance ratio above 0.8 for both local and delayed update schemes. Thus, we adopt the first strategy to set $n_d$, as the total number of proposed update moves in a single-step delayed update.

To conclude, the delay technique replaces the multiple vector-vector outer products in local update into a single matrix-matrix multiplication during upgrading the $\mathbf{T}^{\sigma}$ matrix, thereby accelerating the simulation. While this approach slightly complicates the computation of $r_{i,\sigma}$ in Eq.~(\ref{eq:DelayRatio}) and the vectors $\mathbf{x}$, $\mathbf{y}^{\rm T}$ in Eq.~(\ref{eq:DelayRgenk}), leading to a marginal increase in their computational cost, the overall delayed update scheme can still achieve a significant speedup over the conventional local update.

In contrast, the delayed update for PQMC algorithm presented in Ref.~\cite{Sun2024} is a direct implementation of the correspondence in finite-temperature DQMC. In that approach, the determinant ratio $r_{i,\sigma}$ is evaluated from the single-particle Green's function matrix $\mathbf{G}^{\sigma}$ [see Eq.~(\ref{eq:GrFMatNew})], and the delay technique is similarly applied to $\mathbf{G}^{\sigma}$ as $(\mathbf{G}^{\sigma})_{[n_d]} = (\mathbf{G}^{\sigma})_{[0]} - \mathbf{X}\mathbf{Y}$, where $\mathbf{X}$ and $\mathbf{Y}$ are $N_s\times n_d$ and $n_d \times N_s$ matrices, respectively. Consequently, its computational cost scales as $\mathcal{O}(N_s^3)$, given that $\sum n_d$$\sim$$N_s$ holds for a single time slice. Therefore, our delayed update scheme is evidently more advantageous.

\subsection{The block force-bias update}
\label{sec:BlockForcebias}

In the full force-bias update described in Sec.~\ref{sec:LocalFrcbias}, the acceptance ratio can be significantly suppressed with increasing fermion filling $n$ and system size $N_s$. This dependence is implicitly embedded in the factor $f_{\mathbf{x}}$, as revealed in Eqs.~(\ref{eq:RatioApprox}) and (\ref{eq:AcceptRatio}). A straightforward way to fix this issue is to reduce $f_{\mathbf{x}}$, which can be achieved by dividing the full force-bias update, that contains all $N_s$ lattice sites on a given time slice, into multiple smaller attempts, each involving $n_b$ sites. This leads to the block force-bias update scheme, where $n_b$ is the block size. When $n_b = N_s$, the block scheme reduces to the full force-bias update, but with a difference in the computational effort.

To illustrate the algorithmic details, we focus on the first block force-bias update on the $\ell$-th time slice, which involves lattice sites indexed from $i=1$ to $i=n_b$. With the same $\langle\psi_l|$ as that in Eq.~(\ref{eq:FullFrcbsPhiLPhiR}) and a modified $|\psi_r\rangle$ as
\begin{equation}\begin{aligned}
\label{eq:BlockFrcbsPhiLPhiR}
|\psi_r\rangle = [\hat{B}_I^{(b)}(\mathbf{x}_{\ell})]^{-1} \hat{B}(\mathbf{x}_{\ell})\cdots\hat{B}(\mathbf{x}_2)\hat{B}(\mathbf{x}_1)|\psi_T\rangle,
\end{aligned}\end{equation}
the weight ratio for the update can be formulated as 
\begin{equation}\begin{aligned}
\label{eq:BlockFrcbsRatio}
\frac{W(\mathbf{X}^{\prime})}{W(\mathbf{X})}
= \frac{\langle\psi_l|\hat{B}_I^{(b)}(\mathbf{x}_{\ell}^{\prime})|\psi_r\rangle}{\langle\psi_l|\psi_r\rangle} \times \Big(\frac{\langle\psi_l|\hat{B}_I^{(b)}(\mathbf{x}_{\ell})|\psi_r\rangle}{\langle\psi_l|\psi_r\rangle}\Big)^{-1},
\end{aligned}\end{equation}
and the block propagator $\hat{B}_I^{(b)}(\mathbf{x}_{\ell})$ is given by
\begin{equation}\begin{aligned}
\hat{B}_I^{(b)}(\mathbf{x}_{\ell})
= {\rm exp}\Big[{\rm i}\gamma\sum_{i=1}^{n_b} x_{\ell,i}(\hat{n}_{i\uparrow}-\hat{n}_{i\downarrow})\Big].
\end{aligned}\end{equation}
Similar to Eq.~(\ref{eq:RatioApprox}), we can evaluate the ratio in Eq.~(\ref{eq:BlockFrcbsRatio}) to the first order of $\Delta\tau$ as
\begin{equation}\begin{aligned}
\label{eq:BlkFrcBsErr}
\frac{\langle\psi_l|\hat{B}_I^{(b)}(\mathbf{x})|\psi_r\rangle}{\langle\psi_l|\psi_r\rangle} = e^{{\rm i}\gamma\sum_{i=1}^{n_b} x_{i}(\bar{n}_{i\uparrow}-\bar{n}_{i\downarrow})} + \mathcal{O}[\Delta\tau U f_{\mathbf{x}}^{(b)}],
\end{aligned}\end{equation}
where $\bar{n}_{i\sigma}=\langle\hat{n}_{i\sigma}\rangle=\langle\psi_l|\hat{n}_{i\sigma}|\psi_r\rangle/\langle\psi_l|\psi_r\rangle$ is the force bias, and the factor $f_{\mathbf{x}}^{(b)}=(\langle\hat{v}_b^2\rangle-\langle\hat{v}_b\rangle^2)/2$ ($\mathbf{x}=\mathbf{x}_{\ell}$ or $\mathbf{x}_{\ell}^{\prime}$) with $\hat{v}_b={\rm i}\sum_{i=1}^{n_b} x_i(\hat{n}_{i\uparrow}-\hat{n}_{i\downarrow})$. Accordingly, the prior probability $\mathcal{P}(\mathbf{X}\to\mathbf{X}^{\prime})=e^{{\rm i}\gamma\sum_{i=1}^{n_b} x_{\ell,i}^{\prime}(\bar{n}_{i\uparrow}-\bar{n}_{i\downarrow})}$ is applied to propose the updates $x_{\ell,i}\to x_{\ell,i}^{\prime}$ ($1\le i\le n_b$) by sampling each $x_{\ell,i}^{\prime}$. To proceed and decide whether the block update should be accepted, we still need to evaluate $\bar{n}_{i\sigma}$ and $W(\mathbf{X}^{\prime})/W(\mathbf{X})$ [or equivalently, the determinant ratio $r_{\sigma}$ in Eq.~(\ref{eq:DetRatio10})]. The force bias can be computed via $\bar{n}_{i\sigma}=[\mathbb{U}_{R}^{\sigma}(\mathbf{U}_{L}^{\sigma}\mathbb{U}_{R}^{\sigma})^{-1}\mathbf{U}_{L}^{\sigma}]_{ii}$ with $\mathbb{U}_{R}^{\sigma}=(\mathbf{1}_{N_s}+\boldsymbol{\Delta}_b^{\sigma})\mathbf{U}_{R}^{\sigma}$, where the diagonal matrix $\boldsymbol{\Delta}_b^{\sigma}$ hosts $(\boldsymbol{\Delta}_b^{\sigma})_{i,i}=e^{-{\rm i}\gamma f_{\sigma}x_{\ell,i}}-1$ ($f_{\uparrow}=+1$ and $f_{\downarrow}=-1$) for $1\le i\le n_b$ and zero otherwise. The inverse matrix $(\mathbf{U}_{L}^{\sigma}\mathbb{U}_{R}^{\sigma})^{-1}$ can be efficiently computed from $\mathbf{T}^{\sigma}=(\mathbf{U}_{L}^{\sigma}\mathbf{U}_{R}^{\sigma})^{-1}$ using Eq.~(\ref{eq:UpdateLTRMat}) as
\begin{equation}\begin{aligned}
\label{eq:BlkFrcBsTnew1}
(\mathbf{U}_{L}^{\sigma}\mathbb{U}_{R}^{\sigma})^{-1} = \mathbf{T}^{\sigma} - (\mathbf{T}^{\sigma}\mathbf{V})[\mathbf{1}_{n_b}+(\mathbf{U}\mathbf{T}^{\sigma})\mathbf{V}]^{-1}(\mathbf{U}\mathbf{T}^{\sigma}),
\end{aligned}\end{equation}
where $\mathbf{V}$ and $\mathbf{U}$ are $N_{\sigma}\times n_b$ and $n_b\times N_{\sigma}$ matrices, respectively, constructed from the $n_b$ column and row vectors of $\mathbf{v}_{i}=(\boldsymbol{\Delta}_b^{\sigma})_{i,i}(\mathbf{U}_L^{\sigma})_{i{\rm-col}}$ and $\mathbf{u}_i^{\rm T}=(\mathbf{U}_R^{\sigma})_{i{\rm-row}}$ (with $1\le i\le n_b$) involved in the block update. The matrix version of Sherman-Morrison formula has been employed in achieving Eq.~(\ref{eq:BlkFrcBsTnew1}) (see Appendix~\ref{sec:AppendixC}). Combining these calculations, the computational complexity for evaluating $\bar{n}_{i\sigma}$ is $\mathcal{O}(4n_bN_{\sigma}^2+2n_b^2N_{\sigma}+n_b^3+3n_bN_{\sigma})$. Besides, the determinant ratio for the block update follows the second equality of Eq.~(\ref{eq:DetRatio10}) as $r_{\sigma}={\rm det}(\mathbf{1}_{N_{\sigma}}+\mathbf{U}_L^{\sigma}\boldsymbol{\Delta}^{\sigma}\mathbf{U}_R^{\sigma}\mathbf{T}^{\sigma})$ with the diagonal matrix $\boldsymbol{\Delta}^{\sigma}$ hosting $\Delta_{i,i}^{\sigma}=e^{+{\rm i}\gamma f_{\sigma}(x_{\ell,i}^{\prime}-x_{\ell,i})}-1$ for $1\le i\le n_b$ and zero otherwise. Since the newly sampled field $x_{\ell,i}^{\prime}$ [drawn from $\mathcal{P}(\mathbf{X}\to\mathbf{X}^{\prime})$] is independent of $x_{\ell,i}$, we denote $n_{\rm F}$ (satisfying $n_{\rm F}\le n_b$) as the the number of flipped fields ($x_{\ell,i}^{\prime}\ne x_{\ell,i}$) in the block update, with their corresponding lattice sites indexed as $i_k$ with $1\le k\le n_{\rm F}$. Consequently, $\boldsymbol{\Delta}^{\sigma}$ has only $n_{\rm F}$ nonzero diagonal elements, allowing the ratio to be further simplified as $r_{\sigma}={\rm det}[\mathbf{1}_{n_{\rm F}}+(\mathbf{U}^{\prime}\mathbf{T}^{\sigma})\mathbf{V}^{\prime}]$, where $\mathbf{U}^{\prime}$ and $\mathbf{V}^{\prime}$ are constructed respectively from the $n_{\rm F}$ column and row vectors of $\mathbf{v}_{i_k}=\boldsymbol{\Delta}^{\sigma}_{i_k,i_k}(\mathbf{U}_L^{\sigma})_{i_k{\rm-col}}$ and $\mathbf{u}_{i_k}^{\rm T}=(\mathbf{U}_R^{\sigma})_{i_k{\rm-row}}$. If the block update is accepted, the $\mathbf{T}^{\sigma}$ matrix needs to be upgraded by
\begin{equation}\begin{aligned}
\label{eq:BlkFrcBsTnew2}
(\mathbf{T}^{\sigma})^{\prime} = \mathbf{T}^{\sigma} - (\mathbf{T}^{\sigma}\mathbf{V}^{\prime})[\mathbf{1}_{n_{\rm F}}+(\mathbf{U}^{\prime}\mathbf{T}^{\sigma})\mathbf{V}^{\prime}]^{-1}(\mathbf{U}^{\prime}\mathbf{T}^{\sigma}).
\end{aligned}\end{equation}
It is evident that $\mathbf{U}^{\prime}$ and $\mathbf{V}^{\prime}$ are submatrices of $\mathbf{U}$ and $\mathbf{V}$ in Eq.~(\ref{eq:BlkFrcBsTnew1}), except for the difference of $(\boldsymbol{\Delta}_b^{\sigma})_{i_k,i_k}$ in $\mathbf{V}$ and $\boldsymbol{\Delta}^{\sigma}_{i_k,i_k}$ in $\mathbf{V}^{\prime}$, which can be easily handled in practical calculations. Thus, $\mathbf{T}^{\sigma}\mathbf{V}^{\prime}$ and $\mathbf{U}^{\prime}\mathbf{T}^{\sigma}$ can be directly obtained from $\mathbf{T}^{\sigma}\mathbf{V}$ and $\mathbf{U}\mathbf{T}^{\sigma}$ evaluated in Eq.~(\ref{eq:BlkFrcBsTnew1}) without requiring additional computations. This results in a complexity of $\mathcal{O}(n_{\rm F}N_{\sigma}^2+n_{\rm F}^2N_{\sigma}+n_{\rm F}^3)$ for calculating $r_{\sigma}$ and upgrading $\mathbf{T}^{\sigma}$. Combining these two parts, the overall and leading computational cost for a single block update is $\mathcal{O}[(4n_b+n_{\rm F})N_{\sigma}^2+(2n_b^2+n_{\rm F}^2)N_{\sigma}]$.

After completing the first block force-bias update, the second one on the $\ell$-th time slice follows, involving lattice sites from $i=n_b+1$ to $i=2n_b$, and the same procedures are repeated to compute $\bar{n}_{i\sigma}$, $r_{\sigma}$ and upgrade $\mathbf{T}^{\sigma}$. This process continues iteratively, with successive block updates applied sequentially until all fields on the $\ell$-th time slice have been visited. Taking all block updates into account, the leading computational complexity for updating all auxiliary fields on a given time slice is $\mathcal{O}(N_sN_{\sigma}^2)$, consistent with the local, full force-bias, and delayed update schemes. Meanwhile, the factor $f_{\mathbf{x}}^{(b)}$ in Eq.~(\ref{eq:BlkFrcBsErr}), which governs the acceptance ratio of the block force-bias update, depends on $n_b$ rather than the system size for fixed physical parameters (interaction $U$ and filling $n$). Therefore, $n_b$ can be tuned to achieve high acceptance ratio even for large system sizes. 

\section{Numerical results}
\label{sec:Results}

In this section, we focus on PQMC results for the models (\ref{eq:2DHubbard}) and (\ref{eq:2DSocHubbard}) applying different update schemes. In Secs.~\ref{sec:SpeedDelayed} and~\ref{sec:SpeedFrcBs}, we present the efficiency tests for the delayed update and block force-bias update in the standard Hubbard model (\ref{eq:2DHubbard}), highlighting the speedups achieved over the local update as a function of system size at representative fermion fillings. In Sec.~\ref{sec:SOCHubbardTest}, we further test these update schemes in the SOC-Hubbard model (\ref{eq:2DSocHubbard}). Then in Sec.~\ref{sec:PhysicalResults}, we display consistent physical results obtained using different update schemes in PQMC simulations for the 2D attractive Hubbard model, providing benchmarks that may be useful for future studies. 

\begin{figure}[t]
\centering
\includegraphics[width=0.981\linewidth]{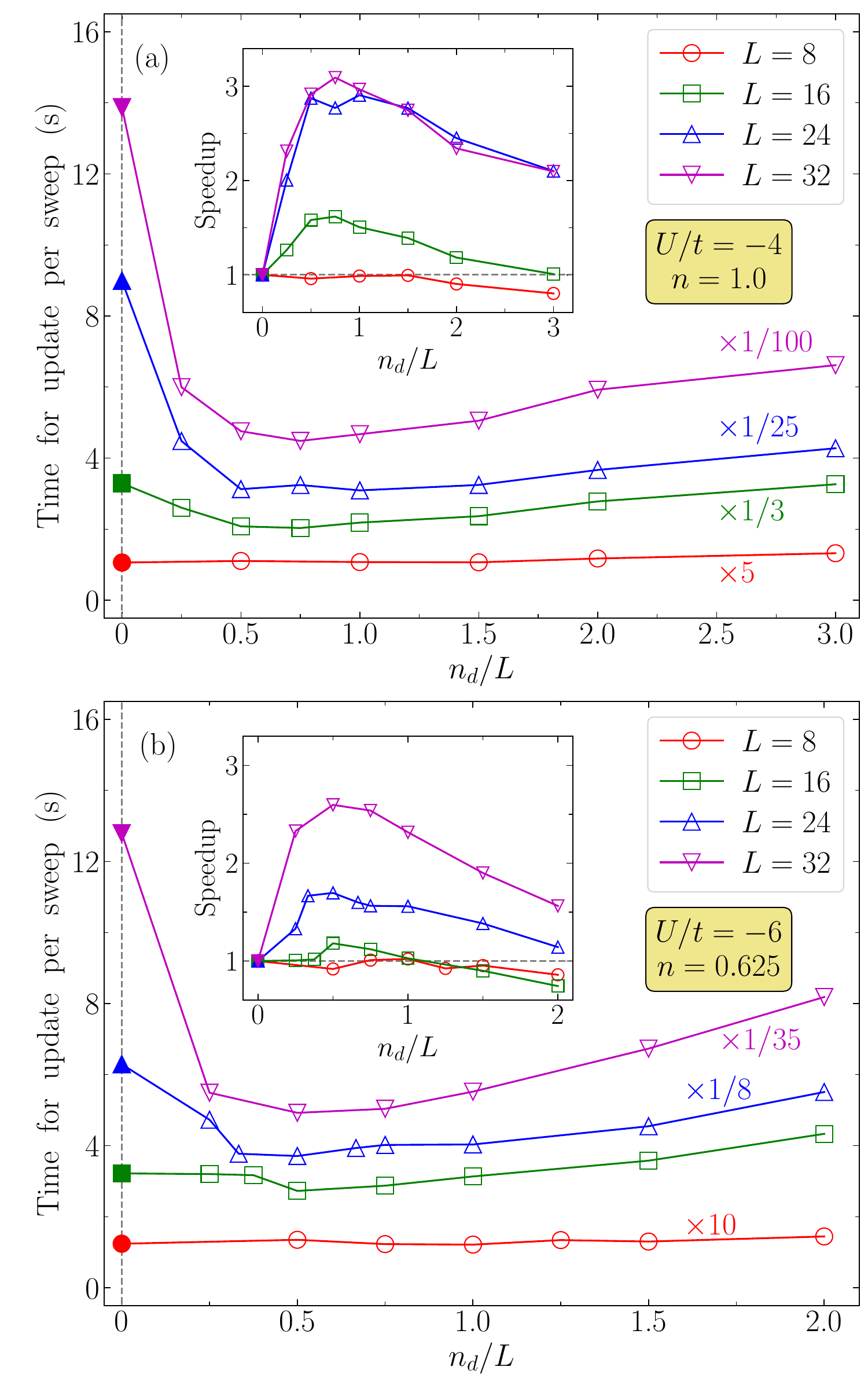}
\caption{The average time for update per sweep (in seconds) using the delayed update as a function of $n_d/L$ (with $n_d$ as the predetermined delay rank), in PQMC simulations of the 2D Hubbard model~(\ref{eq:2DHubbard}) with (a) $U/t=-4,n=1.0$, and (b) $U/t=-6,n=0.625$. The time data are rescaled by factors $\times 5,\times 1/3,\times 1/25,\times 1/100$ for $L=8,16,24,32$ in (a), and by $\times 10,\times 1/8,\times 1/35$ for $L=8,24,32$ in (b). The leftmost data points (solid symbols) at $n_d/L=0$ are results from the conventional local update scheme. The insets in both panels illustrate corresponding speedups achieved by the delayed update compared to the local update. The timing data are summarized in Tables~\ref{Table:A1} and \ref{Table:A2} in Appendix~\ref{sec:AppendixE}. }
\label{fig:Delayedndtest}
\end{figure}

We characterize the efficiency of the delayed and block force-bias update schemes by measuring the average time consumed for the update per sweep. Here, a sweep refers to a complete forward and backward scan over all auxiliary fields from $\ell = 1$ to $\ell = M$ time slices. The speedup relative to the local update is quantified as the ratio of the respective consumed time. All the simulations were performed on Intel CPUs (see Appendix~\ref{sec:AppendixD}), and additional tests using AMD CPUs produced comparable speedups.

\subsection{Efficiency tests for the delayed update}
\label{sec:SpeedDelayed}

\begin{figure}[t]
\centering
\includegraphics[width=0.990\linewidth]{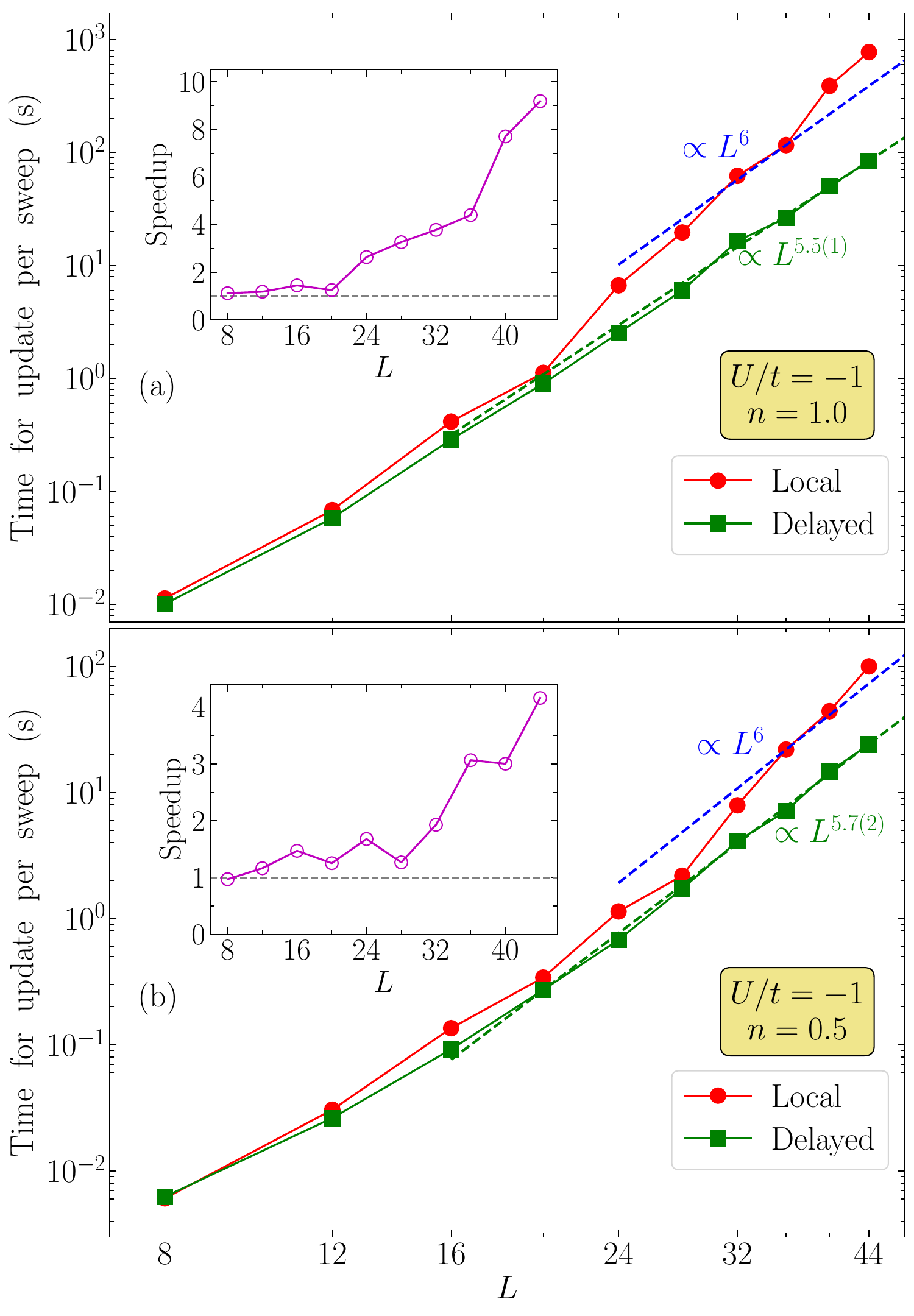}
\caption{Comparison of the average time for update per sweep (in seconds) between the local and delayed updates, as a function of $L$ in PQMC simulations of the 2D Hubbard model~(\ref{eq:2DHubbard}) with (a) $U/t=-1,n=1.0$, and (b) $U/t=-1,n=0.5$ (with simulation parameters $2\Theta t=40$ and $\Delta\tau t=0.10$ for both cases). The values of $n_d/L=1.0$ and $n_d/L=0.5$ are used in (a) and (b), respectively. Blue dashed lines plot the theoretical computational complexity ($\propto L^6$), while green dashed lines show algebraic fits ($\propto L^{\alpha}$) to the consumed time of delayed update, yielding fitted exponents $\alpha$ slightly below $6$. The insets in both panels illustrate corresponding speedups achieved by the delayed update compared to the local update. }
\label{fig:DelayedLtest}
\end{figure}

As discussed in Sec.~\ref{sec:Delayed}, the delay rank $n_d$ is a tunning parameter in the delayed update scheme. Previous studies on Hirsch-Fye QMC~\cite{Nukala2009} and finite-temperature DQMC~\cite{Sun2024} have shown that the efficiency of the delayed update depends on $n_d$. In our PQMC simulations, we treat $n_d$ as an input parameter and set it as the number of proposed update moves within a single-step delayed update. In the following, we first test the $n_d$ dependence of the efficiency for the delayed update, and then demonstrate its speedup relative to the local update versus the system size. 

In Fig.~\ref{fig:Delayedndtest}, we present the average time per sweep consumed by the delayed update, in comparison with the local update, in PQMC simulations of the 2D Hubbard model (\ref{eq:2DHubbard}). Two representative parameter sets are considered, $(U/t=-4,n=1.0)$ and $(U/t=-6,n=0.625)$, for which the number of fermions reads as $N_{\sigma}=nN_s/2$. We observe that the delayed update starts to show the advantage over the local update at $L=16$, whereas its efficiency is limited for $L < 16$ due to relatively small matrix size. Then for $L\ge 16$, we find that the consumed time initially decreases with increasing $n_d$, then rises again, resulting in peak speedup over the local update around $n_d/L=0.5$$\sim$$1.0$, as shown in the insets. This non-monotonic behavior arises from the competition between the leading computational complexity $\mathcal{O}[(2n_d+n_{\rm A})N_{\sigma}^2]$ associated with matrix multiplications and the subleading cost $\mathcal{O}[2n_{\rm A}(n_{\rm A}-1)N_{\sigma}]$ stemming from calculations of $\sum_{m}c_m\mathbf{y}_m^{\rm T}$ and $\sum_{m}d_m\mathbf{x}_m$ in Eq.~(\ref{eq:DelayRgenk}). Note that the latter is absent in local update. As $n_d$ increases, the subleading cost can eventually overtake the leading one when the condition $2n_{\rm A}(n_{\rm A}-1)\simeq(2n_d + n_{\rm A})N_{\sigma}$ is met. Accordingly, the additional cost (for $\sum_{m}c_m\mathbf{y}_m^{\rm T}$ and $\sum_{m}d_m\mathbf{x}_m$) in delayed update cancels out its acceleration gained from replacing vector operations in local update with matrix multiplications. In the extreme case of $n_d=N_s$, the delayed update becomes entirely inefficient, as the additional cost scaling as $\mathcal{O}(N_s^2N_{\sigma})$ (considering $n_{\rm A}=Rn_d$) dominates the update process. Therefore, choosing an intermediate value of $n_d$ is crucial for the delayed update to achieve high efficiency. The optimal $n_d$ clearly depends on $N_{\sigma}$ and the acceptance ratio $R=n_{\rm A}/n_d$ of the update. As plotted in insets of Fig.~\ref{fig:Delayedndtest}, the highest speedups ($\sim$$3$ and $\sim$$2.5$ for $L=32$) are achieved with $n_d \simeq 0.75L$ for $n=1.0$ case, and $n_d \simeq 0.50L$ for $n=0.625$, indicating a decreasing trend in the optimal $n_d$ towards the lower filling regime.

Building on the above $n_d$ test, we next examine the efficiency of the delayed update as a function of system size. In Fig.~\ref{fig:DelayedLtest}, we compare the time consumed per sweep by the local and delayed updates, for the 2D Hubbard model (\ref{eq:2DHubbard}) with $(U/t=-1,n=1.0)$ and $(U/t=-1,n=0.5)$. Our PQMC simulations adopt $2\Theta t=40,\Delta\tau=0.10$, and cover an extended range of $L$ from $L=8$ to $L=44$. It is evident that, as the system size increases, the speedup of the delay update relative to local update generally grows. Furthermore, this speedup tends to become more pronounced at higher fermion filling $n$ (or larger $N_{\sigma}$). Specifically, in the $L=44$ system, the speedup reaches $\sim$$9.2$ for $n=1.0$ and $\sim$$4.3$ for $n=0.5$. Considersing that the local update occupies $\sim$$95\%$ of the total computation time, the overall speedups of the PQMC simulations are $\sim$$6.5$ and $\sim$$3.7$ for the $n=1.0$ and $n=0.5$ cases (for $L=44$), respectively. These values are reasonably close to the speedups measured in the delayed update scheme of finite-temperature DQMC~\cite{Sun2024}, indicating comparable efficiency between the two methods. Besides, the results in Fig.~\ref{fig:DelayedLtest} show that the actual scaling of computational complexity is faster than $L^6$ for local update and slower than $L^6$ for delayed update [Note the theoretical computational complexity is $\mathcal{O}(N_sN_{\sigma}^2)\propto\mathcal{O}(L^6)$, since $N_s=L^2$ and $N_{\sigma}=nN_s/2=nL^2/2$]. This also highlights the significantly enhanced efficiency achieved by replacing the vector operations (in local update) with matrix multiplications (in delayed update).

\begin{figure}[t]
\centering
\includegraphics[width=0.987\linewidth]{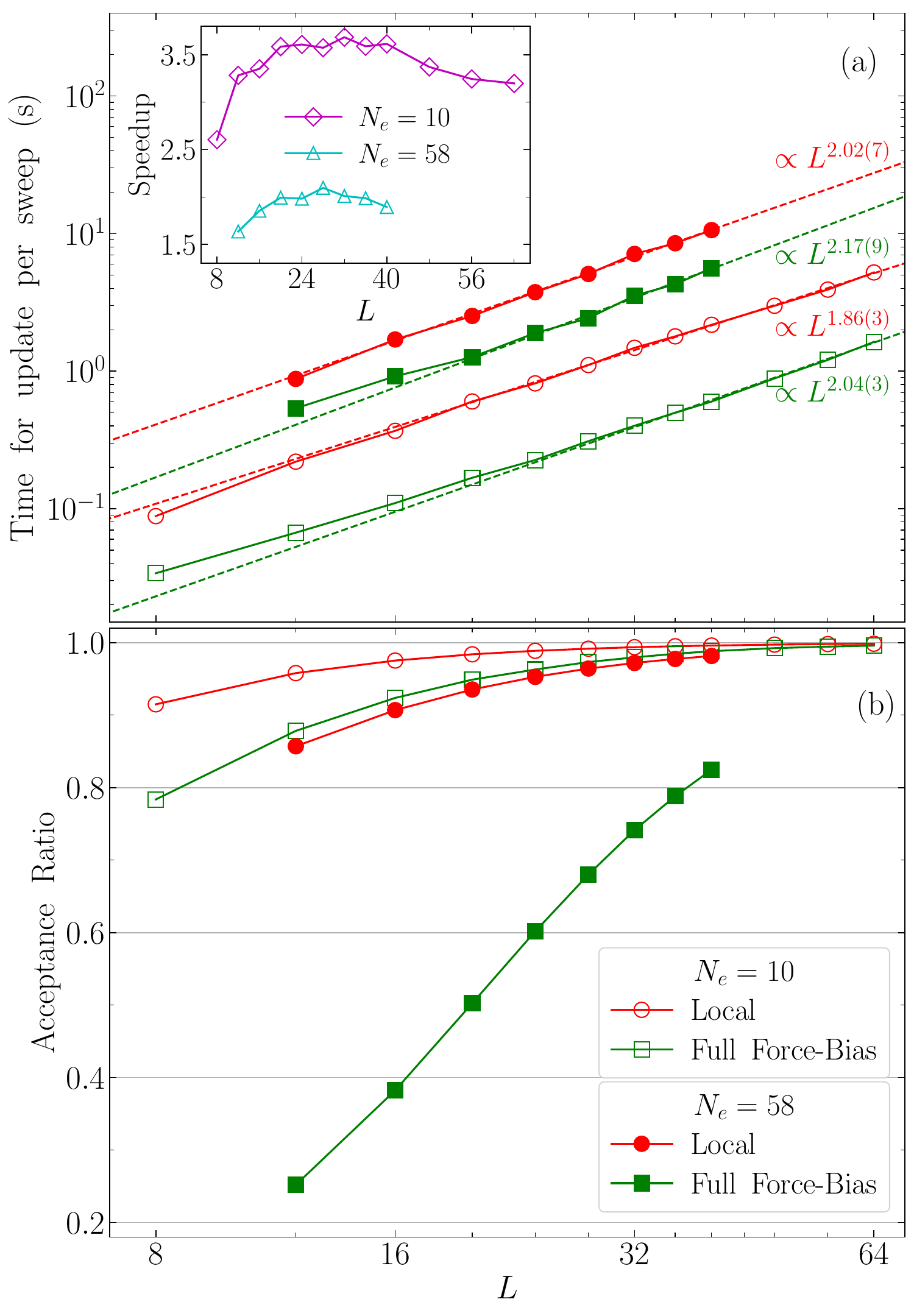}
\caption{Comparisons of (a) the average time for update per sweep (in seconds), and (b) the acceptance ratio, between the local update and full force-bias update in PQMC simulations of the 2D interacting Fermi gas in the crossover regime [$\log(k_Fa)=+0.50$]. Results are shown for two cases, $N_e=10$ and $N_e=58$. (a), (b) Show log-log plot and semilog plot, respectively. The dashed lines in (a) represent algebraic fits ($\propto L^{\alpha}$) to the consumed time, yielding fitted exponents $\alpha$ all close to 2 as expected. The inset of (a) illustrate corresponding speedups achieved by the full force-bias update compared to the local update. }
\label{fig:FrcBsFermiGas}
\end{figure}

\subsection{Efficiency tests for the force-bias update}
\label{sec:SpeedFrcBs}

In this subsection, we focus on the efficiency tests of both the full and block force-bias update schemes. While the full force-bias update mostly works well in dilute systems (such as interacting Fermi gas), it tends to suffer from low acceptance ratios in lattice models at dense fillings. The block force-bias update solves this issue, and offers a more robust approach in dense filling regimes.

\subsubsection{The full force-bias update}
\label{sec:SpeedFullFrcBs}

\begin{figure}[t]
\centering
\includegraphics[width=0.99\linewidth]{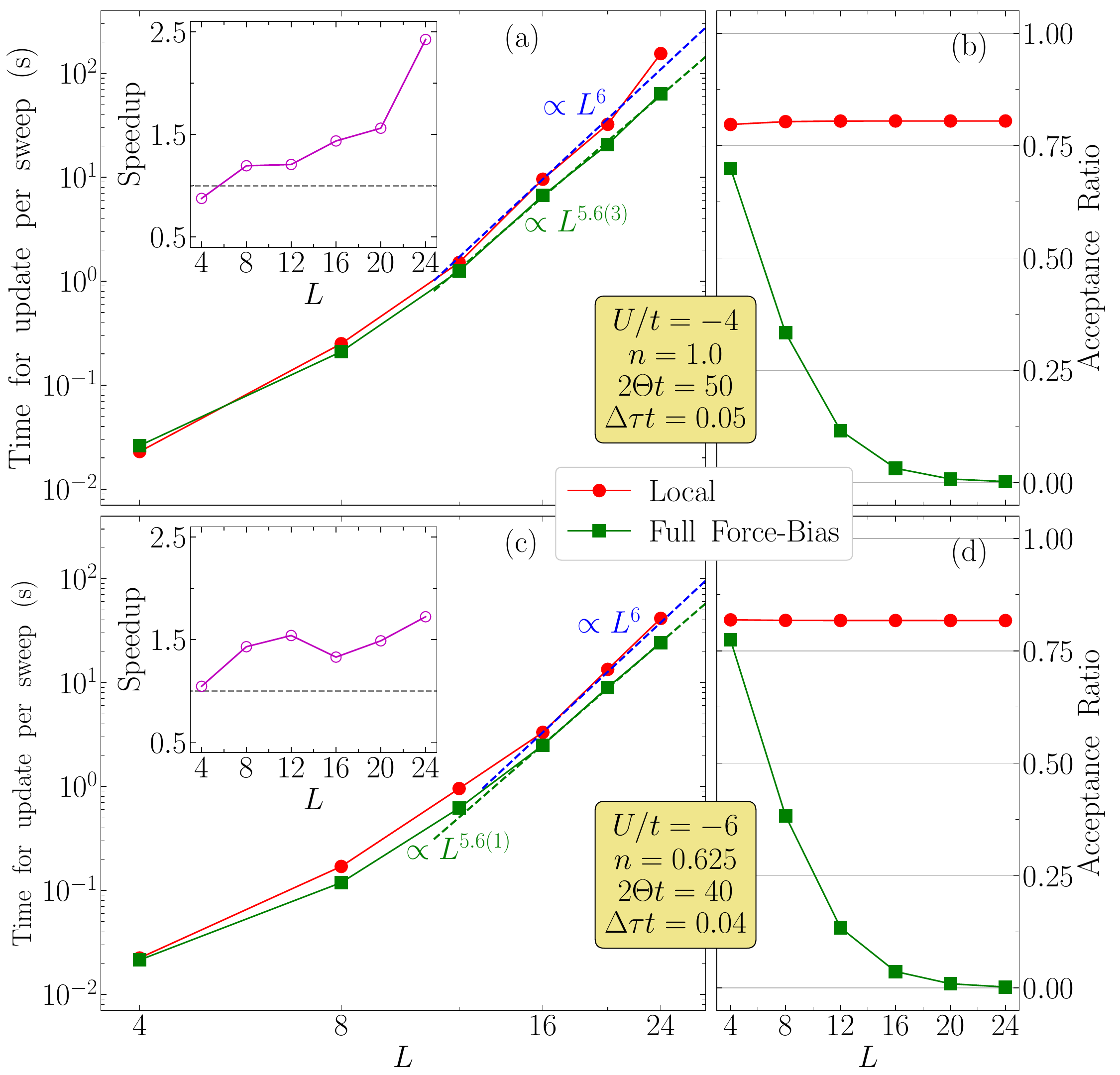}
\caption{Comparisons of the average time for update per sweep (in seconds) and the acceptance ratio between the local update and full force-bias update in PQMC simulations of the 2D Hubbard model~(\ref{eq:2DHubbard}). (a), (b) Show results for the model with $U/t=-4,n=1.0$ (with parameters $2\Theta t$ and $\Delta\tau t$ included), while (c) and (d) are for $U/t=-6,n=0.625$. In (a) and (c), blue dashed lines plot the theoretical scaling ($\propto L^{6}$), and green dashed lines show the algebraic fits ($\propto L^{\alpha}$) to the consumed time of full force-bias update, yielding fitted exponents $\alpha$ slightly below $6$. Insets in (a) and (c) illustrate corresponding speedups achieved by the full force-bias update compared to the local update.}
\label{fig:FrcBsModel}
\end{figure}

Here we first present the application of the full force-bias update scheme to the 2D Fermi gas with contact attraction~\cite{Shihao2015,Ettore2017,Yuanyao2022}, and then we demonstrate its limitation in the 2D Hubbard model at dense fillings. 

The essential physics of interacting Fermi gas includes fermionic superfluidity and BCS-BEC crossover~\cite{Shihao2015,Yuanyao2022}. 
In PQMC simulations, we model the uniform 2D Fermi gas by the attractive Hubbard model in Eq.~(\ref{eq:2DHubbard}) with a modified kinetic dispersion $\varepsilon^{H}_{\mathbf{k}}=4t-2t(\cos k_x + \cos k_y)$. The interaction strength $U$ is determined from the dimensionless parameter $\log(k_Fa)$ [with $k_F=\sqrt{2\pi n}$ as the Fermi vector ($n$ is the fermion filling), and $a$ as the 2D scattering length] as
\begin{equation}\begin{aligned}
\label{eq:Uovt}
\frac{U}{t}=-\frac{4\pi}{\log(k_Fa)-\log(\mathcal{C}\sqrt{n})},
\end{aligned}\end{equation} 
with $\mathcal{C}=0.49758$ for above Hubbard dispersion $\varepsilon^{H}_{\mathbf{k}}$~\cite{Shihao2015}. The dilute gas limit is typically accessed via first reaching the continnum limit ($n=N_e/N_s\to0$) with fixed $N_e$, and then the thermodynamic limit ($N_e\to\infty$)~\cite{Shihao2015,Yuanyao2022}. 

\begin{figure}[t]
\centering
\includegraphics[width=0.98\linewidth]{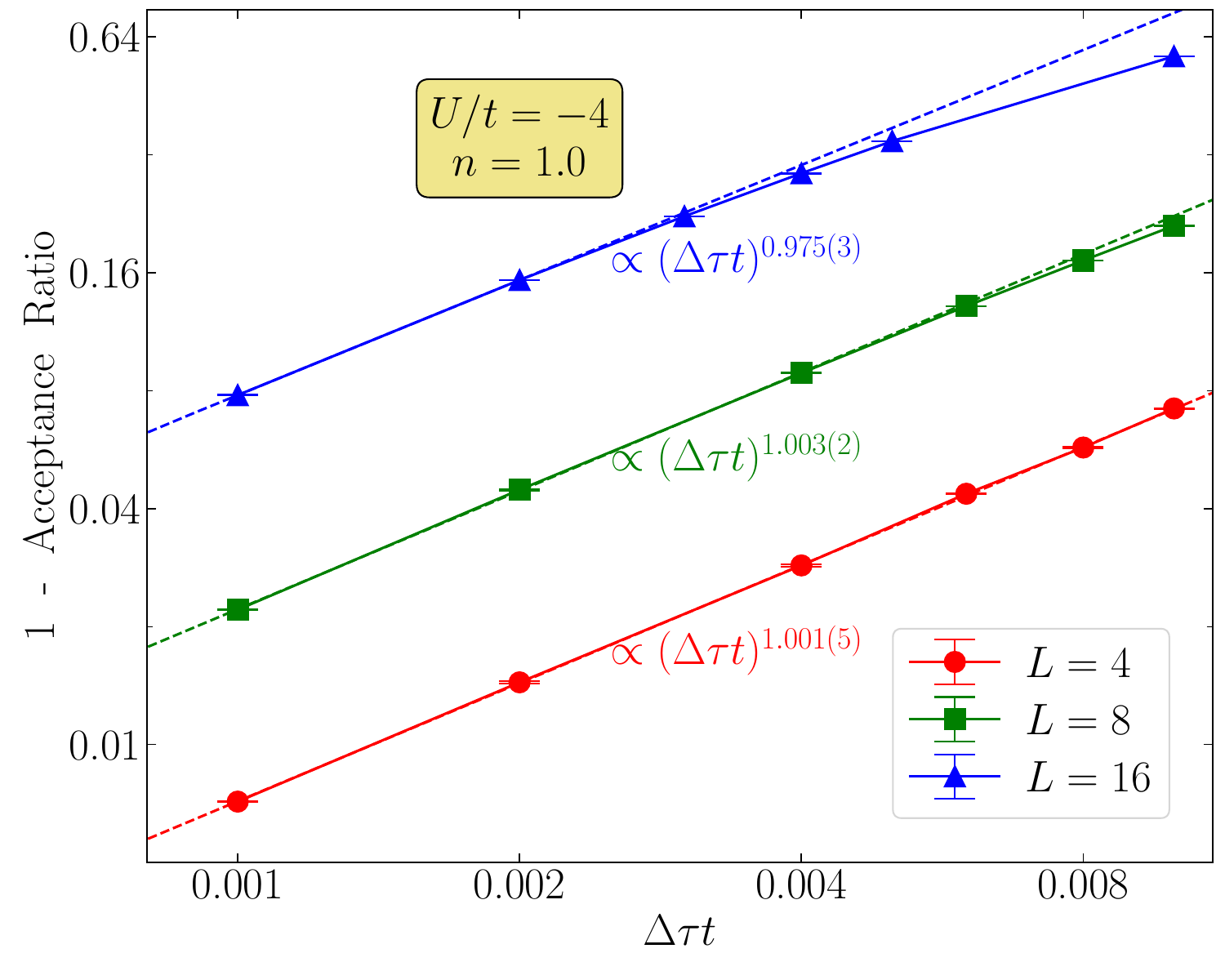}
\caption{The log-log plot of $(1-R)$ (with $R$ denoting the acceptance ratio) versus $\Delta\tau t$ for the full force-bias update in PQMC simulations of the 2D Hubbard model~(\ref{eq:2DHubbard}) with $U/t=-4,n=1.0$. Results are shown for $L=4,8,16$. The dashed lines represent algebraic fits as $(1-R)\propto(\Delta\tau t)^{\eta}$, with fitted exponents $\eta$ close to 1, suggesting a linear dependence.}
\label{fig:FrcBsAcptRatio}
\end{figure}

\begin{figure*}[t]
\centering
\includegraphics[width=0.98\linewidth]{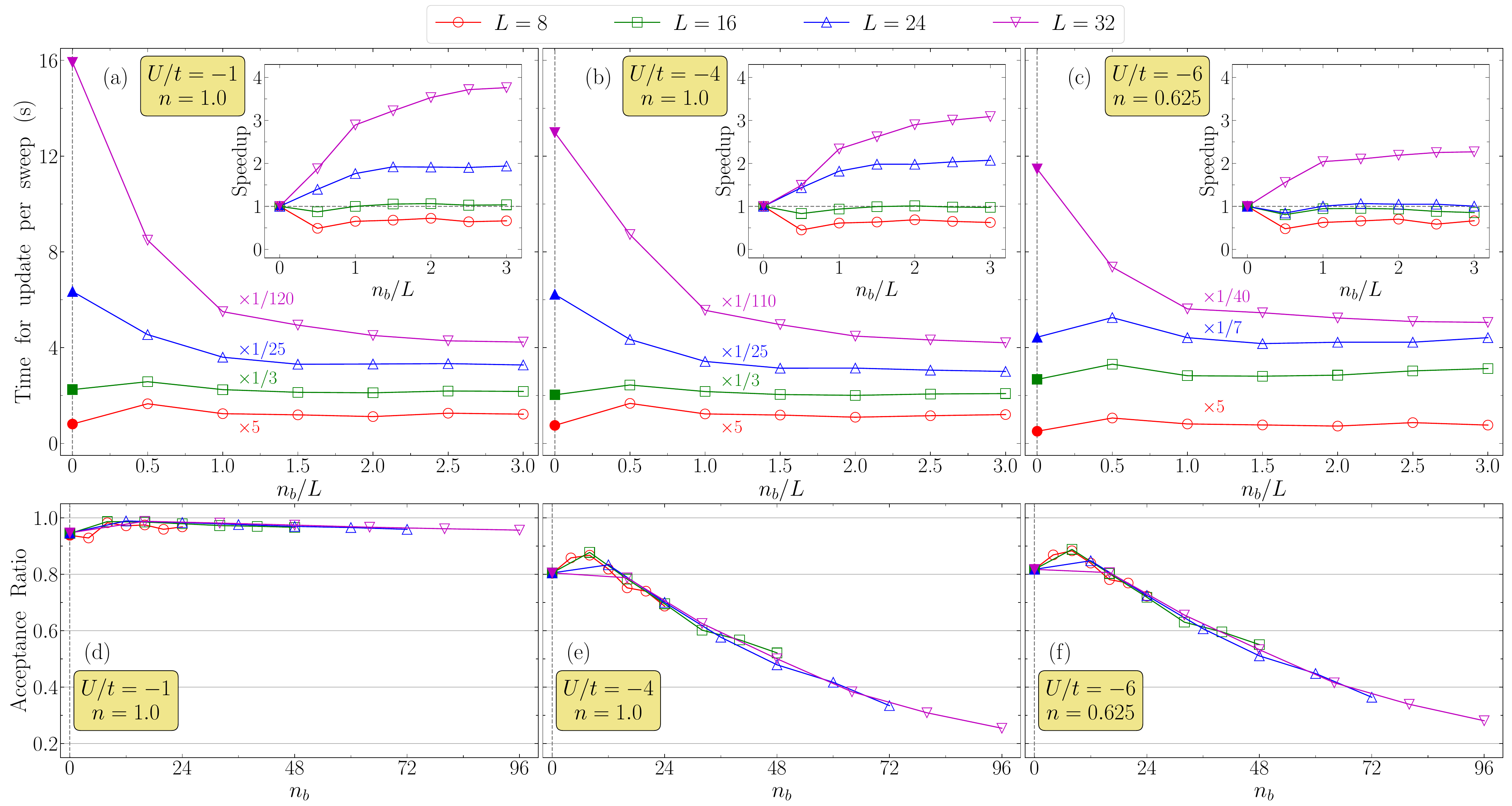}
\caption{The average time for update per sweep (in seconds) using the block force-bias update as a function of $n_b/L$ (with $n_b$ as the block size), in PQMC simulations of the 2D Hubbard model~(\ref{eq:2DHubbard}) under three parameter sets: (a) $U/t=-1,n=1.0$; (b) $U/t=-4,n=1.0$; (c) $U/t=-6,n=0.625$. The corresponding acceptance ratios (versus $n_b$) for these three cases are presented in (d)-(f). The leftmost data points (solid symbols) at $n_b/L=0.0$ [in (a)-(c)] or at $n_b=0$ [in (d)-(f)] are results from the local update. In (a)-(c), the rescaling factors for the consumed time are included (with the same color as the symbol), and the insets plot the corresponding speedups achieved by the block force-bias update compared to the local update. The timing data in (a)-(c) are summarized respectively in Tables~\ref{Table:B1}, \ref{Table:B2}, and \ref{Table:B3} in Appendix~\ref{sec:AppendixE}.}
\label{fig:BlkFrcBsNbtest}
\end{figure*}

Our comparison results of the average time for update per sweep and the acceptance ratio between local and full force-bias updates, for the 2D Fermi gas at $\log(k_Fa)=+0.50$ with $N_e=10$ and $58$, are shown in Fig.~\ref{fig:FrcBsFermiGas} versus $L$. The full force-bias update clearly accelerates the simulation, and its speedup relative to local update becomes more pronounced as $N_e$ decreases. Most of the speedup values are greater than 3 for $N_e=10$ and $1.5$ for $N_e=58$ [see the inset of Fig.~\ref{fig:FrcBsFermiGas}(a)]. The nonmonotonic speedup versus $L$ likely results from the decreasing efficiency of matrix operations in the full force-bias update as the involved matrices $\mathbf{U}_L^{\sigma},\mathbf{U}_R^{\sigma}$ become increasingly rectangular. The scalings of the consumed time for both update schemes still conform approximately with $\mathcal{O}(L^2)$, as expected. Meanwhile, as shown in Fig.~\ref{fig:FrcBsFermiGas}(b), the acceptance ratios are consistently high and increase toward unity with growing $L$. This trend is primarily attributed to the decreasing $|U|/t$ from Eq.~(\ref{eq:Uovt}), i.e., from $U/t=-7.60361$ at $L=12$ to $U/t=-4.39898$ at $L=40$ for $\log(k_Fa)=+0.50$ with $N_e=58$. In full force-bias update, the $f_{\mathbf{x}}$ factor in Eqs.~(\ref{eq:RatioApprox}) and (\ref{eq:AcceptRatio}) is also expected to decrease as the filling $n$ becomes smaller, which further contributes to the observed trend. The speedup and rising acceptance ratio of the full force-bias update highlight its high efficiency for dilute Fermi gas. However, at $L=12$, the acceptance ratio is only $\sim$$0.25$ for $N_e=58$. The situation gets even worse with increasing $N_e$ or towards stronger interactions [smaller $\log(k_Fa)$ on the BEC side], where $|U|/t$ becomes larger. These results indicate that the full force-bias update can also be inefficient, even in dilute systems, when applied to finite-size PQMC simulations. 

We then turn to the 2D Hubbard model~(\ref{eq:2DHubbard}) at dense fillings. In Fig.~\ref{fig:FrcBsModel}, we present the analogous comparisons to those in Fig.~\ref{fig:FrcBsFermiGas}, for the model with $(U/t=-4,n=1.0)$ and $(U/t=-6,n=0.625)$. As shown in Figs.~\ref{fig:FrcBsModel}(a) and \ref{fig:FrcBsModel}(c), the full force-bias update achieves noticeable accelerations over the local update in both cases, with the speedup generally increasing with $L$. On the other hand, the local update maintains acceptance ratios above 0.8, showing negligible dependence on $L$ [Figs.~\ref{fig:FrcBsModel}(b) and \ref{fig:FrcBsModel}(d)]. In contrast, the acceptance ratio of full force-bias update rapidly declines with increasing $L$, and almost vanishes at $L=24$. As indicated by Eqs.~(\ref{eq:RatioApprox}) and (\ref{eq:AcceptRatio}), the rapid suppression of the acceptance ratio in these dense-filling regimes should be driven by the growing $f_{\mathbf{x}}$ factor with increasing $L$. These observations explicitly render the full force-bias update scheme effectively impractical for PQMC simulations at larger system sizes. 

The low acceptance ratio issue in full force-bias update as discussed above can be alleviated by using smaller $\Delta\tau t$ for larger $L$ to compensate the growth of $f_{\mathbf{x}}$, as suggested by Eqs.~(\ref{eq:RatioApprox}) and (\ref{eq:AcceptRatio}). Furthermore, these equations indicate that the acceptance ratio $R$ approaches unity at a rate linear in $\Delta\tau$, i.e., $(1-R)\propto\Delta\tau t$. We indeed confirm such a behavior in small-$\Delta\tau$ regime, as illustrated in Fig.~\ref{fig:FrcBsAcptRatio}. With fixed $U/t$ and $n$, the linear relation is less prominent at larger $L$, due to the enhanced $f_{\mathbf{x}}$ factor. Nevertheless, raising $R$ and thereby improving the efficiency of full force-bias update by applying smaller $\Delta\tau t$ in PQMC simulations is impractical, as it substantially increases the computational cost. This approach inevitably fails as the system approaches the thermodynamic limit (i.e., with increasing $L$). 

\subsubsection{The block force-bias update}
\label{sec:SpeedBlockFrcBs}

\begin{figure}[t]
\centering
\includegraphics[width=1.00\linewidth]{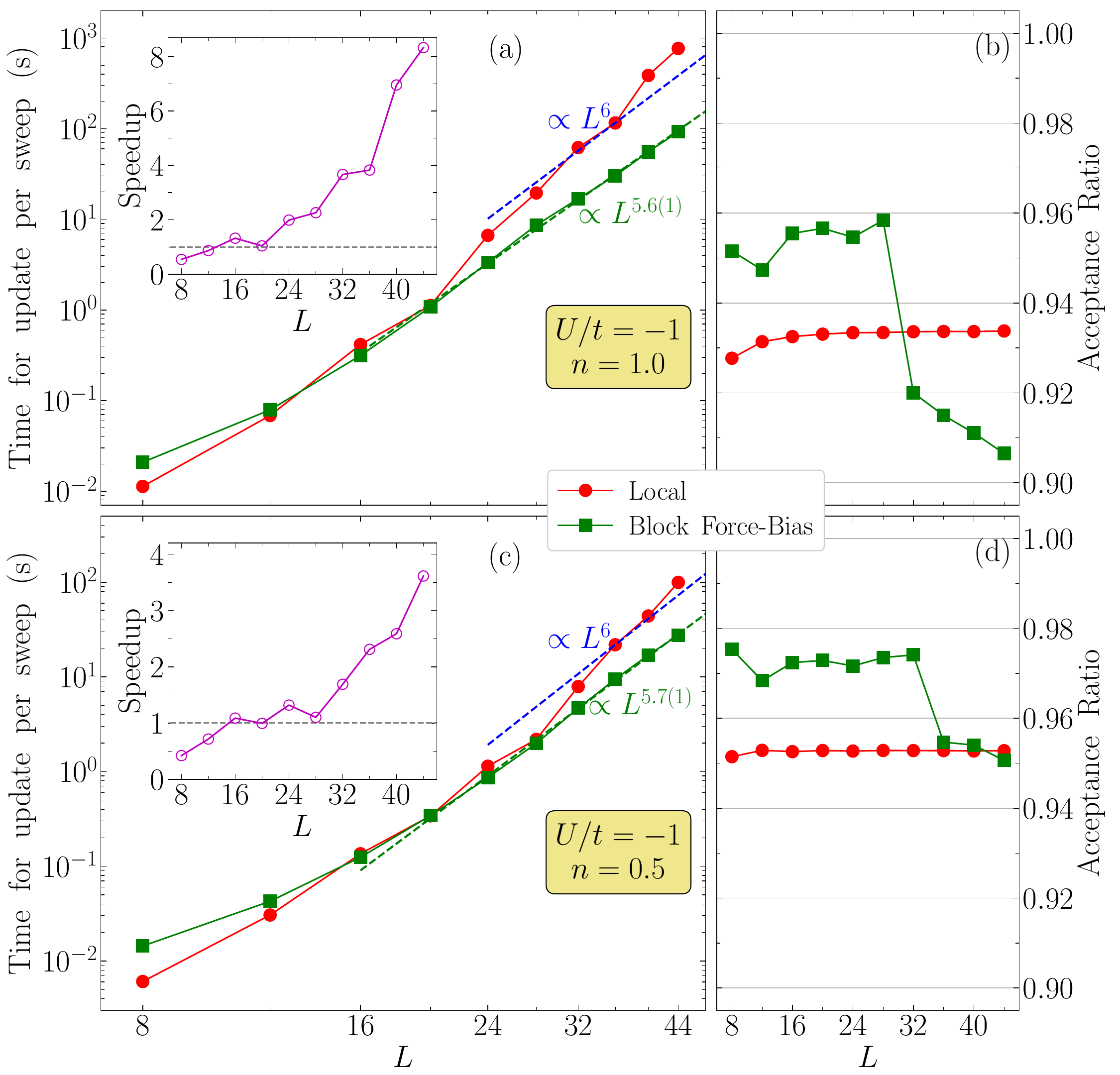}
\caption{Comparisons of the average time for update per sweep (in seconds) and the acceptance ratio between the local update and block force-bias update in PQMC simulations of the 2D Hubbard model~(\ref{eq:2DHubbard}). (a), (b) Show results for the model with $U/t=-1,n=1.0$, while (c) and (d) are for $U/t=-1,n=0.5$. The simulation parameters $2\Theta t=40$ and $\Delta\tau t=0.10$ are applied for both cases. In (a) and (b), $n_b=48$ for $L<32$ and $n_b=4L$ for $L\ge 32$ are used, whereas in (c) and (d), $n_b=48$ for $L<36$ and $n_b=4L$ for $L\ge 36$ are adopted. In (a) and (c), blue dashed lines plot the theoretical scaling ($\propto L^{6}$), and green dashed lines show the algebraic fits ($\propto L^{\alpha}$) to the consumed time of block force-bias update, yielding fitted exponents $\alpha$ slightly below $6$. Insets in (a) and (c) illustrate corresponding speedups achieved by the block force-bias update compared to the local update.}
\label{fig:BlkFrcBsLtest}
\end{figure}

Following the discussion in Sec.~\ref{sec:SpeedFullFrcBs}, an alternative strategy to resolve the low acceptance ratio issue in the full force-bias update is to reduce the $f_{\mathbf{x}}$ factor, according to Eqs.~(\ref{eq:RatioApprox}) and (\ref{eq:AcceptRatio}). This reduction can be achieved using a block variant of the force-bias update, which introduces the block size $n_b$ as a tunable parameter. Notably, the extreme case of $n_b=N_s$ is equivalent to the full force-bias update. In the following, we first test the $n_b$ dependence of the efficiency for the block force-bias update scheme, and then demonstrate its speedup relative to the local update versus the system size. 

In Fig.~\ref{fig:BlkFrcBsNbtest}, we show the average time per sweep consumed by the block force-bias update in PQMC simulations of the 2D Hubbard model~(\ref{eq:2DHubbard}). Three sets of model parameters are considered: ($U/t=-1,n=1.0$), ($U/t=-4,n=1.0$), and ($U/t=-6,n=0.625$). These allow for crosschecks of the efficiency across different interaction strengths and fillings. For all three cases, the block force-bias update clearly accelerates the simulations applying local update for $L>16$ [see Figs.~\ref{fig:BlkFrcBsNbtest}(a)-\ref{fig:BlkFrcBsNbtest}(c)]. We also observe that the consumed time mostly decreases monotonically with increasing $n_b$ in the range of $n_b/L<3.0$. For larger values of $n_b/L$, however, the time cost is expected to increase again, and eventually surpass that of the local update (not shown). This is attributed to the computational overhead of evaluating the force bias $\bar{n}_{i\sigma}$, as the first part of the block force-bias update, which already involves a leading cost of $\mathcal{O}(4n_bN_{\sigma}^2+2n_b^2N_{\sigma})$ (see the computational complexity analysis in Sec.~\ref{sec:BlockForcebias}). As $n_b$ approaches $N_s$, this cost scales as $\mathcal{O}(N_s^2 N_{\sigma})$, exceeding the conventional $\mathcal{O}(N_sN_{\sigma}^2)$ scaling. Accordingly, the acceptance ratios of all three cases [see Figs.~\ref{fig:BlkFrcBsNbtest}(d)-\ref{fig:BlkFrcBsNbtest}(f)] constantly decrease with increasing $n_b$. Moreover, the data from systems of different sizes nearly collapse onto a single curve for each case, indicating that the acceptance ratio in block force-bias update primarily depends on $n_b$ rather than $L$. This behavior is embedded in the definition of the $f_{\mathbf{x}}^{(b)}$ factor in Eq.~(\ref{eq:BlkFrcBsErr}). For the case of $(U/t=-1, n=1.0)$, the acceptance ratio remains above 0.9 even at $n_b/L=3$, and the corresponding speedup over the local update reaches $\sim$$4$ for $L=32$. This performance surpasses that of the delayed update, as shown in Fig.~\ref{fig:Delayedndtest}(a). For comparison, the acceptance ratio $R$ for $(U/t=-4, n=1.0)$ is significantly suppressed due to the larger $|U|/t$, with $R$$\sim$$0.5$ at $n_b=48$. The corresponding speedup is $\sim$$3$ for $L=32$, comparable to that achieved by the delayed update. For $(U/t=-6, n=0.625)$, the reduction of $f_{\mathbf{x}}^{(b)}$ due to the lower $n$ partially offsets the effect of the larger $|U|/t$, leading to a slightly higher acceptance ratio compared to the $(U/t=-4, n=1.0)$ case. Nevertheless, the speedup is more modest, as $\sim$$2.3$ for $L=32$ at $n_b=48$. These results illustrate that the block force-bias update is more effective in regimes with weak interaction and intermediate to high filling.

Based on the above $n_b$ tests, we proceed to examine the efficiency of the block force-bias update versus the system size. In Fig.~\ref{fig:BlkFrcBsLtest}, we present the comparisons of the time consumed per sweep and the acceptance ratios by the local and block force-bias updates, for the 2D Hubbard model (\ref{eq:2DHubbard}) with $(U/t=-1,n=1.0)$ and $(U/t=-1,n=0.5)$. For $L\le 16$, the advantage of matrix multiplications in block force-bias update over the vector operations in local update is limited by the relatively small matrix size, resulting in the lack of accelerations. Afterwards for $L>16$, the speedup becomes pronounced, and it constantly grows with increasing $L$. Specifically, at $L=44$, the speedup reaches $8.2$ and $3.6$ for $n=1.0$ and $n=0.5$, respectively, which are comparable to the performance of delayed update (see Fig.~\ref{fig:DelayedLtest}). On the other hand, the specific choices of $n_b$ in these calculations ensure high acceptance ratios, i.e. above $0.9$, for the block force-bias update, while the ratio is almost constant in the local update. Note that the drops of acceptance ratio at $L=32$ for the $n=1.0$ case and at $L=36$ for the $n=0.5$ case result from reasonable adjustments to $n_b$ (see the caption of Fig.~\ref{fig:BlkFrcBsLtest}). These adjustments lead to the favorable scalings ($\propto L^{\alpha}$ with $\alpha$ slightly below $6$) of the consumed time, as shown by the green dashed lines in Figs.~\ref{fig:BlkFrcBsLtest}(a) and \ref{fig:BlkFrcBsLtest}(c).

As discussed in Sec.~\ref{sec:BlockForcebias} and illustrated by the results in Figs.~\ref{fig:BlkFrcBsNbtest} and \ref{fig:BlkFrcBsLtest}, both the computational time and acceptance ratio explicitly depend on $n_b$ in the block force-bias update. Therefore, in practical simulations with various model parameters, it is important to choose an $n_b$ via numerical tests, that balances computational time with a satisfactory acceptance ratio (for example, higher than $0.5$) within the Markov chain process. 

\subsection{Tests on the SOC-Hubbard model}
\label{sec:SOCHubbardTest}

\begin{figure*}[t]
\centering
\includegraphics[width=0.990\linewidth]{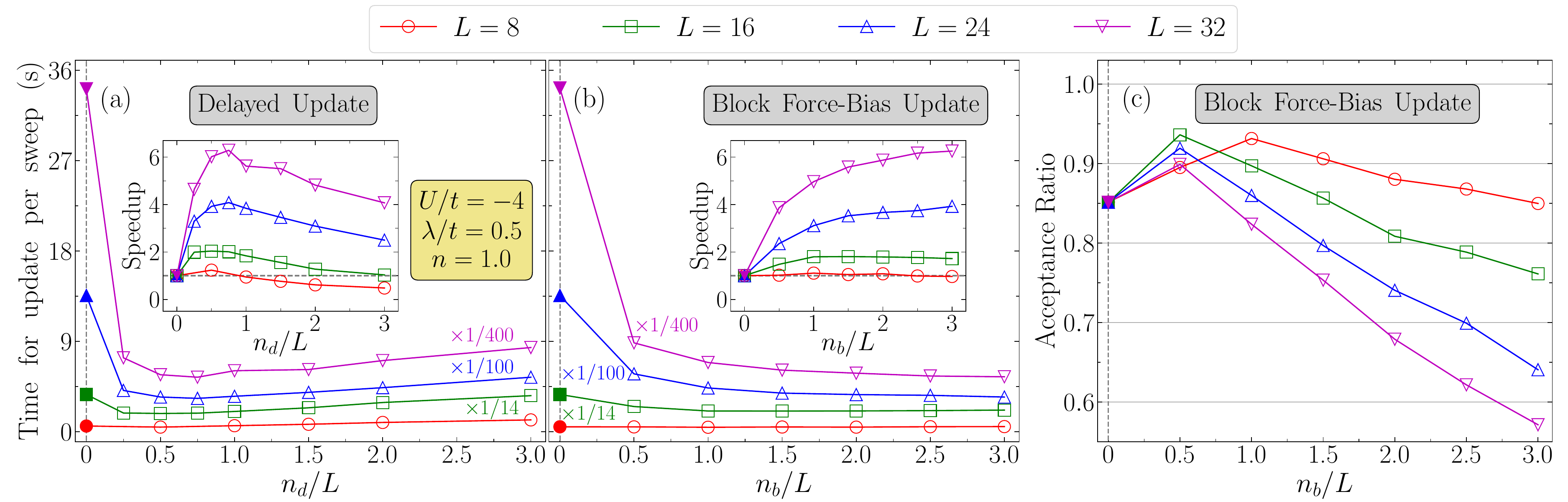}
\caption{The average time for update per sweep (in seconds) using (a) the delayed update (versus $n_d/L$), and (b) the block force-bias update (versus $n_b/L$), in PQMC simulations of the 2D SOC-Hubbard model~(\ref{eq:2DSocHubbard}) with $U/t=-4,n=1.0$ and SOC strength $\lambda/t=0.5$. The rescaling factors for the consumed time are included (with the same color as the symbol), and the insets plot the corresponding speedups achieved by the delayed update [in (a)] and the block force-bias update [in (b)] compared to the local update. (c) Shows the results of acceptance ratio versus $n_b/L$ for the block force-bias update. In (a)-(c), the leftmost data points (solid symbols) at $n_d/L=0$ or $n_b/L=0$ are results from the local update scheme. }
\label{fig:DelayFrcBsSOC}
\end{figure*}

In our PQMC simulations for the 2D SOC-Hubbard model in Eq.~(\ref{eq:2DSocHubbard}), we apply an alternative HS transformation that decouples the on-site Hubbard interaction into spin-$\hat{s}^y$ channel (denoted as HS-$\hat{s}^y$) as
\begin{equation}\begin{aligned}
\label{eq:HSspinSyDecomp}
e^{-\Delta\tau U \big(\hat{n}_{i\uparrow} \hat{n}_{i\downarrow} - \frac{\hat{n}_{i\uparrow} + \hat{n}_{i\downarrow}}{2}\big) }
= \frac{1}{2}\sum_{x_{i}=\pm1}e^{\gamma x_i(c_{i\uparrow}^+ c_{i\downarrow}^{} - c_{i\downarrow}^+ c_{i\uparrow}^{} )},
\end{aligned}\end{equation}
with the coupling constant $\gamma=\cos^{-1}(e^{\Delta\tau U/2})$ for $U<0$. Then it is evident that the spin-up and -down sectors are coupled in both the noninteracting term $\hat{H}_0$ and the effective Hamiltonian in the $\hat{B}_I(\mathbf{x}_{\ell})$ operator [see Eq.~(\ref{eq:ExpDtH})]. The framework proceeds within the generalized Hartree-Fock basis as $\mathbf{c}^+=(c_{1\uparrow}^+,c_{2\uparrow}^+,\cdots,c_{N_s\uparrow}^+,c_{1\downarrow}^+,c_{2\downarrow}^+,\cdots,c_{N_s\downarrow}^+)$. As a result, the trial wave function involves an $2N_s\times N_e$ matrix $\mathbf{P}$, while in Eq.~(\ref{eq:GrFMatNew}), the $\mathbf{U}_R$ ($\mathbf{U}_L$) is an $2N_s\times N_e$ ($N_e\times 2N_s$) matrix, and the $\mathbf{T}$ matrix is of $N_e\times N_e$. The local update in PQMC for this spin-coupled system is of the rank-2 update~\cite{Sun2024}, in which the $\mathbf{x}$ and $\mathbf{y}^{\rm T}$ vectors in Eq.~(\ref{eq:UpdateLTRlocal}) are now $2\times N_e$ and $N_e\times 2$ matrices. Correspondingly, in the delayed update, the vectors $\mathbf{x}_m$ and $\mathbf{y}_m^{\rm T}$ in Eqs.~(\ref{eq:DelayTMat00}) and (\ref{eq:DelayRgenk}) become $N_e\times2$ and $2\times N_e$ matrices. In the force-bias update, according to the HS transformation in Eq.~(\ref{eq:HSspinSyDecomp}), we need to evaluate the force bias $\langle c_{i\uparrow}^+ c_{i\downarrow}^{}\rangle$ and $\langle c_{i\downarrow}^+ c_{i\uparrow}^{}\rangle$ [where $\langle\hat{o}\rangle=\langle\psi_l|\hat{o}|\psi_r\rangle/\langle\psi_l|\psi_r\rangle$ with $|\psi_r\rangle$ and $|\psi_l\rangle$ in Eq.~(\ref{eq:FullFrcbsPhiLPhiR}) or~(\ref{eq:BlockFrcbsPhiLPhiR})], using the similar formulas as those discussed in Secs.~\ref{sec:LocalFrcbias} and~\ref{sec:BlockForcebias}. The dimensions of all the matrices involved in the calculations should also be enlargerd accordingly. We further summarize other algorithmic details of the PQMC formalism and the update schemes for the model~(\ref{eq:2DSocHubbard}) in Appendix~\ref{sec:AppendixA}. 

We then focus on the efficiency tests of both the delayed update and block force-bias update in the 2D SOC-Hubbard model~(\ref{eq:2DSocHubbard}). In Fig.~\ref{fig:DelayFrcBsSOC}, we show the average time per sweep consumed by the delayed update and the block force-bias update, as well as their acceptance ratios, in PQMC simulations for the model. We use the parameters $U/t=-4,n=1.0$ with the SOC strength $\lambda=0.5$. 

As shown in Fig.~\ref{fig:DelayFrcBsSOC}(a), the delayed update achieves accelerations over the local update for $L\ge 16$, and the corresponding speedups are larger than those recorded for the standard Hubbard model~(\ref{eq:2DHubbard}) [see Fig.~\ref{fig:Delayedndtest}(a)]. At $L=32$, the maximal speedup achieved at $n_d/L=0.75$ is above $6$, about twice of that shown in Fig.~\ref{fig:Delayedndtest}(a). This improvement in the model~(\ref{eq:2DSocHubbard}) is due to the enlarged matrix sizes in PQMC simulations of the spin-coupled system, as discussed above. The nonmonotonic behavior of the consumed time by the delayed update with increasing $n_d$ can be understood in the same way as explained for Fig.~\ref{fig:Delayedndtest}(a). Moreover, the acceptance ratio is consistently around $0.85$ for all systems studied [solid symbols at $n_b/L=0$ in Fig.~\ref{fig:DelayFrcBsSOC}(c)] in the delayed update. (Note that the acceptance ratio is exactly the same for the local and delayed updates.)

As a comparison, the block force-bias update scheme also exhibits great performance for the model~(\ref{eq:2DSocHubbard}), as illustrated by the results in Figs.~\ref{fig:DelayFrcBsSOC}(a) and \ref{fig:DelayFrcBsSOC}(b). Its speedup over the local update increases notably with $L$, and it also reaches $\sim$$6$ at $L=32$ for $n_b/L=2$$\sim$$3$. In this regime, the acceptance ratio remains above $0.55$, which is reasonably high. Even when the acceptance ratio is constrained to that of the delayed update (around $0.85$), the block force-bias update still delivers 
substantial speedups of $\sim$$3$ for $L=24$ and $\sim$$4.6$ for $L=32$. Furthermore, by comparing with the results in Figs.~\ref{fig:BlkFrcBsNbtest}(b) and \ref{fig:BlkFrcBsNbtest}(e), we find that the presence of SOC significantly enhances the acceptance ratio, particularly in the regime of relatively large $n_b$ (i.e., $n_b/L>1$). 

\begin{figure*}[t]
\centering
\includegraphics[width=0.99\linewidth]{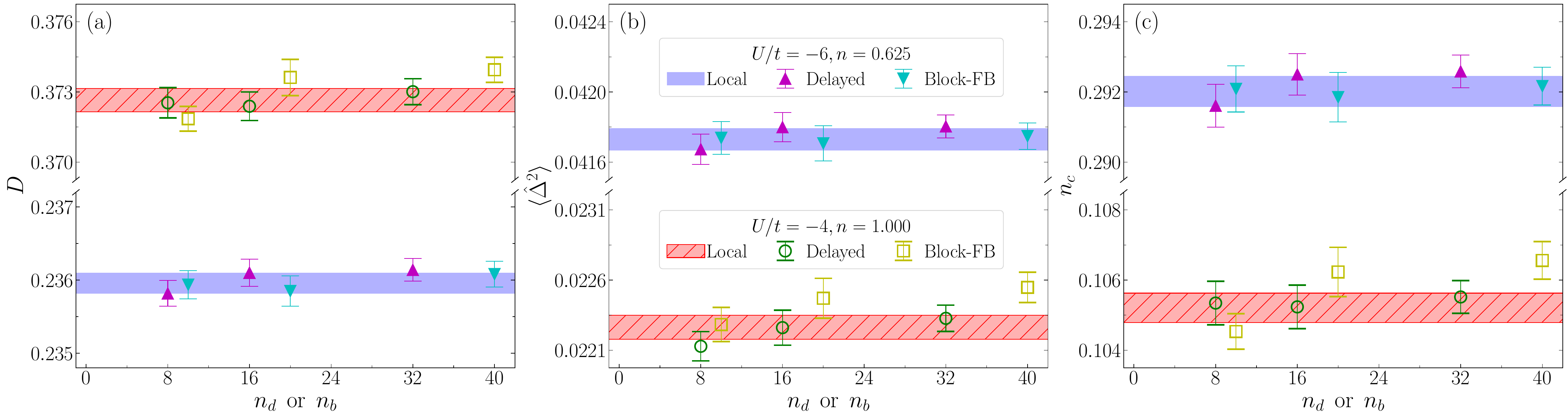}
\caption{Benchmark results of (a) double occupancy $D$, (b) on-site spin-singlet pairing correlator $\langle\Delta^2\rangle$, and (c) condensate fraction $n_c$, from PQMC simulations applying local, delayed, and block force-bias (denoted as ``Block-FB'') updates, for 2D attractive Hubbard model~(\ref{eq:2DHubbard}) with $L=20$. Two sets of the model parameters are considered, $(U/t=-4,n=1.0)$ and $(U/t=-6,n=0.625)$, for which the simulation parameters $(2\Theta t=46,\Delta\tau t=0.05)$ and $(2\Theta t=36,\Delta\tau t=0.04)$ are used, respectively. The results from the local update are plotted with shaded bands. For the latter two update schemes, the data are shown as a function of $n_d$ (the predetermined delay rank in delayed update, with $n_d=8,16,32$) or $n_b$ (the block size in block force-bias update, with $n_b=10,20,40$). All the raw data in this plot are summarized in Table~\ref{Table:C1} in Appendix~\ref{sec:AppendixE}.}
\label{fig:PhyResults}
\end{figure*}

The numerical results in Fig.~\ref{fig:DelayFrcBsSOC} clearly show that both the delayed update and block force-bias update schemes are highly efficient in PQMC simulations in the SOC-Hubbard model~(\ref{eq:2DSocHubbard}). The speedups compared to the local update are actually more pronounced than those achieved for the standard Hubbard model~(\ref{eq:2DHubbard}). Here we simply bypass the explicit efficiency tests versus the system size (similar to those presented in Figs.~\ref{fig:DelayedLtest} and~\ref{fig:BlkFrcBsLtest}) for the model~(\ref{eq:2DSocHubbard}). Based on the results for $L=8,16,24,32$ in Fig.~\ref{fig:DelayFrcBsSOC}, it is reasonable to expect that the speedups at $L=44$ for both update schemes in the model~(\ref{eq:2DSocHubbard}) would surpass those in the model~(\ref{eq:2DHubbard}), likely exceeding one order of magnitude.

The efficiency tests on the SOC-Hubbard model~(\ref{eq:2DSocHubbard}) also highlight that our delayed and block force-bias updates are broadly applicable to extended interactions beyond the standard Hubbard model. This generality arises from the fact that arbitrary off-site two-body interactions in fermionic systems can be decoupled into fermion hopping terms coupled to auxiliary fields via HS transformations~\cite{Shiwei2003,Shihao2021,Yuanyao2016,Qin2017}, as in the HS-$\hat{s}^y$ decomposition of Eq.~(\ref{eq:HSspinSyDecomp}) used for the model~(\ref{eq:2DSocHubbard}). Except for the possible differences on the specific hopping terms and the auxiliary fields (of discrete or continuous type), the process for extended interactions in both update schemes follows the same procedures as those for the model~(\ref{eq:2DSocHubbard}). Representative examples can range from the spinless $t$-$V$ model, as studied in Ref.~\cite{Sun2024}, to the realistic long-range Coulomb interaction in quantum chemistry~\cite{Shiwei2003,Shihao2021}.

\subsection{Benchmark results of physical observables in the Hubbard model}
\label{sec:PhysicalResults}

In this subsection, we benchmark the numerical results of physical observables obtained from PQMC simulations using the local, delayed, and block force-bias update schemes. The calculation of observables in PQMC can be referred in Eqs.~(\ref{eq:GSObsQMC}) and (\ref{eq:Oxformula}). We employ the 2D attractive Hubbard model as a testbed, and consider two sets of model parameters: $(U/t=-4,n=1.0)$ and $(U/t=-6,n=0.625)$. An intermediate system with $L=20$ is adopted for simplicity to be accessed, for which the results can be easily verified by the following studies. 

We concentrate on three observables in the 2D attractive Hubbard model. The first observable is the double occupancy, defined as $D=N_s^{-1}\sum_{i}\langle\hat{n}_{i\uparrow}\hat{n}_{i\downarrow}\rangle$. The second one is the on-site spin-singlet pairing correlator as $\langle\Delta^2\rangle=L^{-2}\sum_{\mathbf{r}}P(\mathbf{r})$, with $P(\mathbf{r})=L^{-2}\sum_{\mathbf{i}}\langle\hat{\Delta}_{\mathbf{i}}^+\hat{\Delta}_{\mathbf{i}+\mathbf{r}}+\hat{\Delta}_{\mathbf{i}}\hat{\Delta}_{\mathbf{i}+\mathbf{r}}^+\rangle/4$ as the $\mathbf{r}$-space pairing correlation and $\hat{\Delta}_{\mathbf{i}}^+=c_{\mathbf{i}\uparrow}^+c_{\mathbf{i}\downarrow}^+$ as the on-site pairing operator. The third one is the condensate fraction $n_c$, defined as the proportion of Bose-condensed fermion pairs~\cite{Shihao2015,Yuanyao2022,Yuanyao2025}. It can be computed from the leading eigenvalue $\lambda_{\rm max}$ of the $\mathbf{k}$-space pairing matrix $M_{\mathbf{k}\mathbf{k}^{\prime}}=\langle\hat{\Delta}_{\mathbf{k}}^+\hat{\Delta}_{\mathbf{k}^{\prime}}\rangle-\delta_{\mathbf{k}\mathbf{k}^{\prime}}\langle c_{\mathbf{k}\uparrow}^+c_{\mathbf{k}\uparrow}\rangle\langle c_{-\mathbf{k}\downarrow}^+c_{-\mathbf{k \downarrow}}\rangle$ with the operator $\hat{\Delta}_{\mathbf{k}}^+=c_{\mathbf{k}\uparrow}^+c_{-\mathbf{k}\downarrow}^+$, using $n_c=\lambda_{\rm max}/(N_e/2)$. As recently shown in Ref.~\cite{Yuanyao2025}, this quantity and its finite-size scaling behaviors can provide methods to accurately determine the Berezinskii-Kosterlitz-Thouless (BKT) transition temperature for 2D superconductivity and superfluidity. In the context of the attractive Hubbard model, both $\langle\Delta^2\rangle$ and $n_c$ can characterize the pairing properties, and they capture complementary aspects of fermion pairing from the $\mathbf{r}$-space and $\mathbf{k}$-space perspectives~\cite{Yuanyao2025}.

The benchmark results for $D$, $\langle\Delta^2\rangle$, and $n_c$ are shown in Fig.~\ref{fig:PhyResults}. In our PQMC simulations, we have verified that the chosen parameters $2\Theta$ and $\Delta\tau t$ are appropriate for reaching the ground state in the projection and for eliminating the Trotter error within statistical uncertainty, respectively. We adopt $n_d=8,16,32$ for the delayed update, and $n_b=10,20,40$ for the block force-bias update for comparisons. For each observable in both sets of the model parameters, the outcomes obtained using the three different update schemes have comparable error bars, and they are all well consistent within two standard deviations. Moreover, for the specific systems with $L=20$ in Fig.~\ref{fig:PhyResults}, the simulations employing the delayed and block force-bias updates achieve overall speedups of about $1$$\sim$$2$ compared to the local update. In addition, while the acceptance ratios for the local (and delayed) updates remain above $0.8$, those for the block force-bias update are all higher than $0.5$. These numerical results further verify the efficiency tests in Figs.~\ref{fig:Delayedndtest} and \ref{fig:BlkFrcBsNbtest}, and validate the correctness of the two update schemes in PQMC simulations. The raw data of $D$, $\langle\Delta^2\rangle$, and $n_c$ in Fig.~\ref{fig:PhyResults} can be referred in Table~\ref{Table:C1} in Appendix~\ref{sec:AppendixE}.

\section{Discussions}
\label{sec:discussion}

In the preceding sections, we have introduced the delayed and block force-bias update schemes within the PQMC framework, and presented simulation results for the 2D Hubbard models that are free of the minus sign problem. Here, we briefly discuss the application of these two update schemes to the ground-state CPQMC method in Sec.~\ref{sec:ApplyCPQMC}, and provide an application diagram illustrating the use of different update schemes in ground-state AFQMC simulations for general interacting fermion systems in Sec.~\ref{sec:ApplyDiagram}.

\subsection{Applications to ground-state CPQMC method}
\label{sec:ApplyCPQMC}

The ground-state CPQMC method~\cite{Shiwei1995,Shiwei1997,Shiwei2003} shares many fundamental ingredients with PQMC formalism. It also obtains the many-body ground state via imaginary-time projection starting from an initial wave function $|\phi^{(0)}\rangle$, i.e., $|\Psi_g\rangle\propto (e^{-\Delta\tau\hat{H}})^M|\phi^{(0)}\rangle$. For each $e^{-\Delta\tau\hat{H}}$ operator, the Trotter-Suzuki decomposition [Eq.~(\ref{eq:AsymTrot})], and the HS transformation [Eq.~(\ref{eq:ExpDtH})] are successively applied to transform the many-boby quantum problem into effectively free-fermion systems coupled to classical auxiliary fields. Instead of Markov chain sampling in PQMC method, CPQMC applies branching random walk formalism to sample the auxiliary-field configurations. During this process, a trial wave function $|\psi_T\rangle$ is used to guide the random walk, and to eliminate the sign problem via the constrained-path approximation~\cite{Shiwei1995,Shiwei1997} or the phase problem via the phaseless approximation~\cite{Shiwei2003,Shihao2021}. In general, the wave function after applying the projection operator $(e^{-\Delta\tau\hat{H}})^{\ell-1}$ can be written as~\cite{Shihao2013}
\begin{equation}\begin{aligned}
\label{eq:CPWvfct1}
|\Psi_g^{(\ell-1)}\rangle = \sum_{k=1}^{N_w} \omega_k^{(\ell-1)}\frac{|\phi_k^{(\ell-1)}\rangle}{\langle\psi_T|\phi_k^{(\ell-1)}\rangle},
\end{aligned}\end{equation}
where $N_w$ is the number of random walkers, and $|\phi_k^{(\ell-1)}\rangle$ and $\omega_k^{(\ell-1)}$ denote the Slater determinant wave function and the corresponding weight of the $k$-th walker, respectively. We typically set $|\phi_k^{(0)}\rangle=|\phi^{(0)}\rangle$ and $\omega_k^{(0)}=1$. The sampling of auxiliary fields is revealed during applying another $e^{-\Delta\tau\hat{H}}$ operator to $|\Psi_g^{(\ell)}\rangle$, leading to
\begin{equation}\begin{aligned}
\label{eq:CPWvfct2}
|\Psi_g^{(\ell)}\rangle = \sum_{k=1}^{N_w} \omega_k^{(\ell-1)}\sum_{\mathbf{x}_k^{(\ell)}}\tilde{p}(\mathbf{x}_k^{(\ell)})\frac{\hat{B}(\mathbf{x}_k^{(\ell)})|\phi_k^{(\ell-1)}\rangle}{\langle\psi_T|\hat{B}(\mathbf{x}_k^{(\ell)})|\phi_k^{(\ell-1)}\rangle},
\end{aligned}\end{equation}
where Eq.~(\ref{eq:ExpDtH}) is used, and $\tilde{p}(\mathbf{x}_k^{(\ell)})$ takes the form
\begin{equation}\begin{aligned}
\label{eq:ptilde}
\tilde{p}(\mathbf{x}_k^{(\ell)}) = p(\mathbf{x}_k^{(\ell)})\frac{\langle\psi_T|\hat{B}(\mathbf{x}_k^{(\ell)})|\phi_k^{(\ell-1)}\rangle}{\langle\psi_T|\phi_k^{(\ell-1)}\rangle}.
\end{aligned}\end{equation}
The constrained-path approximation is implemented by enforcing the condition $\langle\psi_T|\phi_k^{(\ell)}\rangle>0$ for all $k$ and $\ell$. During the projection, the weight $\omega_k^{(\ell)}$ can fluctuate significantly across different random walkers (as some walkers can have very large or small weights). A population control procedure~\cite{Calandra1998} is usually employed to overcome this issue, and to render the weight distribution more uniform, which can effectively reduce the statistical error of physical observables. 

It is evident that one can use $\tilde{p}(\mathbf{x}_k^{(\ell)})$ in Eq.~(\ref{eq:ptilde}) as the probability function to sample $\mathbf{x}_k^{(\ell)}$, and accordingly to evaluate the summation (or integral) over $\mathbf{x}_k^{(\ell)}$ in Eq.~(\ref{eq:CPWvfct2}). In the following, we continue to refer to this process in CPQMC method as an update of the auxiliary fields, although it is more like a direct sampling (but still with importance sampling). More specifically for the Hubbard model~(\ref{eq:2DHubbard}), we can either sample every discrete field $x_{k,i}^{(\ell)}$ (with $i$ as the site index) in $\mathbf{x}_k^{(\ell)}$ sequentially, or generalize all its components via the force
bias. These actually correspond to the local update and the full force-bias update in CPQMC method. 

In the local update scheme, we can compute a probability $\lambda(x_{k,i}^{(\ell)}=\pm1)$ from $\tilde{p}(\mathbf{x}_k^{(\ell)})$, construct the normalized probability $\bar{\lambda}(x_{k,i}^{(\ell)})=\lambda(x_{k,i}^{(\ell)})/\mathcal{N}_{k,i}$ [with $\mathcal{N}_{k,i}=\lambda(x_{k,i}^{(\ell)}=+1)+\lambda(x_{k,i}^{(\ell)}=-1)$], and then sample $x_{k,i}^{(\ell)}$ via a heat-bath approach using $\bar{\lambda}(x_{k,i}^{(\ell)})$. Meanwhile, the weight $\omega_k^{(\ell-1)}$ should be multiplied by the factor $\prod_i\mathcal{N}_{k,i}$ and carried over into $\omega_k^{(\ell)}$, which serves as the major contribution to the fluctuation of the weight. The local update process also occupies more than $80\%$ of the overall computational time in CPQMC simulations. In the full force-bias update, we simply take $\langle\psi_l|$ and $|\psi_r\rangle$ in Eq.~(\ref{eq:FullFrcbsPhiLPhiR}) as $\langle\psi_l|=\langle\psi_T|$ and $|\psi_r\rangle=e^{-\Delta\tau\hat{H}_0}|\phi_k^{(\ell-1)}\rangle$, compute the force bias $\bar{n}_{i\sigma}$, and then sample all components in $\mathbf{x}_k^{(\ell)}$ using the distribution $\tilde{p}(\mathbf{x}_k^{(\ell)})$ in Eq.~(\ref{eq:ptilde}). This process also involves a heat-bath sampling (see Sec.~\ref{sec:LocalFrcbias}), which similarly produces a normalization factor $\prod_i\mathcal{N}_{k,i}$ that transforms $\omega_k^{(\ell-1)}$ to $\omega_k^{(\ell)}$. For both update schemes, the matrix $\mathbf{T}^{\sigma}=(\mathbf{U}_{L}^{\sigma}\mathbf{U}_{R}^{\sigma})^{-1}$, similar to that in PQMC method, still plays the central role in CPQMC, for which the initial $(\mathbf{U}_{L}^{\sigma}\mathbf{U}_{R}^{\sigma})$ corresponds to the matrix involved in $\langle\psi_T|e^{-\Delta\tau\hat{H}_0}|\phi_k^{(\ell-1)}\rangle$. The upgrade of $\mathbf{T}^{\sigma}$ follows the same procedures presented in Sec.~\ref{sec:LocalFrcbias} for the PQMC method. 

The delayed update and block force-bias update developed in this work can then be straightforwardly incorporated into the update process of the CPQMC method. For delayed update, we still use Eq.~(\ref{eq:DelayRgenk}) to construct the vectors $\mathbf{x}_m$ and $\mathbf{y}_m^{\rm T}$ (without the need to compute $r_{i_k,\sigma}$), and apply Eq.~(\ref{eq:FinalDelay}) to replace multiple vector-vector outer products by a single matrix multiplication. The only key difference is that, since CPQMC does not involve an acceptance ratio, the matrix $\mathbf{X}$ ($\mathbf{Y}$) in Eq.~(\ref{eq:FinalDelay}) takes the size $N_{\sigma} \times n_d$ ($n_d \times N_{\sigma}$), differing from that in PQMC. Or, equivalently, we can simply take $n_{\rm A}=n_d$ in Eq.~(\ref{eq:FinalDelay}) for a single-step delayed update in CPQMC method, with the computational cost accordingly scaling as $\mathcal{O}(3n_dN_{\sigma}^2)$. For full force-bias update, although its direct implementation in CPQMC method no longer suffers from low acceptance ratios, it can instead induce severe fluctuations in the weights of the random walkers. This is evidenced by the fact that, as the system size grows, the normalization factor $\prod_{i=1}^{N_s}\mathcal{N}_{k,i}$ arising from the heat-bath sampling tends to become more widely scattered across different walkers. Besides, due to the global generalization of all components in $\mathbf{x}_k^{(\ell)}$, the zero overlap condition (i.e., $\langle\psi_T|\phi_k^{(\ell)}\rangle=0$) is more likely to occur. Both issues necessitate a larger number of walkers $N_w$ or more frequent population control procedures, thereby significantly increasing the overall computational cost. An alternative way to overcome these issues is to apply the block force-bias update, as presented in Sec.~\ref{sec:BlockForcebias}. In the block update scheme for the CPQMC method, a single-step update involving only $n_b$ (also as the block size) auxiliary fields produces a normalization factor $\prod_{i=1}^{n_b}\mathcal{N}_{k,i}$ that barely depends on the system size [similar to the factor $f_{\mathbf{x}}^{(b)}$ in Eq.~(\ref{eq:BlkFrcBsErr}) for PQMC]. Comparing to the full scheme, the reduced factor can alleviate the fluctuation in the weights of the random walkers. In practical CPQMC simulations, the calculation of $(\mathbf{U}_{L}^{\sigma}\mathbb{U}_{R}^{\sigma})^{-1}$ in Eq.~(\ref{eq:BlkFrcBsTnew1}) for the force bias and the upgrade of $\mathbf{T}^{\sigma}$ matrix using Eq.~(\ref{eq:BlkFrcBsTnew2}) after a block update, are actually merged into a single formula similar to Eq.~(\ref{eq:BlkFrcBsTnew1}), but with a slightly different $\boldsymbol{\Delta}_b^{\sigma}$ matrix now hosting $(\boldsymbol{\Delta}_b^{\sigma})_{i,i}=e^{+{\rm i}\gamma f_{\sigma}x_{k,i}^{(\ell)}}-1$ for $1\le i\le n_b$ and zero otherwise. Consequently, the leading computational complexity remains $\mathcal{O}(n_b N_{\sigma}^2)$.

Since the vector and matrix operations involved are similar in each update scheme for PQMC and CPQMC methods, we expect the delayed and block force-bias updates to yield comparable speedups over the local update via appropriate choices of $n_d$ and $n_b$ in CPQMC simulations, as observed in PQMC. Moreover, as discussed in Sec.~\ref{sec:SOCHubbardTest}, both update schemes are applicable to the simulations involving arbitrary off-site two-body interactions in fermionic systems. For example, for the interaction term $\hat{h}_I=c_i^+c_j^{}c_k^+c_l^{}+h.c.=(\hat{T}_1^2+\hat{T}_2^2)/2$ where $\hat{T}_1=(c_i^+c_j^{}+c_k^+c_l^{})+h.c.$ and $\hat{T}_2={\rm i}(c_i^+c_j^{}+c_k^+c_l^{})+h.c.$, we can transform the $e^{-\Delta\tau\hat{h}_I}$ operator to single-particle propagators $e^{\gamma_1 x_1\hat{T}_1}$ and $e^{\gamma_2 x_2\hat{T}_2}$ (with $\gamma_1,\gamma_2$ and $x_1,x_2$ as the coupling coefficients and auxiliary fields, respectively) using the HS transformation with continuous auxiliary field~\cite{Shihao2013} or with four-component auxiliary field~\cite{Assaad1998,WangDa2014}. Since $\hat{T}_1$ and $\hat{T}_2$ only contain several hopping terms resembling the situation of SOC-Hubbard model discussed in Sec.~\ref{sec:SOCHubbardTest}, our delayed and block force-bias update schemes can be similarly implemented. Representative examples that have been explored previously include the NN density interaction in the spinless $t$-$V$ model~\cite{Zixiang2015,Sun2024} and the interlayer interactions in the bilayer lattice models~\cite{Yuanyao2016,Wu2016,Qin2017}. For the long-range Coulomb interaction in realistic systems~\cite{Shihao2021}, the phase problem generally arises and the use of the CPQMC method is necessary [the computational cost of the update process scales as $\mathcal{O}(N_s^2N_e^2)$ for long-range interactions]. We note that~\cite{Shihao2021} the Cholesky decomposition is typically applied to the interaction term as $\sum_{ijkl}V_{ijkl}c_i^+c_j^+c_k^{}c_l^{}=\sum_{\eta}\hat{L}_{\eta}^2$, and the HS transformation yields the single-particle propagator $e^{x_{\eta}\sqrt{-\Delta\tau}\hat{L}_{\eta}}$, with $\hat{L}_{\eta}=\sum_{il}(\mathbf{L}_{\eta})_{il}c_i^+c_l^{}$ as a general hopping bilinear term. Then sampling of the auxiliary fields is predominantly carried out using the full force-bias update. Alternatively, if the local update is used, it must operate on full matrices rather than vectors [such as $\mathbf{x}$ and $\mathbf{y}^{\rm T}$ in Eq.~(\ref{eq:UpdateLTRlocal}) for the on-site Hubbard interaction] because the representation matrix for $e^{x_{\eta}\sqrt{-\Delta\tau}\hat{L}_{\eta}}$ is generally dense. In this situation, the delayed update may no longer be necessary. Nevertheless, our block force-bias update scheme can still be employed to help alleviate the strong weight fluctuation across the random walkers during the projection that might arise in the full force-bias update process. Therefore, the delayed and block force-bias updates developed in this work may find even broader applicability in ground-state CPQMC simulations than in PQMC, particularly for a wide range of correlated fermionic systems.

\subsection{The application diagram for the update schemes}
\label{sec:ApplyDiagram}

Based on the efficiency tests of the delayed and force-bias update schemes as well as the related discussions in Secs.~\ref{sec:Results} and~\ref{sec:ApplyCPQMC}, we now present an application diagram of these update schemes from a broader perspective, within the unified ground-state AFQMC framework (including both PQMC and CPQMC) for general correlated fermion systems. The diagram is constructed in the parameter space spanned by the interaction strength and fermion filling. Given that the local update typically becomes less efficient as the system size increases, we focus here on the delayed update, as well as the full and block force-bias updates. 

\begin{figure}[t]
\centering
\includegraphics[width=0.99\linewidth]{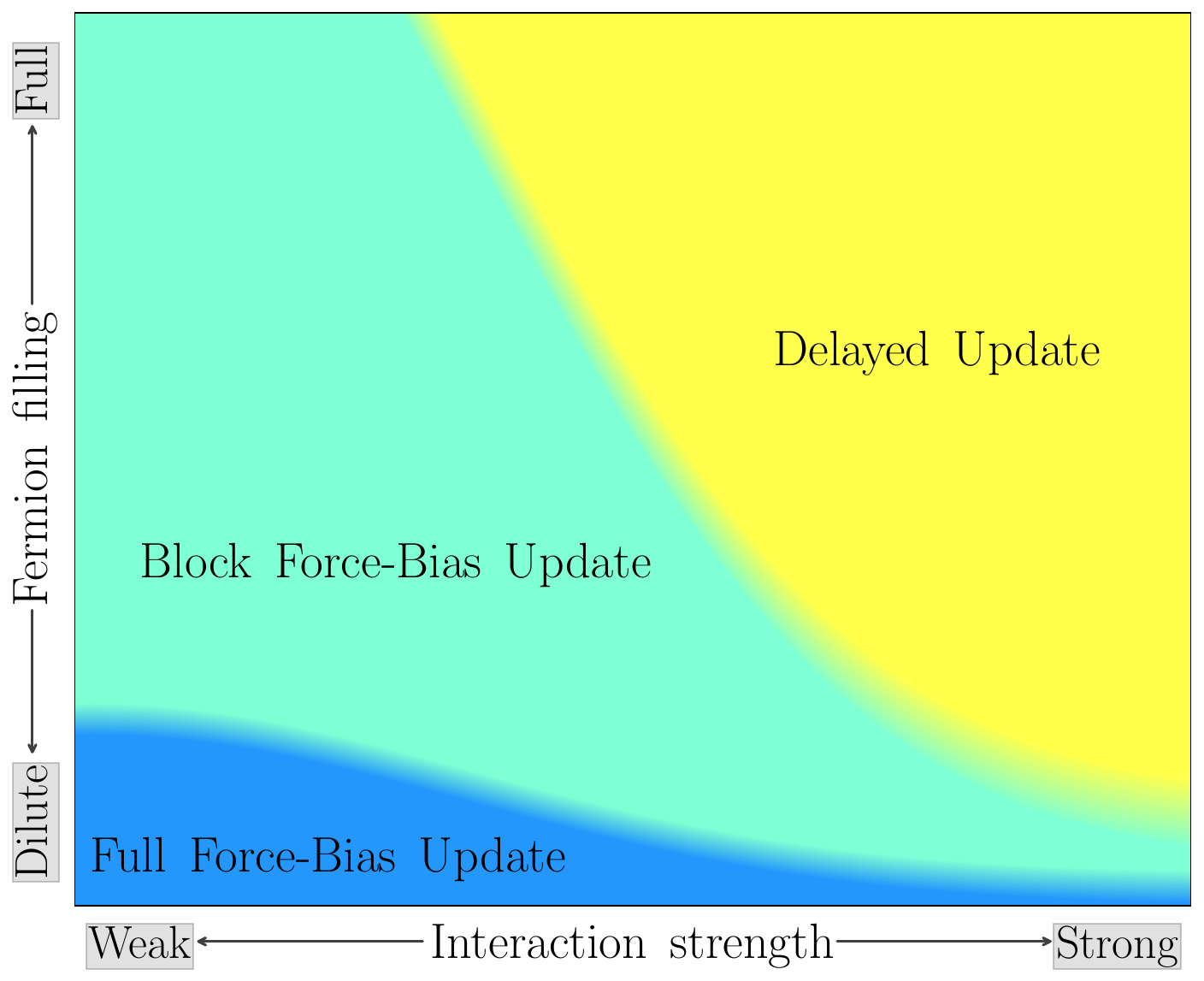}
\caption{The schematic application diagram of the delayed update, along with the full and block force-bias updates, for the ground-state AFQMC (including both PQMC and CPQMC) simulations of general correlated fermion systems, mapped over the parameter space of interaction strength and fermion filling. The shaded regions highlight where each update scheme achieves the highest efficiency, i.e., light blue for the full force-bias update, light green for the block force-bias update, and yellow for the delayed update. Note that the optimal application regions for delayed and block force-bias updates have overlap (see the accompanying text). }
\label{fig:ApplyDiagram}
\end{figure}

The schematic application diagram is shown in Fig.~\ref{fig:ApplyDiagram}. {\it First}, the full force-bias update is highly efficient in dilute systems, but its efficiency is suppressed as the interaction strength increases. As a result, it maintains the efficiency with slightly higher filling under weak interactions, while it is restricted to more dilute filling towards stronger interactions, resulting in the light blue shaded region in Fig.~\ref{fig:ApplyDiagram}. These points can be readily seen from Eq.~(\ref{eq:FrcBias}), the numerical results in Fig.~\ref{fig:FrcBsFermiGas}, and the discussions in Sec.~\ref{sec:LocalFrcbias}. {\it Second}, the block force-bias update expands the applicability of the full scheme into the dense filling regime. It achieves high efficiency for weak interactions across dilute to intermediate fillings. Similar to the full scheme, stronger interaction and higher filling continue to have competing effects on its efficiency, leading to the light green shaded region in Fig.~\ref{fig:ApplyDiagram}. These conclusions are supported by Eq.~(\ref{eq:BlkFrcBsErr}), the numerical results in Figs.~\ref{fig:BlkFrcBsNbtest} and \ref{fig:BlkFrcBsLtest}, and the discussions in Sec.~\ref{sec:BlockForcebias}. {\it Third}, the delayed update occupies the rest area (the yellow shaded region) in the plot. As demonstrated by the formalism in Sec.~\ref{sec:Delayed} and the numerical results in Figs.~\ref{fig:Delayedndtest} and \ref{fig:DelayedLtest}, the efficiency of the delayed update is primarily governed by the fermion filling. In contrast, its dependence on the interaction strength is rather weak, as evidenced by only minor variations in the acceptance ratio across different interaction strengths. Notably, the scheme achieves its highest efficiency at high fillings. Thus, the optimal application region in Fig.~\ref{fig:ApplyDiagram} for delayed update actually has overlap with the block force-bias update. 

We also note that the application diagram is not fixed. These update schemes, particularly the delayed and block force-bias updates, can be applied to other regions of the parameter space beyond their respective shaded areas. The key distinction lies in the achievable speedup compared to the local update. In practical AFQMC simulations for specific fermion systems, reasonable tests for the efficiency of these update schemes are always meaningful and essential.

\section{Conclusions}
\label{sec:Summary}

We have presented two highly efficient update schemes, the {\it delayed update} and {\it block force-bias update}, for the ground-state AFQMC simulations of correlated fermion systems. These update schemes preserves the same computational scaling as the local update [i.e., $\mathcal{O}(N_sN_e^2)$ for systems with local interactions] but with reduced prefactors, leading to accelerated simulations. We illustrated the efficiency of these updates with a set of benchmark calculations, comparing the consumed time with the local update, for the Hubbard models~(\ref{eq:2DHubbard}) and (\ref{eq:2DSocHubbard}) using the PQMC method. We showed that, compared to the local update, both update schemes can achieve significant speedups, which increase with growing system size and already reach $\sim$$8$ for systems with $1600$ lattice sites.

We have also discussed the implementations of these update schemes to the ground-state CPQMC method, which can deal with a broader class of interacting fermion systems suffering from the sign problem. Combining this with the PQMC method used in the efficiency tests, we have established an application diagram for the delayed and force-bias updates that applies to general correlated fermion systems, within the unified AFQMC framework. In practical applications, our proposed update schemes enable the AFQMC simulations of larger systems compared to conventional local update, considering the speedups. This enhanced computational efficiency allows for more reliable thermodynamic limit extrapolations through finite-size scaling. Therefore, we expect these update schemes to significantly change the prospect for zero-temperature many-body computations in various correlated fermion systems.

\begin{acknowledgments}
Y.-Y.H. acknowledges X. Y. Xu for valuable discussions. This work was supported by the National Natural Science Foundation of China (under Grants No. 12247103 and No. 12204377), the Innovation Program for Quantum Science and Technology (under Grant No. 2021ZD0301900), and the Youth Innovation Team of Shaanxi Universities.
\end{acknowledgments}

\appendix

\section{The PQMC formalism for spin-coupled systems}
\label{sec:AppendixA}

We consider the general lattice model for spin-$1/2$ fermions, described by the Hamiltonian $\hat{H}=\hat{H}_0+\hat{H}_I$, where $\hat{H}_0$ is the noninteracting part and $\hat{H}_I$ represents the two-body interactions. We denote $N_s$ as the number of lattice sites and $N_e$ as the number of fermions. With the notation $\mathbf{c}^+=(c_{1\uparrow}^+,c_{2\uparrow}^+,\cdots,c_{N_s\uparrow}^+,c_{1\downarrow}^+,c_{2\downarrow}^+,\cdots,c_{N_s\downarrow}^+)$, we can express $\hat{H}_0=\sum_{ij,\sigma\sigma^{\prime}}(\mathbf{H}_0)_{i\sigma,j\sigma^{\prime}} c_{i\sigma}^+c_{j\sigma^{\prime}}=\mathbf{c}^+\mathbf{H}_0\mathbf{c}$, with $\mathbf{\mathbf{H}_0}=\{(\mathbf{H}_0)_{i\sigma,j\sigma^{\prime}}\}$ is $2N_s\times 2N_s$ hopping matrix. The spin-orbit coupling (SOC) term are included in $\hat{H}_0$ implicitly. 

The PQMC method for spin-coupled systems works in the generalized Hartree-Fock basis, as illustrated by the above $\mathbf{c}^+$ operator. First of all, the initial wave function $|\psi_T\rangle$ in Eq.~(\ref{eq:TrialWvfc}) should be replaced by
\begin{equation}\begin{aligned}
\label{eq:TrialWvfcCouple}
|\psi_{T}\rangle = \prod_{m=1}^{N_e}(\boldsymbol{c}^+\mathbf{P})_m|0\rangle,
\end{aligned}\end{equation}
where $\mathbf{P}$ is a $2N_s\times N_e$ matrix. In the HS transformation, $e^{-\Delta\tau\hat{H}_I}=\sum_{\mathbf{x}}p(\mathbf{x})\hat{B}_I(\mathbf{x})$, the single-particle propagator $\hat{B}_I(\mathbf{x}) = \exp\big\{ \sum_{ij,\sigma\sigma^{\prime}}[\mathbf{H}_I(\mathbf{x})]_{i\sigma,j\sigma^{\prime}} c_{i\sigma}^+c_{j\sigma^{\prime}}\big\}=e^{\mathbf{c}^+\mathbf{H}_I(\mathbf{x})\mathbf{c}}$, where $\mathbf{H}_I(\mathbf{x})=\{[\mathbf{H}_I(\mathbf{x})]_{i\sigma,j\sigma^{\prime}}\}$ is a $2N_s\times 2N_s$ Hermitian or anti-Hermitian matrix. Then the configuration weight [corresponding to Eq.~(\ref{eq:ConfgWeight})] takes the form
\begin{equation}\begin{aligned}
W(\mathbf{X}) 
= P(\mathbf{X})\cdot{\rm det}(\mathbf{P}^+\mathbf{B}_{M}\cdots\mathbf{B}_{\ell}\cdots\mathbf{B}_{1}\mathbf{P}),
\end{aligned}\end{equation}
where $\mathbf{B}_{\ell}=e^{\mathbf{H}_I(\mathbf{x}_{\ell})}e^{-\Delta\tau\mathbf{H}_0}$ is the $2N_s\times 2N_s$ propagator matrix. For the measurements, the static single-particle Green's function matrix [corresponding to Eq.~(\ref{eq:GrFMat})] is now expressed as $\mathbf{G}(\ell\Delta\tau,\ell\Delta\tau)=\mathbf{1}_{2N_s}-\mathbf{R}(\mathbf{L}\mathbf{R})^{-1}\mathbf{L}$ with $\mathbf{L}=\mathbf{P}^+\mathbf{B}_{M}\mathbf{B}_{M-1}\cdots\mathbf{B}_{\ell+1}$ as an $N_e\times 2N_s$ matrix, $\mathbf{R}=\mathbf{B}_{\ell}\cdots\mathbf{B}_2\mathbf{B}_1\mathbf{P}$ is an $2N_s\times N_e$ matrix. Accordingly, the matrices $\mathbf{U}_L$, $\mathbf{U}_R$ and $\mathbf{T}=(\mathbf{U}_L\mathbf{U}_R)^{-1}$ after the numerical stablization procedure are of dimensions $N_e\times 2N_s$, $2N_s\times N_e$ and $N_e\times N_e$, respectively. So we have $\mathbf{G}(\ell\Delta\tau,\ell\Delta\tau)=\mathbf{1}_{2N_s}-\mathbf{U}_R\mathbf{T}\mathbf{U}_L$. 

For the SOC-Hubbard model~(\ref{eq:2DSocHubbard}) with the Hubbard interaction $\hat{H}_I=U\sum_{i}[\hat{n}_{i\uparrow}\hat{n}_{i\downarrow}-(\hat{n}_{i\uparrow}+\hat{n}_{i\downarrow})/2]$, we adopt the discrete HS transformation decoupled into the spin-$\hat{s}^y$ channel~\cite{Xie2025} (denoted as HS-$\hat{s}^y$) reading
\begin{equation}\begin{aligned}
\label{eq:HSspinSyDep}
e^{-\Delta\tau U \big(\hat{n}_{i\uparrow} \hat{n}_{i\downarrow} - \frac{\hat{n}_{i\uparrow} + \hat{n}_{i\downarrow}}{2}\big) }
= \frac{1}{2}\sum_{x_{i}=\pm1}e^{\gamma x_i(c_{i\uparrow}^+ c_{i\downarrow}^{} - c_{i\downarrow}^+ c_{i\uparrow}^{})},
\end{aligned}\end{equation}
with the coupling constant $\gamma=\cos^{-1}(e^{\Delta\tau U/2})$ for $U<0$. The the corresponding propagator $\hat{B}_I(\mathbf{x}_{\ell})$ is given by
\begin{equation}\begin{aligned}
\label{eq:BIOpForSy}
\hat{B}_I(\mathbf{x}_{\ell})
= {\rm exp}\Big[\gamma\sum_{i=1}^{N_s} x_{\ell,i}(c_{i\uparrow}^+ c_{i\downarrow}^{} - c_{i\downarrow}^+ c_{i\uparrow}^{})\Big]
= \prod_{i=1}^{N_s}\hat{b}_i(x_{\ell,i}), 
\end{aligned}\end{equation}
with $\hat{b}_i(x_{\ell,i})=e^{\gamma x_{\ell,i}(c_{i\uparrow}^+ c_{i\downarrow}^{} - c_{i\downarrow}^+ c_{i\uparrow}^{})}$. Evaluating the trace in partition function, we can reach the matrix form for $\hat{B}_I(\mathbf{x}_{\ell})$ operator practically as $e^{\mathbf{H}_I(\mathbf{x}_{\ell})}=\prod_{i=1}^{N_s}\mathbf{b}_i(x_{\ell,i})$, where a effective $2\times2$ matrix is involved in $\mathbf{b}_i(x_{\ell,i})$ as
\begin{equation}\begin{aligned}
\label{eq:A2times2}
\mathbf{A}(x_{\ell,i}) = e^{\gamma x_{\ell,i}\cdot{\rm i}\sigma_y} = 
\begin{pmatrix}
{\rm cos}(\gamma x_{\ell,i}) & {\rm sin}(\gamma x_{\ell,i}) \\
-{\rm sin}(\gamma x_{\ell,i}) & {\rm cos}(\gamma x_{\ell,i})
\end{pmatrix},
\end{aligned}\end{equation}
with $\sigma_x$ as the first Pauli matrix. Then the $\mathbf{b}_i(x_{\ell,i})$ matrix has only four nonzero elements as $[\mathbf{b}_i(x_{\ell,i})]_{i,i}=A_{11}$, $[\mathbf{b}_i(x_{\ell,i})]_{i,N_s+i}=A_{12}$, $[\mathbf{b}_i(x_{\ell,i})]_{N_s+i,i}=A_{21}$, and $[\mathbf{b}_i(x_{\ell,i})]_{N_s+i,N_s+i}=A_{22}$. For the general update process on the $\ell$-th time slice as $\mathbf{x}_{\ell}\to\mathbf{x}_{\ell}^{\prime}$, the determinant ratio in Eq.~(\ref{eq:DetRatio10}) is replaced by 
\begin{equation}\begin{aligned}
\label{eq:DetRatioSOC}
R = \frac{{\rm det}[\mathbf{U}_L(\mathbf{1}_{N_s}+\boldsymbol{\Delta})\mathbf{U}_R]}{{\rm det}(\mathbf{U}_L\mathbf{U}_R)}
= {\rm det}(\mathbf{1}_{N_e}+\mathbf{U}_L\boldsymbol{\Delta}\mathbf{U}_R\mathbf{T}),
\end{aligned}\end{equation}
with $\boldsymbol{\Delta}=e^{\mathbf{H}_I(\mathbf{x}_{\ell}^{\prime})}e^{-\mathbf{H}_I(\mathbf{x}_{\ell})}-\mathbf{1}_{2N_s}$ as a $2N_s\times 2N_s$ matrix. The corresponding upgrade of $\mathbf{U}_R$ and $\mathbf{T}$ matrices upon an accepted update [corresponding to Eq.~(\ref{eq:UpdateLTRMat})] then reads
\begin{equation}\begin{aligned}
\label{eq:UpdateLTRSOC}
(\mathbf{U}_{R})^{\prime} 
&= (\mathbf{1}_{2N_s}+\boldsymbol{\Delta})\mathbf{U}_{R}, \\
\mathbf{T}^{\prime} 
&= \mathbf{T}(\mathbf{1}_{N_e}+\mathbf{U}_L\boldsymbol{\Delta}\mathbf{U}_R\mathbf{T})^{-1}.
\end{aligned}\end{equation}

For the local update $x_{\ell,i}\to x_{\ell,i}^{\prime}$, the above $\boldsymbol{\Delta}$ matrix has only four nonzero elements, whose evaluations involve an effective $2\times2$ matrix as $\boldsymbol{\delta}_i = \mathbf{A}(x_{\ell,i}^{\prime})[\mathbf{A}(x_{\ell,i})]^{-1}-\mathbf{1}_2$ with $\mathbf{A}(x_{\ell,i})$ computed from Eq.~(\ref{eq:A2times2}). Then we can reach $\Delta_{i,i}=(\boldsymbol{\delta}_i)_{11}$, $\Delta_{i,N_s+i}=(\boldsymbol{\delta}_i)_{12}$, $\Delta_{N_s+i,i}=(\boldsymbol{\delta}_i)_{21}$, and $\Delta_{N_s+i,N_s+i}=(\boldsymbol{\delta}_i)_{22}$. Based on the sparsity of $\boldsymbol{\Delta}$, the calculation of the determinant ratio $R$ in Eq.~(\ref{eq:DetRatioSOC}) and the upgrade of $\mathbf{T}$ matrix in Eq.~(\ref{eq:UpdateLTRSOC}) can be simplified as
\begin{equation}\begin{aligned}
\label{eq:LocalSOC}
R 
&= {\rm det}\big(\mathbf{1}_{N_e} + \mathbf{V}\mathbf{Y}\big) = {\rm det}\big(\mathbf{1}_2 + \mathbf{Y}\mathbf{V}\big), \\
\mathbf{T}^{\prime} 
&= \mathbf{T}\big(\mathbf{1}_{N_e} + \mathbf{V}\mathbf{Y}\big)^{-1}
= \mathbf{T} - \mathbf{X}\mathbf{Y},
\end{aligned}\end{equation}
where $\mathbf{V}=(\mathbf{v}_i|\mathbf{v}_{N_s+i})\times\boldsymbol{\delta}_i$ is an $N_e\times 2$ matrix, $\mathbf{X}=\mathbf{T}\mathbf{V}/R$ is also an $N_e\times 2$ matrix, and $\mathbf{Y}=(\mathbf{u}_i|\mathbf{u}_{N_s+i})^{\rm T}\times\mathbf{T}$ is a $2\times N_e$ matrix, with the row vector $\mathbf{u}_i^{\rm T}=(\mathbf{U}_R)_{i{\rm-row}}$ and the column vector $\mathbf{v}_i=(\mathbf{U}_L)_{i{\rm-col}}$. Comparing to the formulas for the spin-decoupled case in Eqs.~(\ref{eq:DetRatiolocal}) and (\ref{eq:UpdateLTRlocal}), where single-vector operations are used, the spin-coupled case requires bi-vector calculations, as shown in Eq.~(\ref{eq:LocalSOC}).

For the full force-bias update, the equations (\ref{eq:RatioApprox}) and (\ref{eq:FrcBias}) should be accordingly changed with the new force bias related to the HS transformation in Eq.~(\ref{eq:HSspinSxDep}) as $\langle c_{i\uparrow}^+ c_{i\downarrow}^{}\rangle$ and $\langle c_{i\downarrow}^+ c_{i\uparrow}^{}\rangle$, regarding the wave functions $\langle\psi_l|$ and $|\psi_r\rangle$. These measurements are no longer the diagonal elements of the single-particle Green's function, but they can also be evaluated similarly, e.g., $\langle c_{i\uparrow}^+ c_{i\downarrow}^{}\rangle=[\mathbb{U}_{R}(\mathbf{U}_{L}\mathbb{U}_{R})^{-1}\mathbf{U}_{L}]_{i,N_s+i}$ with $\mathbb{U}_{R}=e^{-\mathbf{H}_I(\mathbf{x}_{\ell})}\mathbf{U}_{R}$ (the calculation of $\mathbb{U}_{R}$ can be integrated into the propagation of $\mathbf{U}_{R}$ matrix). The following calculation of determinant ratio $R$ and the upgrade of $\mathbf{T}$ matrix are similar to that for the spin-decoupled case as described in Sec.~\ref{sec:LocalFrcbias}. 

An alternative HS transformation decoupled into spin-$\hat{s}^x$ channel~\cite{Xie2025} (denoted as HS-$\hat{s}^x$) can also be used for the Hubbard interaction as
\begin{equation}\begin{aligned}
\label{eq:HSspinSxDep}
e^{-\Delta\tau U \big(\hat{n}_{i\uparrow} \hat{n}_{i\downarrow} - \frac{\hat{n}_{i\uparrow} + \hat{n}_{i\downarrow}}{2}\big) }
= \frac{1}{2}\sum_{x_{i}=\pm1}e^{{\rm i}\gamma x_i(c_{i\uparrow}^+ c_{i\downarrow}^{} + c_{i\downarrow}^+ c_{i\uparrow}^{})},
\end{aligned}\end{equation}
with the coupling constant $\gamma=\cos^{-1}(e^{\Delta\tau U/2})$ for $U<0$. Accordingly, the $\hat{B}_I(\mathbf{x}_{\ell})$ operator in Eq.~(\ref{eq:BIOpForSy}) and the $\mathbf{A}(x_{\ell,i})$ matrix in Eq.~(\ref{eq:A2times2}) needs to be adjusted. Specifically, for HS-$\hat{s}^x$, the $\mathbf{A}(x_{\ell,i})$ matrix takes the following form as
\begin{equation}\begin{aligned}
\label{eq:A2times2Sx}
\mathbf{A}(x_{\ell,i}) = e^{\gamma x_{\ell,i}\cdot{\rm i}\sigma_x} = 
\begin{pmatrix}
{\rm cos}(\gamma x_{\ell,i}) & {\rm i}\cdot{\rm sin}(\gamma x_{\ell,i}) \\
{\rm i}\cdot{\rm sin}(\gamma x_{\ell,i}) & {\rm cos}(\gamma x_{\ell,i})
\end{pmatrix}.
\end{aligned}\end{equation}
The rest calculations stay unchanged, i.e., the same as that for HS-$\hat{s}^y$.

For the delayed update, the vectors $\mathbf{x}_m$ and $\mathbf{y}_m^{\rm T}$ in Eqs.~(\ref{eq:DelayTMat00}) and (\ref{eq:DelayRgenk}) for spin-decoupled case, respectively evolves into $N_e\times2$ and $2\times N_e$ matrices for the spin-couple case. In the delayed upgrade of $\mathbf{T}$ matrix in Eq.~(\ref{eq:FinalDelay}), the $\mathbf{X}$ and $\mathbf{Y}$ matrices accordingly changes to the $N_e\times2n_{\rm A}$ and $2n_{\rm A}\times N_e$ matrices, respectively, containing both spin-up and spin-down column (row) vectors. Similar modifications for the dimensions of the vectors and matrices apply to the block force-bias update scheme [corresponding to Eqs.~(\ref{eq:BlkFrcBsTnew1}) and (\ref{eq:BlkFrcBsTnew2})] for the spin-coupled case, with the new force bias as discussed above.

\section{The implementation with symmetric Trotter-Suzuki decomposition}
\label{sec:AppendixB}

The symmetric Trotter-Suzuki decomposition reads
\begin{equation}\begin{aligned}
e^{-\Delta\tau\hat{H}}=e^{-\Delta\tau\hat{H}_0/2}e^{-\Delta\tau\hat{H}_I}e^{-\Delta\tau\hat{H}_0/2}+\mathcal{O}[(\Delta\tau)^3].
\end{aligned}\end{equation}
It modifies the propagator $\hat{B}(\mathbf{x}_{\ell})=\hat{B}_I(\mathbf{x}_{\ell})e^{-\Delta\tau\hat{H}_0}$ in Eq.~(\ref{eq:ExpDtH}) to a new form as $e^{-\Delta\tau\hat{H}_0/2}\hat{B}_I(\mathbf{x}_{\ell})e^{-\Delta\tau\hat{H}_0/2}$, with $e^{-\Delta\tau\mathbf{H}_0^{\sigma}/2}e^{\mathbf{H}_I^{\sigma}(\mathbf{x}_{\ell})}e^{-\Delta\tau\mathbf{H}_0^{\sigma}/2}$ as its corresponding matrix representation. In practical simulations, repeated multiplication by the matrix $e^{\pm\Delta\tau\mathbf{H}_0^{\sigma}/2}$ matrix during the propagation of $\mathbf{U}_R$ and $\mathbf{U}_L$ [as in computing $\mathbf{B}_{\ell+1}^{\sigma}\mathbf{U}_R$ and $\mathbf{U}_L(\mathbf{B}_{\ell-1}^{\sigma})^{-1}$] can be avoided. This is achieved by precomputing the rescaled wave functions $e^{+\Delta\tau\hat{H}_0/2}|\psi_{T}\rangle$ and $\langle\psi_{T}|e^{-\Delta\tau\hat{H}_0/2}$ (accordingly, the matrices $e^{+\Delta\tau\mathbf{H}_0^{\sigma}/2}\mathbf{P}$ and $\mathbf{P}^+e^{-\Delta\tau\mathbf{H}_0^{\sigma}/2}$), before the core PQMC simulation. Then, the propagation of $\mathbf{U}_R$ and $\mathbf{U}_L$ along imaginary-time slices can proceed in the same manner as in the formalism employing the asymmetric Trotter-Suzuki decomposition in Eq.~(\ref{eq:AsymTrot}). For example, the Slater determinant wave functions can be formulated as
\begin{equation}\begin{aligned}
\langle\psi_l| &= \langle\psi_T|e^{-\Delta\tau\hat{H}_0/2}\hat{B}(\mathbf{x}_M)\cdots\hat{B}(\mathbf{x}_{\ell+2})\hat{B}(\mathbf{x}_{\ell+1}),  \\
|\psi_r\rangle &= \hat{B}(\mathbf{x}_{\ell})\cdots\hat{B}(\mathbf{x}_2)\hat{B}(\mathbf{x}_1)e^{+\Delta\tau\hat{H}_0/2}|\psi_T\rangle,
\end{aligned}\end{equation}
with $\hat{B}(\mathbf{x}_{\ell})=\hat{B}_I(\mathbf{x}_{\ell})e^{-\Delta\tau\hat{H}_0}$. Then the field update process can work with these wave functions, $\langle\psi_l|$ and $|\psi_r\rangle$, applying the update schemes outlined in the main text. Nevertheless, when measuring observables, we need to perform the additional multiplications as $\mathbf{U}_L e^{+\Delta\tau\mathbf{H}_0^{\sigma}/2}$ and $e^{-\Delta\tau\mathbf{H}_0^{\sigma}/2}\mathbf{U}_R$, since the static measurement of $\hat{O}$ operator follows
\begin{equation}\begin{aligned}
\langle\hat{O}\rangle_{\mathbf{X}} 
&= \frac{\langle\psi_l|e^{+\Delta\tau\hat{H}_0/2}\ \hat{O}\ e^{-\Delta\tau\hat{H}_0/2}|\psi_r\rangle}{\langle\psi_l|e^{+\Delta\tau\hat{H}_0/2}\cdot e^{-\Delta\tau\hat{H}_0/2}|\psi_r\rangle}.
\end{aligned}\end{equation}
As a result, the additional computational cost of the symmetric decomposition is negligible in practical PQMC simulations compared to the asymmetric one. Moreover, both Trotter-Suzuki decompositions can be implemented within the same PQMC codebase, enabling a flexible switch between them.

\section{The Sherman-Morrison formulas and proof for Eqs.~(\ref{eq:DetRatio10}), (\ref{eq:UpdateLTRMat}), (\ref{eq:DetRatiolocal}), (\ref{eq:UpdateLTRlocal}), (\ref{eq:BlkFrcBsTnew1}), and (\ref{eq:RatioApprox})}
\label{sec:AppendixC}

In both ground-state and finite-temperature AFQMC algorithms, the Sherman-Morrison formula and its generalized matrix form play a central role in the auxiliary-field update process. In this appendix, we first present the formula in both its vector and matrix forms, and then apply them to derive several key results in the main text. 

The vector version of the Sherman-Morrison formula states that, for an invertible $N\times N$ matrix $\mathbf{A}$, and colum vectors $\mathbf{x}$, $\mathbf{y}$ of dimension $N$, satisfying $(1+\mathbf{y}^{\rm T}\mathbf{A}^{-1}\mathbf{x})\ne0$, then the following relation holds
\begin{equation}\begin{aligned}
\label{eq:ShermanVector}
(\mathbf{A}+\mathbf{x}\mathbf{y}^{\rm T})^{-1}
= \mathbf{A}^{-1} - \frac{\mathbf{A}^{-1}\mathbf{x}\mathbf{y}^{\rm T}\mathbf{A}^{-1}}{1+\mathbf{y}^{\rm T}\mathbf{A}^{-1}\mathbf{x}}.
\end{aligned}\end{equation}

The matrix version of the Sherman-Morrison formula states that, for an invertible $N\times N$ matrix $\mathbf{A}$, and $N\times M$ matrices $\mathbf{U}$ and $\mathbf{V}$, provided that $(\mathbf{1}_M+\mathbf{V}^{\rm T}\mathbf{A}^{-1}\mathbf{U})$ is invertible, then the following relation holds
\begin{equation}\begin{aligned}
\label{eq:ShermanMatrix}
&(\mathbf{A}+\mathbf{U}\mathbf{V}^{\rm T})^{-1}
= \mathbf{A}^{-1}  \\
&\hspace{1.0cm} - \mathbf{A}^{-1}\mathbf{U}(\mathbf{1}_M+\mathbf{V}^{\rm T}\mathbf{A}^{-1}\mathbf{U})^{-1}\mathbf{V}^{\rm T}\mathbf{A}^{-1}.
\end{aligned}\end{equation}

The general configuration update framework within a given imaginary-time slice discussed in Sec.~\ref{sec:GeneralUpdate} is the starting point for other detailed update schemes. Equation (\ref{eq:DetRatio10}) is then evident from the fundamental property of the matrix determinant as
\begin{equation}\begin{aligned}
\label{eq:DetRatioApp}
r_{\sigma} 
&= \frac{{\rm det}\big[\mathbf{U}_L^{\sigma}(\mathbf{1}_{N_s}+\boldsymbol{\Delta}^{\sigma})\mathbf{U}_R^{\sigma}\big]}{{\rm det}\big(\mathbf{U}_L^{\sigma}\mathbf{U}_R^{\sigma}\big)} \\
&= {\rm det}\big[\mathbf{U}_L^{\sigma}\mathbf{U}_R^{\sigma}+\mathbf{U}_L^{\sigma}\boldsymbol{\Delta}^{\sigma}\mathbf{U}_R^{\sigma}\big]\cdot{\rm det}\big[(\mathbf{U}_L^{\sigma}\mathbf{U}_R^{\sigma})^{-1}\big] \\
&= {\rm det}\big(\mathbf{1}_{N_{\sigma}}+\mathbf{U}_L^{\sigma}\boldsymbol{\Delta}^{\sigma}\mathbf{U}_R^{\sigma}\mathbf{T}^{\sigma}\big), 
\end{aligned}\end{equation}
where we adopt $\mathbf{T}^{\sigma}=(\mathbf{U}_L^{\sigma}\mathbf{U}_R^{\sigma})^{-1}$. Once the update is accepted, the new $\mathbf{T}^{\sigma}$ matrix is expressed as 
\begin{equation}\begin{aligned}
\label{eq:UpdateTApp}
(\mathbf{T}^{\sigma})^{\prime}
&= \big[\mathbf{U}_L^{\sigma}(\mathbf{1}_{N_s}+\boldsymbol{\Delta}^{\sigma})\mathbf{U}_R^{\sigma}\big]^{-1} \\
&= \big\{\big[\mathbf{1}_{N_s}+\mathbf{U}_L^{\sigma}\boldsymbol{\Delta}^{\sigma}\mathbf{U}_R^{\sigma}(\mathbf{U}_L^{\sigma}\mathbf{U}_R^{\sigma})^{-1}\big](\mathbf{U}_L^{\sigma}\mathbf{U}_R^{\sigma})\big\}^{-1} \\
&= (\mathbf{U}_L^{\sigma}\mathbf{U}_R^{\sigma})^{-1}\big[\mathbf{1}_{N_s}+\mathbf{U}_L^{\sigma}\boldsymbol{\Delta}^{\sigma}\mathbf{U}_R^{\sigma}(\mathbf{U}_L^{\sigma}\mathbf{U}_R^{\sigma})^{-1}\big]^{-1} \\
&= \mathbf{T}^{\sigma}\big(\mathbf{1}_{N_{\sigma}}+\mathbf{U}_L^{\sigma}\boldsymbol{\Delta}^{\sigma}\mathbf{U}_R^{\sigma}\mathbf{T}^{\sigma}\big)^{-1},
\end{aligned}\end{equation}
which is exactly the upgrading formula for $\mathbf{T}^{\sigma}$ in Eq.~(\ref{eq:UpdateLTRMat}). 

We can then make simplifications for the inverse matrix $(\mathbf{1}_{N_{\sigma}}+\mathbf{U}_L^{\sigma}\boldsymbol{\Delta}^{\sigma}\mathbf{U}_R^{\sigma}\mathbf{T}^{\sigma})^{-1}$ in the above expression by applying the Sherman-Morrison formulas, which can be categorized into several cases depending on the $\boldsymbol{\Delta}^{\sigma}$ matrix. 

\begin{table}[h]
\caption{The timing data for Fig.~\ref{fig:Delayedndtest}(a), as the average time for update per sweep (in seconds) by the local and delay updates versus various values of the delay rank $n_d$ and linear system size $L$, in PQMC simulations of the model~(\ref{eq:2DHubbard}) with $U/t=-4$ and $n=1.0$.}
\setlength{\tabcolsep}{8pt}
\centering 
\resizebox{\linewidth}{!}{ 
\begin{tabular}{|c|c|c|c|c|}  
\hline
\diagbox{$n_d/L$}{$L$} & $8$ & $16$ & $24$ & $32$\\ 
\hline  
Local & 0.213 & 9.88 & 225 & 1387\\  
\hline  
$0.25$ & & 7.82 & 112 & 600\\  
\hline  
$0.50$ & 0.221 & 6.24 & 78.2 & 475\\ 
\hline  
$0.75$ & & 6.10 & 81.1 & 448\\   
\hline  
$1.00$ & 0.215 & 6.56 & 77.3 & 467\\  
\hline  
$1.50$ & 0.214 & 7.10 & 81.1 & 505\\  
\hline  
$2.00$ & 0.235 & 8.34 & 91.7 & 592\\  
\hline   
$3.00$ & 0.265 & 9.80 & 107 & 661\\  
\hline  
\end{tabular}
}
\label{Table:A1}
\end{table}

\begin{table}[h]
\caption{The timing data for Fig.~\ref{fig:Delayedndtest}(b), as the average time for update per sweep (in seconds) by the local and delay updates versus various values of the delay rank $n_d$ and linear system size $L$, in PQMC simulations of the model~(\ref{eq:2DHubbard}) with $U/t=-6$ and $n=0.625$.}
\setlength{\tabcolsep}{9pt}
\centering
\resizebox{\linewidth}{!}{   
\begin{tabular}{|c|c|c|c|c|}  
    \hline
    \diagbox{$n_d/L$}{$L$} & $8$ & $16$ & $24$ & $32$\\ 
    \hline  
    Local & 0.124 & 3.22 & 50.3 & 447\\  
    \hline  
    $0.25$ & & 3.20 & 37.8 & 192\\  
    \hline  
    $0.33$ & & & 30.2 &\\
    \hline  
    $0.50$ & 0.135 & 2.73 & 29.7 & 172\\ 
    \hline  
    $0.67$ & & & 31.5 &\\  
    \hline  
    $0.75$ & 0.123 & 2.88 & 32.2 & 176\\   
    \hline  
    $1.00$ & 0.122 & 3.14 & 32.3 & 193\\
    \hline 
    $1.50$ & 0.130 & 3.58 & 36.4 & 235\\  
    \hline  
    $2.00$ & 0.145 & 4.33 & 44.1 & 286\\  
    \hline  
\end{tabular}
}
\label{Table:A2}
\end{table}

{\it First}, for a general $\boldsymbol{\Delta}^{\sigma}$, we can directly apply Eq.~(\ref{eq:ShermanMatrix}) for the inverse by setting $\mathbf{A}=\mathbf{1}_{N_s}$, $\mathbf{U}=\mathbf{U}_L^{\sigma}\boldsymbol{\Delta}^{\sigma}$, and $\mathbf{V}^{\rm T}=\mathbf{U}_R^{\sigma}\mathbf{T}^{\sigma}$, which produces 
\begin{equation}\begin{aligned}
&(\mathbf{1}_{N_{\sigma}}+\mathbf{U}_L^{\sigma}\boldsymbol{\Delta}^{\sigma}\mathbf{U}_R^{\sigma}\mathbf{T}^{\sigma})^{-1}
= \mathbf{1}_{N_{\sigma}} \\
&\hspace{1.0cm} - \mathbf{U}_L^{\sigma}\boldsymbol{\Delta}^{\sigma}(\mathbf{1}_{N_{\sigma}}+\mathbf{U}_R^{\sigma}\mathbf{T}^{\sigma}\mathbf{U}_L^{\sigma}\boldsymbol{\Delta}^{\sigma})^{-1}\mathbf{U}_R^{\sigma}\mathbf{T}^{\sigma}.
\end{aligned}\end{equation}

{\it Second}, for on-site Hubbard interaction applying HS-$\hat{s}^z$ [see Eq.~(\ref{eq:HSspinDecomp})], the inverse $(\mathbf{1}_{N_{\sigma}}+\mathbf{U}_L^{\sigma}\boldsymbol{\Delta}^{\sigma}\mathbf{U}_R^{\sigma}\mathbf{T}^{\sigma})^{-1}$ can be reformulated depending on the update scheme. In the local update, the $\boldsymbol{\Delta}^{\sigma}$ matrix is diagonal and it has only one nonzero element, saying $\Delta_{i,i}^{\sigma}$. As a result, we can simplify the matrix multiplication in both Eqs.~(\ref{eq:DetRatioApp}) and (\ref{eq:UpdateTApp}) as $\mathbf{U}_L^{\sigma}\boldsymbol{\Delta}^{\sigma}\mathbf{U}_R^{\sigma}\mathbf{T}^{\sigma}=\mathbf{v}_i\mathbf{y}^{\rm T}$, with the column vector $\mathbf{v}_i=\Delta_{i,i}^{\sigma}(\mathbf{U}_L^{\sigma})_{i{\rm-col}}$, and the row vectors $\mathbf{u}_i^{\rm T}=(\mathbf{U}_R^{\sigma})_{i{\rm-row}}$ and $\mathbf{y}^{\rm T}=\mathbf{u}_i^{\rm T}\mathbf{T}^{\sigma}$. Then the determinant ratio in Eq.~(\ref{eq:DetRatioApp}) can be simplified as
\begin{equation}\begin{aligned}
r_{\sigma} 
&= {\rm det}\big(\mathbf{1}_{N_{\sigma}}+\mathbf{U}_L^{\sigma}\boldsymbol{\Delta}^{\sigma}\mathbf{U}_R^{\sigma}\mathbf{T}^{\sigma}\big) \\
&= {\rm det}(\mathbf{1}_{N_{\sigma}}+\mathbf{v}_i\mathbf{y}^{\rm T})
= 1 + \mathbf{y}^{\rm T}\mathbf{v}_i,
\end{aligned}\end{equation}
which thus proves Eq.~(\ref{eq:DetRatiolocal}). Accordingly, the inverse matrix can be evaluated using Eq.~(\ref{eq:ShermanVector}) as
\begin{equation}\begin{aligned}
(\mathbf{1}_{N_{\sigma}}+\mathbf{U}_L^{\sigma}\boldsymbol{\Delta}^{\sigma}\mathbf{U}_R^{\sigma}\mathbf{T}^{\sigma})^{-1}
&= (\mathbf{1}_{N_{\sigma}}+\mathbf{v}_i\mathbf{y}^{\rm T})^{-1} \\
&= \mathbf{1}_{N_{\sigma}} - \frac{\mathbf{v}_i\mathbf{y}^{\rm T}}{1 + \mathbf{y}^{\rm T}\mathbf{v}_i}.
\end{aligned}\end{equation}
Then with $\mathbf{x}=(\mathbf{T}^{\sigma}\mathbf{v}_i)/r_{i,\sigma}$ as a column vector, it subsequently simplifies Eq.~(\ref{eq:UpdateTApp}) as $(\mathbf{T}^{\sigma})^{\prime} = \mathbf{T}^{\sigma} - \mathbf{x}\mathbf{y}^{\rm T}$, which is exactly Eq.~(\ref{eq:UpdateLTRlocal}). For the delayed update, the upgrade of $\mathbf{T}^{\sigma}$ as $(\mathbf{T}^{\sigma})^{\prime} = \mathbf{T}^{\sigma} - \mathbf{x}\mathbf{y}^{\rm T}$ upon accepted update of $x_{\ell,i}$ is recursively applied to achieve the general term formula in Eq.~(\ref{eq:DelayRgenk}). For the full force-bias update, there is no need for the evaluation of the $(\mathbf{1}_{N_{\sigma}}+\mathbf{U}_L^{\sigma}\boldsymbol{\Delta}^{\sigma}\mathbf{U}_R^{\sigma}\mathbf{T}^{\sigma})^{-1}$ matrix. For the block force-bias update, when computing the force bias, we need to first compute $(\mathbf{U}_{L}^{\sigma}\mathbb{U}_{R}^{\sigma})^{-1}$ in Eq.~(\ref{eq:BlkFrcBsTnew1}) as
\begin{equation}\begin{aligned}
(\mathbf{U}_{L}^{\sigma}\mathbb{U}_{R}^{\sigma})^{-1} 
&= [\mathbf{U}_{L}^{\sigma}(\mathbf{1}_{N_s}+\boldsymbol{\Delta}_b^{\sigma})\mathbf{U}_{R}^{\sigma}]^{-1} \\
&= \mathbf{T}^{\sigma}(\mathbf{1}_{N_{\sigma}}+\mathbf{U}_L^{\sigma}\boldsymbol{\Delta}_b^{\sigma}\mathbf{U}_R^{\sigma}\mathbf{T}^{\sigma})^{-1},
\end{aligned}\end{equation}
where the diagonal matrix $\boldsymbol{\Delta}_b^{\sigma}$ only has $n_b$ nonzero elements, i.e., $(\boldsymbol{\Delta}_b^{\sigma})_{i,i}$ with $1\le i\le n_b$. As a result, we can rewrite the matrix multiplication in the inverse matrix as $\mathbf{U}_L^{\sigma}\boldsymbol{\Delta}_b^{\sigma}\mathbf{U}_R^{\sigma}\mathbf{T}^{\sigma}=\mathbf{V}(\mathbf{U}\mathbf{T}^{\sigma})$, where $\mathbf{V}$ and $\mathbf{U}$ are $N_{\sigma}\times n_b$ and $n_b\times N_{\sigma}$ matrices, respectively, constructed from the $n_b$ column and row vectors of $\mathbf{v}_{i}=(\boldsymbol{\Delta}_b^{\sigma})_{i,i}(\mathbf{U}_L^{\sigma})_{i{\rm-col}}$ and $\mathbf{u}_i^{\rm T}=(\mathbf{U}_R^{\sigma})_{i{\rm-row}}$ (with $1\le i\le n_b$) involved in the block update. Then the inverse matrix can be evaluated, using the matrix version of Sherman-Morrison formula in Eq.~(\ref{eq:ShermanMatrix}), as
\begin{equation}\begin{aligned}
\label{eq:BlkFrcBsApp}
&(\mathbf{1}_{N_{\sigma}}+\mathbf{U}_L^{\sigma}\boldsymbol{\Delta}_b^{\sigma}\mathbf{U}_R^{\sigma}\mathbf{T}^{\sigma})^{-1} \\
&\hspace{1.0cm} = [\mathbf{1}_{N_{\sigma}} + \mathbf{V}(\mathbf{U}\mathbf{T}^{\sigma})]^{-1} \\
&\hspace{1.0cm} = \mathbf{1}_{N_{\sigma}} - \mathbf{V}[\mathbf{1}_{n_b}+(\mathbf{U}\mathbf{T}^{\sigma})\mathbf{V}]^{-1}(\mathbf{U}\mathbf{T}^{\sigma}),
\end{aligned}\end{equation}
and thus proves Eq.~(\ref{eq:BlkFrcBsTnew1}). The subsequent upgrade of $\mathbf{T}^{\sigma}$ for the block update, which involves $n_{\rm F}$ flipped fields, follows similar formula as Eq.~(\ref{eq:BlkFrcBsApp}) but with a different diagonal matrix $\boldsymbol{\Delta}^{\sigma}$ that has only $n_{\rm F}$ nonzero elements.

\begin{table}[h]
\caption{The timing data for Fig.~\ref{fig:BlkFrcBsNbtest}(a), as the average time for update per sweep (in seconds) by the local and block-force bias updates versus various values of the block size $n_b$ and linear system size $L$, with $U/t=-1$ and $n=1.0$.}
\setlength{\tabcolsep}{9pt}
\centering 
\resizebox{\linewidth}{!}{   
\begin{tabular}{|c|c|c|c|c|}  
\hline
\diagbox{$n_b/L$}{$L$} & $8$ & $16$ & $24$ & $32$\\ 
\hline  
Local & 0.162 & 6.74 & 159 & 1910\\  
\hline  
$0.50$ & 0.331 & 7.73 & 113 & 1019\\    
\hline  
$1.00$ & 0.248 & 6.74 & 90.0 & 660\\  
\hline  
$1.50$ & 0.238 & 6.30 & 82.7 & 592\\  
\hline  
$2.00$ & 0.224 & 6.34 & 82.9 & 540\\  
\hline  
$2.50$ & 0.252 & 6.57 & 83.3 & 514\\  
\hline   
$3.00$ & 0.244 & 6.51 & 81.9 & 508\\  
\hline  
\end{tabular}
}
\label{Table:B1}
\end{table}

\begin{table}[h]
\caption{The timing data for Fig.~\ref{fig:BlkFrcBsNbtest}(b), as the average time for update per sweep (in seconds) by the local and block-force bias updates versus various values of the block size $n_b$ and linear system size $L$, with $U/t=-4$ and $n=1.0$.}
\setlength{\tabcolsep}{9pt}
\centering  
\resizebox{\linewidth}{!}{  
\begin{tabular}{|c|c|c|c|c|}  
\hline
\diagbox{$n_b/L$}{$L$} & $8$ & $16$ & $24$ & $32$\\ 
\hline  
Local & 0.150 & 6.09 & 156 & 1428\\   
\hline  
$0.50$ & 0.334 & 7.32 & 109 & 959\\ 
\hline  
$1.00$ & 0.246 & 6.50 & 85.6 & 611\\  
\hline  
$1.50$ & 0.236 & 6.12 & 78.6 & 545\\  
\hline  
$2.00$ & 0.219 & 6.02 & 78.6 & 492\\ 
\hline  
$2.50$ & 0.231 & 6.19 & 76.5 & 475\\  
\hline   
$3.00$ & 0.241 & 6.24 & 75.1 & 462\\  
\hline  
\end{tabular}
}
\label{Table:B2}
\end{table}

{\it Third}, for on-site Hubbard interaction applying HS-$\hat{s}^x$ [see Eq.~(\ref{eq:HSspinSxDep})], the corresponding $\boldsymbol{\Delta}$ matrix in Eq.~(\ref{eq:UpdateLTRSOC}) is not diagonal anymore, and evaluation of the inverse matrix $(\mathbf{1}_{N_e}+\mathbf{U}_L\boldsymbol{\Delta}\mathbf{U}_R\mathbf{T})^{-1}$ is a little bit more complicated. The framework for this spin-coupled case is presented in Appendix~\ref{sec:AppendixA}. For the local update, $\boldsymbol{\Delta}$ only has a nonzero $2\times 2$ submatrix $\boldsymbol{\delta}_i = \mathbf{A}(x_{\ell,i}^{\prime})[\mathbf{A}(x_{\ell,i})]^{-1}-\mathbf{1}_2$ with $\mathbf{A}(x_{\ell,i})$ computed from Eq.~(\ref{eq:A2times2}). Then the determinant ratio and the $\mathbf{T}$ matrix upgrade can be reformulated as Eq.~(\ref{eq:LocalSOC}), using the determinant property and matrix version of Sherman-Morrison formula in Eq.~(\ref{eq:ShermanMatrix}). Similarly, in the delayed update and block force-bias update, Eq.~(\ref{eq:ShermanMatrix}) is also applied to upgrade $\mathbf{T}$ matrix and evaluate the $(\mathbf{U}_{L}\mathbb{U}_{R})^{-1}$ matrix to compute the forca bias. 

\begin{table}[h]
\caption{The timing data for Fig.~\ref{fig:BlkFrcBsNbtest}(c), as the average time for update per sweep (in seconds) by the local and block-force bias updates versus various values of the block size $n_b$ and linear system size $L$, with $U/t=-6$ and $n=0.625$.}
\setlength{\tabcolsep}{9.5pt}
\centering  
\resizebox{\linewidth}{!}{
\begin{tabular}{|c|c|c|c|c|}
\hline
\diagbox{$n_b/L$}{$L$} & $8$ & $16$ & $24$ & $32$\\ 
\hline  
Local & 0.101 & 2.67 & 31.0 & 459\\  
\hline  
$0.50$ & 0.212 & 3.31 & 36.8 & 294\\   
\hline  
$1.00$ & 0.162 & 2.83 & 30.9 & 224\\  
\hline  
$1.50$ & 0.154 & 2.80 & 29.2 & 218\\  
\hline  
$2.00$ & 0.144 & 2.85 & 29.6 & 210\\  
\hline  
$2.50$ & 0.173 & 3.03 & 29.6 & 204\\ 
\hline   
$3.00$ & 0.153 & 3.13 & 30.9 & 202\\  
\hline  
\end{tabular}
}
\label{Table:B3}
\end{table}

\begin{table*}
\caption{The raw data for Fig.~\ref{fig:PhyResults}, as the benchmark results of double occupancy $D$, on-site spin-singlet pairing correlator $\langle\Delta^2\rangle$, and condensate fraction $n_c$, from PQMC simulations applying local, delayed, and block force-bias updates, for 2D attractive Hubbard model~(\ref{eq:2DHubbard}) with $L=20$. Two sets of the model parameters are considered, $(U/t=-4,n=1.0)$ and $(U/t=-6,n=0.625)$, for which the simulation parameters $(2\Theta t=46,\Delta\tau t=0.05)$ and $(2\Theta t=36,\Delta\tau t=0.04)$ are used, respectively. For the latter two update schemes, the data are shown as a function of $n_d$ (the predetermined delay rank in delayed update, with $n_d=8,16,32$) or $n_b$ (the block size in block force-bias update, with $n_b=10,20,40$).}
\setlength{\tabcolsep}{6pt}
\centering  
\resizebox{\linewidth}{!}{
\begin{tabular}{|cl||ccc||ccc|}
\hline
\multicolumn{2}{|c||}{\multirow{2}{*}{Observables}} & \multicolumn{3}{c||}{$U/t=4,n=1.0$}  &    \multicolumn{3}{c|}{$U/t=6,n=0.625$}             \\ \cline{3-8} 
\multicolumn{2}{|c||}{}                     & \multicolumn{1}{c|}{$D$} & \multicolumn{1}{c|}{$\langle\Delta^2\rangle$} & $n_c$ & \multicolumn{1}{c|}{$D$} & 
\multicolumn{1}{c|}{$\langle\Delta^2\rangle$} & $n_c$ \\ \hline
\multicolumn{2}{|c||}{Local}                & \multicolumn{1}{c|}{0.3726(5)} & \multicolumn{1}{l|}{0.02226(9)} & 0.1052(4) 
& \multicolumn{1}{c|}{0.2359(1)} & \multicolumn{1}{l|}{0.04173(6)} &  0.2920(4)\\ \hline
\multicolumn{1}{|c|}{\multirow{3}{*}{}} & $n_d=8$ & \multicolumn{1}{c|}{0.3725(6)} & \multicolumn{1}{l|}{0.02213(10)} & 0.1053(6) 
& \multicolumn{1}{c|}{0.2358(2)} & \multicolumn{1}{l|}{0.04167(9)} &  0.2916(6)\\ \cline{2-8} 
\multicolumn{1}{|c|}{Delayed}             & $n_d=16$ & \multicolumn{1}{c|}{0.3724(6)} & \multicolumn{1}{l|}{0.02226(12)} & 0.1052(6) 
& \multicolumn{1}{c|}{0.2361(2)} & \multicolumn{1}{l|}{0.04178(8)} &  0.2925(6)\\ \cline{2-8} 
\multicolumn{1}{|c|}{}                  & $n_d=32$ & \multicolumn{1}{c|}{0.3730(6)} & \multicolumn{1}{l|}{0.02232(9)} & 0.1055(5) 
& \multicolumn{1}{c|}{0.2361(2)} & \multicolumn{1}{l|}{0.04180(6)} &  0.2926(5)\\ \hline
\multicolumn{1}{|c|}{\multirow{3}{*}{Block force-bias}} & $n_b=10$ & \multicolumn{1}{c|}{0.3718(5)} & \multicolumn{1}{l|}{0.02228(12)} & 0.1045(5) 
& \multicolumn{1}{c|}{0.2359(2)} & \multicolumn{1}{l|}{0.04174(9)} &  0.2921(7)\\ \cline{2-8} 
\multicolumn{1}{|c|}{}             & $n_b=20$ & \multicolumn{1}{c|}{0.3736(8)} & \multicolumn{1}{c|}{0.02247(14)} & 0.1062(7) 
& \multicolumn{1}{c|}{0.2358(2)} & \multicolumn{1}{l|}{0.04171(10)} &  0.2918(7)\\ \cline{2-8} 
\multicolumn{1}{|c|}{}                  & $n_b=40$ & \multicolumn{1}{c|}{0.3739(5)} & \multicolumn{1}{l|}{0.02255(11)} & 0.1066(6) 
& \multicolumn{1}{c|}{0.2361(2)} & \multicolumn{1}{l|}{0.04175(8)} &  0.2922(5)\\ \hline
\end{tabular}
}
\label{Table:C1}
\end{table*}

{\it Fourth}, for the general long-range interactions that are usually studied in {\rm ab initio} quantum chemistry and realistic materials~\cite{Shihao2021}, the $\boldsymbol{\Delta}^{\sigma}$ matrix in Eqs.~(\ref{eq:DetRatioApp}) and (\ref{eq:UpdateTApp}) or the $\boldsymbol{\Delta}$ matrix in Eqs.~(\ref{eq:DetRatioSOC}) and (\ref{eq:UpdateLTRSOC}) involved in the update process can be typically characterized by a nonzero submatrix $\boldsymbol{\delta}$ of dimension $d\times d$. Then the calculation of the inverse matrix $(\mathbf{1}_{N_{\sigma}}+\mathbf{U}_L^{\sigma}\boldsymbol{\Delta}^{\sigma}\mathbf{U}_R^{\sigma}\mathbf{T}^{\sigma})^{-1}$ or $(\mathbf{1}_{N_e}+\mathbf{U}_L\boldsymbol{\Delta}\mathbf{U}_R\mathbf{T})^{-1}$, as well as the determinant ratio and upgrading $\mathbf{T}^{\sigma}$ or $\mathbf{T}$ matrix can be tranformed into the expressions similar to that in Eq.~(\ref{eq:BlkFrcBsApp}). 

In both the full and block force-bias update schemes, we need to evaluate the ratio $\langle\psi_l|\hat{B}_I(\mathbf{x})|\psi_r\rangle/\langle\psi_l|\psi_r\rangle$ [see Eqs.~(\ref{eq:RatioApprox}) and (\ref{eq:BlkFrcBsErr})] to get the prior probability. Denoting the operator $\hat{v}={\rm i}\sum_{i=1}^{N_s} x_{\ell,i}(\hat{n}_{i\uparrow}-\hat{n}_{i\downarrow})$, we can apply the Taylor expansion for the $\hat{B}_I(\mathbf{x})$ operator as 
\begin{equation}\begin{aligned}
\hat{B}_I(\mathbf{x}_{\ell})
&= {\rm exp}\Big[{\rm i}\gamma\sum_{i=1}^{N_s} x_{\ell,i}(\hat{n}_{i\uparrow}-\hat{n}_{i\downarrow})\Big] = e^{\gamma\hat{v}} \\
&= 1 + \gamma\hat{v} + \frac{\gamma^2\hat{v}^2}{2} + \mathcal{O}(\gamma^3),
\end{aligned}\end{equation}
which results in the ratio as
\begin{equation}\begin{aligned}
\frac{\langle\psi_l|\hat{B}_I(\mathbf{x})|\psi_r\rangle}{\langle\psi_l|\psi_r\rangle}
= 1 + \gamma\langle\hat{v}\rangle+\frac{\gamma^2}{2}\langle\hat{v}^2\rangle + \mathcal{O}(\gamma^3),
\end{aligned}\end{equation}
where we denote the expectation $\langle\hat{o}\rangle=\langle\psi_l|\hat{o}|\psi_r\rangle/\langle\psi_l|\psi_r\rangle$. Besides, we also have 
\begin{equation}\begin{aligned}
e^{\gamma\langle\hat{v}\rangle} = 1 + \gamma\langle\hat{v}\rangle + \frac{\gamma^2\langle\hat{v}\rangle^2}{2} + \mathcal{O}(\gamma^3).
\end{aligned}\end{equation}
Combining these two relations, we can achieve
\begin{equation}\begin{aligned}
&\frac{\langle\psi_l|\hat{B}_I(\mathbf{x})|\psi_r\rangle}{\langle\psi_l|\psi_r\rangle}
= e^{\gamma\langle\hat{v}\rangle} + \frac{\gamma^2}{2}\big(\langle\hat{v}^2\rangle - \langle\hat{v}\rangle^2\big) + \mathcal{O}(\gamma^3) \\
&= e^{\gamma\langle\hat{v}\rangle} - (\Delta\tau U)\frac{\langle\hat{v}^2\rangle - \langle\hat{v}\rangle^2}{2} + \mathcal{O}[(-\Delta\tau U)^{3/2}],
\end{aligned}\end{equation}
where $\gamma=\cos^{-1}(e^{\Delta\tau U/2})=\sqrt{-\Delta\tau U}+\mathcal{O}[(-\Delta\tau U)^{3/2}]$ is used. The above equation thus proves Eq.~(\ref{eq:RatioApprox}), and similarly Eq.~(\ref{eq:BlkFrcBsErr}). The factor $f_{\mathbf{x}}=(\langle\hat{v}^2\rangle-\langle\hat{v}\rangle^2)/2$ is half the variance of the $\hat{v}$ operator, and it explicitly depends on the field configuration. It is evident that, $f_{\mathbf{x}}$ should increase with the system size $N_s$ due to the summation involved in the $\hat{v}$ operator. This growth contributes to the overall suppression of the acceptance ratio as a function of $N_s$ (and also $\Delta\tau|U|$) in the full force-bias update. Furthermore, since $f_{\mathbf{x}}$ formally behaves as a correlation function, it is also expected to depend on $U$, potentially leading to a suppression of the acceptance ratio that is faster than linear. In contrast, in the block force-bias update scheme, the block factor $f_{\mathbf{x}}^{(b)}$ in Eq.~(\ref{eq:BlkFrcBsErr}) explicitly depends on $n_b$ instead of $N_s$. Thus, with a fixed $n_b$ parameter, $f_{\mathbf{x}}^{(b)}$ should possess a weak (or even negligible) size dependence with increasing $N_s$. Consequently, the overall acceptance ratio of the block force-bias update is mainly controlled by $n_b$, and it can maintain almost constant with fixed $n_b$ versus $N_s$. 

\section{The CPU information and settings of PQMC simulations}
\label{sec:AppendixD}

All the test results from PQMC simulations presented in Sec.~\ref{sec:Results} are conducted on Intel CPUs, whose configuration is [2$\times$Intel Xeon Gold 6132 (2.6GHz, 14 cores, 20MB Intel Cache), memory 96 GB]. We have also performed the efficiency tests of both the delayed update and block force-bias update on AMD CPUs, whose configuration is [2$\times$AMD 7543 32C (2.8GHz, 32 cores, 48MB Intel Cache), memory 256 GB]. Our results show that, although the absolute runtime differs between the two CPUs, the speedups relative to the local update are very similar. This indicates that the performance gains from these new update schemes are robust and largely independent of the specific hardware used.

For the simulations, our PQMC code is written with FORTRAN 90 language, and all the matrix and vector calculations are performed using Intel MKL, mostly the subroutines in LAPACK. The vector-vector outer product (ZGERU) is used in local update, while the matrix-matrix multiplication (ZGEMM) is used in the delayed update and block force-bias update. Additionally, evaluating the matrix inverse and the determinant (ZGETRF) appears in the full force-bias update. Moreover, all the FFT calculations are performed with the FFTW library (also integrated in Intel MKL).

\section{Timing data for Figs.~\ref{fig:Delayedndtest} and \ref{fig:BlkFrcBsNbtest}, and raw data for Fig.~\ref{fig:PhyResults}}
\label{sec:AppendixE}

In this appendix, we summarize all the timing data as plotted in Figs.~\ref{fig:Delayedndtest} and \ref{fig:BlkFrcBsNbtest}, and present the raw data for Fig.~\ref{fig:PhyResults}, which might be used as benchmark data for future studies.

First, in Tables~\ref{Table:A1} and \ref{Table:A2}, we present the consumed time data shown in Figs.~\ref{fig:Delayedndtest}(a) and \ref{fig:Delayedndtest}(b), respectively, as the comparison between the local and delayed updates in PQMC simulations for the Hubbard model~(\ref{eq:2DHubbard}) with $(U/t=-4,n=1.0)$ and $(U/t=-6,n=0.625)$.

Second, in Tables~\ref{Table:B1}, \ref{Table:B2}, and \ref{Table:B3}, we present the consumed time data shown in Figs.~\ref{fig:BlkFrcBsNbtest}(a), \ref{fig:BlkFrcBsNbtest}(b), and \ref{fig:BlkFrcBsNbtest}(c), respectively, as the comparison between the local and block force-bias updates in PQMC simulations for the Hubbard model~(\ref{eq:2DHubbard}) with $(U/t=-1,n=1.0)$, $(U/t=-4,n=1.0)$ and $(U/t=-6,n=0.625)$.

Third, in Table~\ref{Table:C1}, we summarize the raw data for Fig.~\ref{fig:PhyResults}, as the benchmark results of double occupancy $D$, on-site spin-singlet pairing correlator $\langle\Delta^2\rangle$, and condensate fraction $n_c$, from PQMC simulations applying local, delayed, and block force-bias updates, for 2D attractive Hubbard model~(\ref{eq:2DHubbard}) with $(U/t=-4,n=1.0)$ and $(U/t=-6,n=0.625)$. These results are for $L=20$.

\bibliography{PQMCdelayRef}

\begin{thebibliography}{87}
\bibitem [{Sub()}]{SubleadNote}%
  \BibitemOpen
  \href@noop {} {}\bibinfo {note} {To evaluate the vectors $\mathbf{x}_k$ and
  $\mathbf{y}_k^{\rm T}$ in Eq.~(\ref{eq:DelayRgenk}), we need to compute the
  quantities $c_m=(\mathbf{u}_{i_k}^{\rm T}\mathbf{x}_m)$,
  $d_m=(\mathbf{y}_m^{\rm T}\mathbf{v}_{i_k})$, and $c_m\mathbf{y}_m^{\rm T}$,
  $d_m\mathbf{x}_m$. Each of these operations requires
  $\mathcal{O}(N_{\sigma})$ computational cost for a given $m$. Summing over
  all $m$ with $1 \le m \le k-1$, the total cost amounts to
  $\mathcal{O}[4(k-1)N_{\sigma}]$. The additional computation of $\sum_m c_m
  d_m$ only costs $\mathcal{O}(k-1)$ and is therefore neglected. Then, in the
  first strategy which sets $n_d$ as the total number of proposed update moves
  in a single-step delayed update, the index $k$ ranges from $1$ to $n_{\rm A}$
  (the number of accepted update moves), and thus the overall computational
  cost reads $\sum_{k=1}^{n_{\rm A}}\mathcal{O}[4(k-1)N_{\sigma}]\simeq
  \mathcal{O}[2n_{\rm A}(n_{\rm A}-1)N_{\sigma}]$. In the second strategy to
  set $n_d$ as the accumulated number of accepted moves in a single-step
  delayed update, the overall computational cost is accordingly
  $\mathcal{O}[2n_d(n_d-1)N_{\sigma}]$.}\BibitemShut {Stop}%
\end{thebibliography}
\end{document}